# Measurements of molecular size and shape on a chip

Xin Zhu[1]†, Timothy J. D. Bennett[1]†, Konstantin C. Zouboulis[1,2], Dimitrios Soulias[1], Michal Grzybek[3], Justin L. P. Benesch[1,2], Afaf H. El-Sagheer[4,5], Ünal Coskun[3,6-8], Madhavi Krishnan[1,2]*

[1]Physical and Theoretical Chemistry Laboratory, Department of Chemistry, University of Oxford; South Parks Road, Oxford OX1 3QZ, United Kingdom

[2]The Kavli Institute for Nanoscience Discovery, University of Oxford; Sherrington Road, Oxford OX1 3QU, United Kingdom

[3]Center of Membrane Biochemistry and Lipid Research, University Hospital and Faculty of Medicine Carl Gustav Carus, Technical University Dresden; 01307 Dresden, Germany

[4]School of Chemistry and Chemical Engineering, University of Southampton; Highfield, Southampton, SO17 1BJ, United Kingdom

[5]Institute for Life Sciences, University of Southampton; Highfield Campus, Southampton, SO17 1BJ, United Kingdom

[6]Paul Langerhans Institute Dresden (PLID) of the Helmholtz Center Munich, University Hospital and Faculty of Medicine Carl Gustav Carus, Technical University Dresden; 01307 Dresden, Germany

[7]German Center for Diabetes Research (DZD), 85764 Neuherberg, Germany

[8]Max Planck Institute of Molecular Cell Biology and Genetics, 01307 Dresden, Germany

*Corresponding author. Email: madhavi.krishnan@chem.ox.ac.uk

† These authors contributed equally to this work

Size and shape are critical discriminators between molecular species and states. We describe a micro-chip based high-throughput imaging approach offering rapid and precise determination of molecular properties under native solution conditions. Our method detects differences in molecular weight across at least three orders of magnitude, and down to two carbon atoms in small molecules. We quantify the strength of molecular interactions over six orders of magnitude in affinity constant, and track reactions in real-time. Highly parallel measurements on individual molecules serve to characterize sample-state heterogeneity at the highest resolution, offering predictive input to model three-dimensional structure. We further leverage the method's structural sensitivity for diagnostics, exploiting ligand-induced conformational changes in the insulin receptor to sense insulin concentration in serum at the sub-nanoliter and sub-zeptomole scale.

**Main text**

The pursuit of high-resolution biomolecular characterization has driven substantial advances in novel analytical technology development (*1-10*). However, most methods entail the use of strong electrical or optical fields, immobilization and/or tethering of molecules to surfaces, or of matrices. All these factors can perturb the integrity and function of biomolecules and their complexes whose structure and conformation arise from the concerted action of weak inter- and intramolecular forces (*11, 12*). Therefore, characterization methods in the solution phase that offer high-precision readouts whilst preserving molecular structure and properties remain highly sought after.

The friction associated with an object moving through a fluid carries a clear signature of its volume and shape, and causes larger objects to travel a shorter distance on average in a given period of time. The Stokes' or hydrodynamic radius, $r_H$, quantifies this property, and has been extensively used to characterize molecular species in solution (*13-16*). The shape of an object may be further captured in terms of its compactness given a fixed volume, as reflected in the diameter of the smallest sphere, $D_s$, enveloping it. This quantity shares similarities with the maximum spatial extent of the molecule, $D_{max}$, encountered in small-angle x-ray scattering (SAXS) (*17*) (Fig. 1C). Indeed, workhorse laboratory techniques such as size-exclusion chromatography exploit this molecular property in order to separate and sieve mixtures. Precise measurement of the size and shape of molecules – "molecular stereometry" – achieved by direct observation of molecular motion in a suitably tailored landscape thus offers the prospect of highly simplified structure-sensitive readouts traditionally believed to lie exclusively within the realm of high-resolution techniques (Fig. 1A) (*18, 19*).

In this work we leverage the combined advantages of random thermal motion and size-dependent confinement effects at the nanoscale to achieve high measurement sensitivity to molecular conformation under native conditions in the solution phase. Spatially constraining the vertical motion of an object diffusing in a geometrically tailored slit can dramatically enhance the time the object spends in local, deeper, 'pocket' or trap regions, in a size-dependent fashion (Fig. 1A, B, Movie S1). The larger the object, or the greater the degree of confinement in the parallel-plate slit region, the longer it resides in the trap. Engineering the number of states accessible to a molecule in the slit – e.g., translational, configurational and conformational – entropy in its various forms can be exploited at the nanoscale to significantly amplify the impact of size and shape on molecular exit (escape) times. Low-power, wide-field imaging of trapped molecules using a large-area camera detector confers on the method both speed and measurement precision (Movie S1). The ability to tune the magnitude of the measured response using device geometry offers unprecedented sensitivity to molecular conformational properties, enabling the technique to outperform traditional solution-phase methods that rely solely on hydrodynamic friction, as will be demonstrated in this work (*16, 20, 21*). The technique offers wide and tailorable dynamic range in molecular weight readouts from 500 Da to at least 500 kDa, quantitative sensitivity to molecular concentration in solution, the ability to process complex samples, and read out reaction rates and equilibrium constants, all in a single platform, thus addressing a hitherto unmet integration challenge. We term the approach escape-time stereometry (ETs) and demonstrate the broad ramifications of the method's core working element – the 'entropic fluidic trap'– in a range of analytical application areas, including molecular diagnostics (*22*).

**Spatial tailoring of translational entropy enables high throughput measurements and molecular counting**

In our experiments we introduce fluorescent molecules at a concentration of 10 fM–10 nM suspended in phosphate buffered saline (PBS, ionic strength ca. 160 mM) into a series of parallel-plate slits of height $h_1$ that carry arrays of periodic cylindrical indentations of total height $h_2$ (Fig. 1A, B). Similar to previous work, the "pocket" nanostructures act as thermodynamic traps for single molecules in solution (Fig. 1B) (*3, 18, 19, 23*). In contrast to previous work, the use of high salt concentrations renders charge-related electrostatic effects negligible, and the trap is expected to be effectively purely entropic (*22, 24*) (section S1.4, fig. S5). We image the diffusive dynamics of molecules, labelled on average with a single fluorophore, using standard wide-field microscopy (Fig. 1A) (*3*). We measured average escape times, $t_{esc}$, for a variety of molecules from small organic fluorophores to disordered and globular proteins, nucleic acids and protein complexes by imaging the chip for about one minute in each case (Fig. 1D, E and 2A). Typical imaging conditions entail exposure times $t_{exp} = 5$ ms and an imaging rate of 100 Hz, unless stated otherwise (Materials and Methods). For low copy-number counting experiments (fig. S5; section S1.3) and for experiments on the insulin receptor discussed later, slit surfaces were passivated by coating with polyethylene glycol to minimize non-specific molecular adhesion (section S1.2).

For a small sphere whose radius $R \approx r_H \ll h_1$, we have $t_{esc} \propto r_H(h_2/h_1)$, which illustrates that the height modulation in the slit enhances molecular residence times by a factor $h_2/h_1$ compared to the free-diffusion value (red line, Fig. 2C, fig. S8, section S3). This entropic enhancement of residence times in the pocket region is crucial to the method since it fosters high throughput direct imaging of the escape process on the order of 100–1000 molecules in a minute or less of measurement, using both low incident optical power densities (~5 W/mm$^2$) and a large-area detector such as a camera chip. Our statistically dominated measurement imprecision depends on $N$, the number of escape events observed, as $1/\sqrt{N}$. Large values of $N = 10^4 - 10^5$ achieved using highly parallel observation of the escape process at a molecular concentration of about 100 pM, implies $\leq 1\%$ measurement imprecision on $r_H$, and is key to high precision measurement (fig. S7, section S1.1). We currently achieve quantitative detection of a labelled species down to ~10 fM in a 10 min continuous-flow measurement, with opportunity for improvement by at least a factor 100 through optimization of device design (fig. S5).

Measuring $t_{esc}$ on a range of molecules spanning 3 orders in molecular weight, we noted the scaling $t_{esc} \propto r_H \propto M^{1/3}$ as expected for globular molecules, which implies that $< 1\%$ imprecision on $t_{esc}$ translates to $< 3\%$ measurement imprecision in molecular weight (Fig. 2A). By contrast, for disordered proteins, e.g., Prothymosin-α (ProTα) and Starmaker-like protein (Stm-l) we measure larger escape times that are in agreement with the scaling relationship $t_{esc} \propto r_H \propto M^{3/5}$ characteristic of unstructured polymers (Fig. 2A, fig. S9) (*25-27*).

Examining escape times for a range of small-molecule chemical derivatives of the fluorophore ATTO532 (molecular weight, $M \lesssim 1$ kDa), we note that ca. 0.5% measurement imprecision on the escape time, $t_{esc}$, implies the theoretical ability to detect differences of about a single carbon atom between molecular species as shown in Fig. 2B (fig. S9). We demonstrate the ability to clearly distinguish a molecular weight difference of 25 Da between the N-hydroxysuccinimide (NHS) and maleimide derivatives of the fluorophore ATTO532 (fig. S1, fig. S9). The NHS ester of ATTO532 hydrolyses in water to form carboxyl-ATTO532 and a leaving NHS group, with the rate of hydrolysis strongly accelerated under alkaline conditions (*28*). Continuous-flow monitoring of the escape process for an aqueous solution of ATTO532-NHS ($M = 743$ Da, fig. S1) at pH 7 and pH 9 performed as a series of intermittent measurements of 30 sec duration yielded average escape times $t_{av}$ over all recorded escape events of duration $\Delta t$ (Fig. 2B, right panel). A systematic overall reduction of 3% in $t_{av}$ measured over the course of ca. 10 minutes implies a mass change of about 100 Da, commensurate with that of the NHS leaving group. The measured decay rate constant of 2.3 $\times 10^{-3}$ s$^{-1}$ at pH 9 corresponds well to reported half-life values of approximately 5 min (section S4.2) (*28*). Real-time tracking of a change in molecular weight due to hydrolysis illustrates the power of both the speed and precision of the approach.

**Nanoscale confinement offers non-linear sensitivity of the readout to molecular size and shape**

The measured $t_{esc}$ values can be used to quantitatively infer molecular hydrodynamic radii, $r_H$, and bounding sphere diameters, $D_s$. In order to do so we performed Brownian Dynamics (BD) simulations of the escape process for spheres of radius $r_H$ and bounding sphere diameter $D_s = 2r_H$ as described previously (section S5.1) (*29*). Simulated average escape times indeed display good agreement with the simple model outlined in the supplementary text (section S3) and are captured by the following equation:

$$t_{esc} = Ar_H(h_2 - D_s)/(h_1 - D_s) + t_o \tag{1}$$

with the coefficient $A \approx 0.3$ s/µm accounting for the viscosity, device geometry and various aspects of the imaging-based detection process, and an offset $t_o = 5.83$ ms $\approx t_{exp}$ (fig. S8). Thus, in a typical experiment, we first use globular proteins of known $r_H$ to determine $h_1$ from fits of the data to Eq. (1), and then use this calibration function to convert the measured $t_{esc}$ for a test molecular species to hydrodynamic radius $r_H$ (fig. S10, section S5.3). Inferred $r_H$ values for a set of globular proteins measured in slits of calibrated height $h_1 = 23.8$ nm agree well with values computed based on molecular structures using the program HYDROPRO (*14*) (Fig. 2D, table S4, section S5.2).

Eq. (1) does indicate that, owing to rotational diffusion, molecules of non-spherical shape may exhibit a further enhancement of $t_{esc}$ compared to their spherical counterparts of similar hydrodynamic radius. Since rotational times are generally much faster than translational diffusion, we may assume that a molecule that rotates isotropically while translating sweeps out a local volume defined by a sphere of minimum diameter $D_s > 2r_H$ as shown in Fig. 1C. The implications of this high sensitivity to the molecular envelope are illustrated in our ability to clearly distinguish differences of 1 bp (ca. 3.4 Å) between rod-shaped double-stranded (ds) DNA fragments in the size range $n_{bp} = 56 - 60$ bp (ca. 20 nm in length) (Fig. 2E).

To test this ability further we performed measurements on dsDNA and dsRNA of length $n_{bp} = 30$ to 60 basepairs, entailing lengths in the range 7 to 21 nm (section S5.4). Performing measurements of $t_{esc}$ in slits of depth $h_1 \approx 25$ nm we found a progressive non-linear increase in $t_{esc}$ with $r_H$ values of the molecular species compared to the expectations for spheres of the same hydrodynamic radius $r_H$ (fig. S11). Fitting the measured $t_{esc}$ vs. $n_{bp}$ data using Eq. (1), assuming $D_s = bn_{bp}$, and using theoretical expressions for $r_H$ for cylinders we obtained a rise per basepair $b \approx 3.2$ Å for the B-form double helix characteristic of dsDNA and $b \approx 2.3$ Å for the A-form double helix expected for dsRNA (Fig. 3A, fig. S11). These values are remarkably close to those inferred from high resolution structural measurements and previous solution phase structural studies (3.32 $\pm$ 0.19 Å for B-DNA and 2.3 Å for A-DNA) (*18, 30*).

**Inferences on DNA-nanostructure size and shape**

Since we expect ETs to offer strong and tunable shape-based discrimination of molecules that may be effectively identical in all other relevant respects (e.g., in hydrodynamic radius, mass and charge), we studied DNA nanostructures that carry the same number of basepairs but differ in 3D conformation (section S5.5) (*23, 31, 32*). We considered DNA nanostructures composed of 240 bases in states referred to as 'bundle' and 'square-tile' as previously described (*23*) (Fig. 3B), where $D_s \approx 20$ nm for the bundle provides an estimate of the size regime of the measurement. The 'bundle' and 'tile' are indistinguishable in diffusion coefficient measurements by FCS (table S5) (*23*). We performed $t_{esc}$ measurements on both species in slits of height $h_1 \approx 100, 70$ and $40$ nm (Fig. 3B). The relative insensitivity of $t_{esc}$ to molecular conformation in the regime $h_1 \geq 70$ nm largely reflects a response expected based on pure diffusion alone. However, we observe systematically increasing $t_{esc}$ values, with disparities between the two DNA nanostructures reaching a factor 1.2 as $h_1$ decreases and approaches $D_s$. The measurement mode may thus be tuned to enhance the sensitivity of the escape-time readout to molecular 3D conformation, reflected in $D_s$, as suggested by Eq. (1).

According to Eq. (1), measurements performed using two different values of $h_1$ ought to support a readout of two properties of the molecule, namely $r_H$ and $D_s$. We used $t_{esc}$ measurements in slits of two different mean calibrated height values of $h_1 = 41.4$ nm and $73.9$ nm in order to infer these conformational parameters for the 'bundle' and 'tile' (fig. S12). Fig. 3C provides a graphic representation of the procedure behind such an inference. A pair of independent measurements using sufficiently different $h_1$ values yields two curves on a plot of $D_s$ vs. $r_H$ whose point of intersection reflects the $r_H$ and $D_s$ values describing the molecular state. Comparing our inferences with values expected from molecular structural models obtained from coarse grained MD simulations using oxDNA, and verified using SAXS (*32, 33*), indeed reveals good mutual agreement (table in Fig. 3C, fig. S12). For example, whilst the measured $r_H$ values for the 'tile' and 'bundle' are rather similar, their $D_s$ values are different and in fact appear to be entirely responsible for the measured disparity between the two species.

**Resolving mixtures of molecular conformational states by constructing single-molecule spectra**

Escape time data examined at ensemble level as shown in Figs. 1-3 can reveal multiple escape timescales, depending on the sample, (fig. S6; section S2.2). A multitude of timescales generally contains important information on the sample, reflecting, e.g., heterogeneous states of multimerization, or distinguishable 3D conformation states in monomeric molecules. Extracting these time components and their abundances in ensemble measurements can under certain conditions be subject to mathematical difficulties (*34*). Accessing information at the level of individual molecules not only avoids potential ambiguities of multi-exponential fitting, but can also furnish insight into the nature of the underlying states, e.g., whether they stem from individual molecules characterised by distinct 3D conformations that are stable over the observation window, or rather from faster interconversion between different conformational states. Because we observe the hopping dynamics of every individual molecule in the microscope's field of view, we may indeed construct spectra of mean escape times reflecting single-molecule properties in a mixture of different molecular species or states as shown in Fig. 4. Performing measurements at low sample concentrations ( $\sim$10 pM) permits facile recognition of trajectories arising from individual molecules whose average escape time now describes the conformational properties of individual molecular states (Fig. 4A, Movie S2, section S6.1).

We first validated the ability to identify, distinguish and quantify stable single-molecule states by examining an equimolar mixture of 24 bp and 60 bp DNA: two geometrically highly well-defined and conformationally stable species (Fig. 4B, fig. S16; section S6.2). We constructed single-molecule spectra by recording the average escape time $t_{av}$ for a number of single-molecule hopping trajectories each consisting of $N_{hop}$ escape events of varying duration $\Delta t$. Each $t_{av}$ measurement in the resulting molecular escape spectrum thus represents an average value originating from a separate molecule. At present we routinely record about $N_{\text{hop}} = 200 - 300$ hops from a molecule before either the label photobleaches or the molecule diffuses out of the field of view. Such a measurement implies a percentage measurement imprecision on $t_{\text{av}}$ approaching $1/\sqrt{N_{\text{hop}}} \approx 5 - 7\%$ on a single molecule. This quantity also reflects the fractional standard deviation, $\sigma/\mu$, of Gaussian distributed $t_{\text{av}}$ values with mean $\mu$, measured for a number of single molecules, $N_{\text{mol}}$, (as shown in Fig. 4B and D) and governs the ability of the method to resolve two closely separated molecular escape-time states (Fig. 4B, fig. S15). On the other hand, the measurement imprecision on mean $t_{\text{av}}$, is approximated well by the standard error on the mean (s.e.m.), $\sigma/\sqrt{N_{\text{mol}}}$, and can be minimized by measuring a large number of single-molecule trajectories (section S6.1). A wider field-of-view combined with a lower incident power density will support recording of longer single-molecule tracks ($N_{\text{hop}} > 10^3$) which could take the resolution $\sigma$ to ~1-2% (*35*), simultaneously enhancing both molecular throughput ($N_{\text{mol}}$), which currently stands at a few hundred molecules, as well as measurement precision (Fig. 4D). Trajectory-based measurement of the mixture of dsDNA species, indeed displayed two peaks in the molecular escape-time spectrum at mean molecular $t_{\text{av}}$ values given by $\mu_1$ and $\mu_2$ of fractional abundances $m_1 \approx m_2 \approx 50\%$, that were in agreement with the corresponding values obtained from a bi-exponential fit of the same data in the ensemble approach (bottom row, Fig. 4B).

*3D conformation of a riboswitch in aqueous solution*

Like proteins, RNA can fold into well-defined 3D structures to carry out a wide range of cellular functions, but structure determination and prediction for RNA has significantly lagged behind their protein counterparts (*36*). We used ETs to examine the solution phase 3D conformation of the SAM-IV riboswitch, whose structure has been recently solved by cryogenic electron microscopy (cryo-EM) (figs. S13, S17 and S18; section S5.6) (*37*).

Rather than a single timescale corresponding to the predominant cryo-EM structure (70% abundance, Fig. 4C), we found that both ensemble-averaged measurements and single molecule spectra of the fluorescently labelled SAM-IV riboswitch, revealed three distinct escape timescales (Fig. 4D-F, fig. S18). Examination of individual molecular trajectories revealed average molecular escape times falling into three distinct classes over the ca. 6 sec measurement window, indicative of stable molecular conformations rather than rapid interconversion between states (Fig. 4 D-F, fig. S17). In order to benchmark the observed timescales, we performed control measurements on free fluorescent dye, 30 bp dsDNA and a 119 nt single-stranded (ss) DNA all of which entail little or no secondary structure (Fig. 4C, D, lower panels; fig. S18, section S6.3). The smallest timescale, $\mu_1 \approx t_1 = 10$ ms, in both the single-molecule spectrum and the ensemble measurements of SAM-IV sample can be attributed to free fluorescent dye molecules in the mixture which are generally difficult to remove entirely. However, we attribute the second and third measured timescales to two different conformational states, or similar groups of states, of the SAM-IV molecule – one compact and the other more extended. The larger of the molecular timescales, $\mu_3$, is similar to the timescale measured for 30 ds DNA whose value of $D_s \approx 9.6$ nm in fact closely resembles that of the reported structure (*37*) (Fig. 4D). The smaller of the two observed SAM-IV timescales, $\mu_2$, in turn is indicative of a conformation of ~60% abundance, suggesting an overall more compact and globular state. All mean timescales at the ensemble- and single molecule-level are in good mutual agreement, i.e., $t_i \approx \mu_i$ (Fig. 4G). Fig. 4G displays a summary of the conformation analysis where the main conclusions are that (1) the measurements $t_3 \approx \mu_3$ (blue symbol and shaded zone in Fig. 4G) are consistent with that of an oblate ellipsoid characterized by $D_s$ and $r_H$ values that are very similar to those computed for the cryo-EM structure (black cross in Fig. 4G), and (2) the observation $t_2 \approx \mu_2$ indicates a conformation characterized by an ellipsoid of smaller $D_s$ and $r_H$ than the reported structure (red symbols and shaded region).

Upon addition of a high concentration of the SAM ligand, the escape time spectra remained comparatively unchanged indicating the absence of a conformational change in the riboswitch upon ligand binding (fig. S18). This observation is in line with cryo-EM data as well as with observations on other riboswitch aptamers, e.g., SAM-I and lysine riboswitches that do not change conformation upon ligand-binding (*38, 39*).

**Measuring binding affinities and rates of intermolecular association and dissociation**

*Nucleic acid interactions*

We next explore the capability of our approach in making rapid, quantitative measurements of intermolecular interactions in mixtures, as reflected in measured dissociation constants, $K_d$, as well as of on- and off-rates of binding – $k_{on}$ and $k_{off}$ (Fig. 5; sections S7.2 and S7.3). DNA hybridization provides a simple tuneable system where intermolecular binding affinities can be varied substantially using single basepair increments between complementary stretches, supporting comparisons of measurements not only with independent experimental techniques but also with theoretical estimates. We validated the measurement approach using fluorescent molecular 'bait' represented by short strands of ATTO532-labelled ssDNA ranging in length from 7-15 bases hybridizing to the corresponding complementary sequence at the 5' end of an approximately 200 base long "scaffold" of unlabelled ssDNA (Fig. 5A, B). Mixtures of the labelled oligo at a concentration of 0.02-1 nM and unlabelled DNA at various concentrations are incubated to equilibrium, loaded into the trap landscape and escape times recorded over a period of 1-5 min. Histograms of event durations reveal two dominant timescales: a short $t_1 \approx 10$ ms corresponding to the free labelled oligo and a longer $t_2 \approx 30$ ms corresponding to the oligo bound to the 200-base ssDNA (Fig. 5A). The fractional abundances of the two species $m_{1,2}$ are determined using bi-exponential fits and plotted as a function of the concentration of the unlabelled binding partner to determine the $K_d$ value of the interaction (upper panel, Fig 5B; section S7.1, S7.2 and S7.5).

We observed a systematic increase in binding affinity with increasing oligo length, reflected in a progressively decreasing measured $K_d$ values, in good agreement with previous reports (Fig. 5D; section S7.5) (*40*). Although samples are incubated over timescales ranging from several hours to several days in order to ensure that the binding reaction reaches equilibrium, the $t_{1,2}$ measurements corresponding to one point on a plot displayed in Fig. 5B takes about 1-5 min over the range of concentrations of unlabelled binding partner probed (fig. S19; section S7.3). Furthermore, the speed of the readout can be harnessed to perform real-time monitoring of the sample following either the mixing of the two unbound species, or alternatively, following the addition of an unlabelled "chaser" binding partner to the bound complex (lower panel Fig. 5C, fig. S20; section S7.3) (*41*). Measuring the fractional abundances as a function of time, similar to Fig. 2B, permits the measurement of off-rates and corresponding on-rates for the same reaction. We find that these measured rates yield measures of $K_d = k_{off}/k_{on}$ that are consistent with the equilibration-based measurement approach as well as with theoretically expected values (*40, 42*) (Fig. 5C; section S7.5).

*DNA-protein and protein-protein interactions*

We further examined DNA-protein binding affinity measurements performed with a singly labelled dsDNA molecule binding to unlabelled restriction enzymes (Fig. 6A, fig. S22; section S7.6) (*43*). Here, we added the unlabelled binding partner, the restriction enzyme EcoRI, at various concentrations to a fixed concentration of 0.25 nM of fluorescently labelled 24 bp DNA containing the enzyme recognition site. In general, binding systems that produce large shifts in $t_{esc}$ of a factor 2 to 5 upon binding yield $P(\Delta t)$ histograms that are amenable to multi-exponential fitting and the extraction of fractional abundances of the underlying species (Fig. 5A). We determined the bound fraction of DNA from fitting two timescales to the $t_{esc}$ data as described in Fig. 5, and obtained a value of $K_d \approx 1.5$ nM, which is in excellent agreement with the reported value of $K_d \approx 1$ nM (*43*). However, many binding reactions are associated with small changes in size and or overall conformation of the complex, which implies small changes in $t_{esc}$, sometimes as small as 10% or less. Here, the similarity of timescales corresponding to the two molecular states can render robust extraction of two exponential components and their individual fractions from the escape time data untenable (*34*). Therefore, rather than fitting escape time histograms we track the quantity $t_{av}$, which represents a simple average over all escape events of duration, $\Delta t$, in the mixture of the two interacting species (section S7.1). We applied this procedure to examine a well-characterized antigen-antibody interaction between human leukocyte antigen (HLA) isoform A*03:01 (HLA A*03:01) and W6/32, a mouse-derived pan-HLA class 1-reactive monoclonal antibody (figs. S3 and S22; section S7.7) (*44*). Anti-HLA antibodies are important as clinical biomarkers for evaluation and monitoring of allograft rejection risks in cell and organ transplantation (*45*). Binding of the labelled HLA antigen ($M \approx 49$ kDa) to its antibody ($M \approx 150$ kDa) results in a 40% increase in $t_{av}$ relative to that of the free, labelled bait. Analysing the $t_{av}$ data to determine the bound fraction, we obtained $K_d \approx 0.4$ nM for this interaction, in excellent agreement with reported values of $K_d \approx 0.4$ nM and $K_d \approx 0.7$ nM (Fig. 6B, fig. S22) (*21, 46*). Furthermore, our measurement displays sensitivity to the HLA antibody at concentrations as low as < 100 pM, providing a general approach to detecting antibodies in solution. The measurements also shows that $t_{av}$ alone can serve as an excellent quantitative measure of the fraction of fluorescent bait in the bound-state, and this approach is utilized again in investigations involving direct detection of free insulin in solution, described next (fig. S21).

In order to explore the possibility of detecting a small, medically relevant protein analyte such as free insulin in a complex sample, we first examined the use of a fluorescently labelled 31-base insulin-binding aptamer that is expected to bind insulin in the folded G-quadruplex containing state (*47*) (Fig. 6C). Upon formation of the structured state, ETs directly detects a 10% reduction in $t_{av}$ for the aptamer (Fig. 6D, fig. S23). We incubated 5 nM of singly-labelled, folded aptamer with 100 μM unlabelled insulin which is well above the reported $K_d$ of the interaction (*48, 49*). We noted a 15% increase in $t_{av}$ for the folded aptamer which may be expected based on the increase in mass of the complex alone. Importantly we noted no change in $t_{av}$ for the unfolded aptamer in the presence of insulin, pointing to the apparent requirement of secondary structure in the aptamer for insulin binding (*47*). Measuring $t_{av}$ for mixtures containing variable concentrations of free, unlabelled insulin and a fixed low concentration of labelled aptamer 'bait' of 1 nM, we obtained a $K_d \approx 90$ nM, in broad agreement with literature values (section S7.8) (*48, 49*). This measurement demonstrates a straightforward, rapid and general approach to detecting the presence of a small, medically important analyte in solution. However, the relatively modest affinity of the aptamer for insulin is a major limitation for direct insulin sensing at physiologically and biochemically relevant concentrations.

We therefore set out to characterize a higher affinity interaction for insulin using a biologically relevant binding partner that is tuned to respond to low nM physiological concentrations of insulin. The insulin receptor (IR) is a transmembrane receptor tyrosine kinase that is activated by insulin binding to its extracellular ectodomain (ECD) (*50*). We measured the binding affinity between a singly labelled insulin molecule ($M \approx 6$ kDa, fig. S2) and the soluble IR-ECD ($M \approx 280$ kDa) by ETs (section S7.9). The large disparity in timescales between free and bound insulin facilitates a direct measurement of bound fractions using two fixed timescales in the ensemble fit (Fig. 6C, fig. S22). The highest affinity we measured for the interaction was $K_d \approx 7$ nM which is smaller than the previously reported $K_d \approx 30$ nM value measured for the same IR-ECD by nano–differential scanning fluorimetry (*50, 51*), and closer to the sub-nM high affinity binding of the first ligand binding site of the full-length insulin receptor (*52-54*).

**Structural inferences on the 'Apo' and 'Liganded' states of the insulin receptor ectodomain**

We then explored the general potential our platform for membrane protein receptor-based immunoanalysis and diagnostic applications. We used fluorescently labelled IR-ECD as 'bait' to detect the much smaller, unlabelled insulin molecule in solution. IR-ECD is expected to bind up to 4 insulin molecules, entailing an increase in molecular weight of ca. 9% ($\Delta M \approx$ 25 kDa) and a modest concomitant increase of ca. 3% in $t_{esc}$ (or $r_H$) (*51*).

In order to examine the conformational properties of the apo- and liganded IR-ECD we first carried out a structure/conformation analysis of IR-ECD in the insulin-free and bound states using ETs as described in Figs. 3 and 4. Based on structural data from cryo-EM, the soluble ectodomain (ECD) of the insulin receptor is expected to have an approximate spatial extent of 17 nm, a hydrodynamic radius of $r_H = 6$ nm, and can bind up to 4 insulin molecules. Molecular structures obtained using x-ray crystallography and cryo-EM indicate that upon insulin binding, the IR-ECD undergoes a substantial conformational change from an 'inverted-V' apo-state to a fully liganded 'T-state' including intermediate states with a lower number of bound insulin molecules (*51, 55*), as also reported for snail venom insulin:human IR-ECD interactions (*56*) (Fig. 7A). However, the generation of crystal structures of IR-ECD in the 'apo' state has relied on internal conformational stabilization. Indeed the high degree of conformational flexibility of the apo-state compared to its liganded counterpart has hindered identification of a representative 3D structure for the ligand-free state using cryo-EM (*51*).

We performed $t_{av}$ measurements on 1 nM labelled IR-ECD, both in the absence and presence of insulin at a concentration of 100 nM, well above the expected $K_d$ of the interaction (section S5.7). In contrast to the molecular binding measurements discussed in the previous section, all of which entailed an increase in size of the complex and therefore of $t_{av}$ upon binding of the small, labelled ligand, here we found that $t_{av}$ values of the fluorescent complex decreased by about 7% to $21 \pm 0.1$ ms compared to $22.5 \pm 0.3$ ms for the ligand-free species (Fig. 7A). Within our view of the underlying measurement principle, a decrease in $t_{av}$ that occurs despite an increase in mass upon binding indicates a reduction in the overall spatial extent of the complex in the bound state. This strongly suggests that the readout is in fact dominated by conformational compaction of IR-ECD upon insulin binding. Note however that the x-ray structure of the *conformationally stabilized* 'apo' state and the cryo-EM structure of the liganded state suggest very similar $r_H$ and $D_s$ values which would entail negligible differences in $t_{av}$ in our experiment, contrary to the experimental observations (top panel, Fig. 7A, fig. S14).

In order to quantitatively explore the implied conformational difference between the two states of the receptor in solution we repeated $t_{av}$ measurements in slits of systematically smaller height in the range $h_1 \approx 30\text{-}60$ nm (Fig. 7A, lower panel, fig. S14). Similar to the observations on the DNA nanostructures, we noted progressively larger disparities in $t_{av}$ between the 'apo' and liganded states increasing from ca. 5 % in the deepest slits to > 20% in the shallowest (Fig. 7A). The plots in Fig. 7B graphically represent permissible values $r_H$ and $D_s$ for the two molecular states inferred from measurements at four different slit heights (fig. S14). The $r_H$ and $D_s$ values inferred from the intersection of lines representing a pair of measurements corresponding to mean values of $h_1 =$38.7 nm and 62.6 nm show that the two states are well resolved, highlighting systematic differences $\Delta D_s = D_{s,\text{lig.}} - D_{s,\text{apo}} \approx 1.5$ nm and $\Delta r_H \lesssim 0.25$ nm. Note that the uncertainty in determining the calibrated height $h_1$, which can be as high as ca. 3-4 nm in some cases, limits the level of accuracy on absolute values of some of our present readouts of $r_H$ and $D_s$, but this can be improved upon in the future (Fig. 7C). Notwithstanding the current uncertainties in determining absolute values of $r_H$ and $D_s$, we note systematic measured differences between the apo and liganded states that are relatively robust to uncertainties in $h_1$ (Fig. 7D). Furthermore, the analysis readily admits solutions for $r_H$ and $D_s$ that agree reasonably with the corresponding values characterizing the cryo-EM structure of the ligand saturated state (pdb: 6SOF) for values of $h_1$ within the calibration error (Fig. 7C, D, black symbol).

Taken together our results indicate that the liganded state may indeed be more compact than the ligand-free 'apo' state. Whilst the absence of a 3D structure of the apo-state in solution precludes a quantitative assessment of our molecular property readouts against an independent structural model, our measurements do indeed point to an apo-state characterised by greater flexibility than the liganded-state (right panels, Fig. 7C). Thus, in contrast to the binding of the 'fluorescent bait' aptamer to insulin, the smaller $t_{\text{av}}$ value of the IR-ECD 'bait' in the presence of insulin appears to be a direct consequence of conformational compaction as the IR-ECD transitions from the insulin-free apo to the insulin-bound liganded state. This conformational compaction is large enough to offset the effect of a modest increase in mass (up to ca. 25 kDa) that occurs upon binding of up to 4 insulin molecules. The relative insensitivity of the hydrodynamic radius $r_H$ of IR-ECD to the presence or absence of insulin is in broad agreement with diffusion measurements on the same samples using FCS, as well as with computational estimates based on the molecular 3D structures of the stabilized 'apo' and liganded states (Fig. 7C, fig. S14, table S4). Thus, the overall measured response is dominated by the conformational change in the receptor induced by ligand binding, as reflected in the reduced diameter of the molecular bounding sphere, $D_s$ (Fig. 7C, D).

*Affinity based detection of free insulin in serum*

Finally, Fig. 7E demonstrates a route to exploiting conformation changes in a receptor protein in order to detect analytes near physiologically relevant concentrations in complex samples. We first incubated 80 pM labelled IR-ECD with a range of insulin concentrations in PBS buffer for an hour at room temperature. Imaging the chip for about 5 min per incubated sample, we noted a gradual reduction in $t_{\text{av}}$ with increasing insulin concentration, with the maximum and minimum timescales corresponding to those expected for the 'apo' (22.5 ms) and liganded states (21 ms) in a slit of $h_1 \approx 40$ nm. Assuming that $t_{\text{av}}$ is proportional to the amount of IR-ECD in the liganded-state we obtain $K_d \approx 2 - 12$ nM, in good agreement with the literature (*50-54*).

We then titrated insulin into simulated serum containing ca. 100 pM of labelled IR-ECD, incubated the samples for 30 min at RT and assessed the binding, as indicated by $t_{\text{av}}$ (fig. S24; section S7.10). We found $K_d$ values around 2-4 nM, supporting detection of free insulin at $\geq 1$ nM in serum, which lies at the higher end of the typical physiological range ($< 1$ nM). Similar measurements in a sample of true human serum however revealed a substantial conformation change with small amounts of added insulin, indicating an estimated $K_d \approx 100$ pM, and demonstrating that the readout principle is not affected by the complex background of human serum (fig. S24; section S7.10). Genetically modified and conformationally stabilized variants of IR-ECD are known to offer much higher affinities of $K_d \approx 80$ pM, comparable with the full-length IR (*57, 58*). Given an affinity reagent with $K_d \leq 100$ pM, our ability to quantitatively measure $t_{av}$ at femtomolar concentrations offers the sub-picomolar detection sensitivity required for highly quantitative, rapid and sensitive detection of analytes in ultra-small volumes of serum or plasma in the pM regime (Fig. 5B, 6B, fig. S5). The same detection principle may be applied to a host of diagnostic biomarkers, including small molecule hormones and metabolites, and such detection assays may be run in a highly parallel spectrally multiplexed fashion.

Rather often, e.g., when metal-ions or small molecules bind to proteins, molecular weight shifts can be miniscule compared to changes in shape and conformational properties (*59*). Post-translational modifications such as phosphorylation, glycosylations or DNA methylation may also be detected based on changes in mass and/or conformation. ETs presents a versatile chip-based measurement platform that meets a variety of molecular analytical goals via highly conformation-sensitive detection under native conditions in solution. The same platform operated in "charge mode" enables state, structure and chemical discrimination based on molecular effective charge as previously described (*3, 18, 19, 23*). Whilst the spectral and photophysical properties of labels may be further leveraged to foster optical multi-dimensionality in the measurement (*60-62*), detection based on scattered light will support label-free measurements, potentially enabling simultaneous molecular weight inferences (*7*).

In fluorescence detection, the labelled species present at trace concentrations can act as a reporter of molecular or complex-state in the presence of high background concentrations (mM) of unlabelled, interacting matter of any mass, identity and compositional complexity. Thus ETs may be well suited to the detection of multi-subunit fragile macromolecular assemblies that form on account of weak interactions and are challenging to detect by other methods (*63*). Rapid solution-phase direct sensing of complex formation may also benefit investigations of long-standing problems such as pathological protein aggregation (*64*).

Finally, the ability to map measurements of conformation on to molecular 3D structure-based computed quantities suggests that the ETs readout could aid machine learning approaches to molecular structural modelling, validation and inference, contributing to tackling frontier problems in structural studies, e.g., the conformation of disordered proteins and RNA, including the ability to characterize rare states.

**Materials and Methods**

Commercially obtained samples

Recombinant human insulin (INS, 91077C) was obtained from SAFC Biosciences, Andover, UK. Ubiquitin from bovine erythrocytes (Ub, U6253), Thioredoxin from Escherichia coli (TRX, T0910), Ribonuclease A from bovine pancreas (RNase A, R6513), Myoglobin from equine heart (MB, M1882), β-Lactoglobulin A from bovine milk (LGB, L7880), Carbonic anhydrase (CA, C7025), human apo-Transferrin (TF, T2036) and Apoferritin (FER, A3660) were all from Sigma-Aldrich, Gillingham, UK. Human insulin detector antibody (IgG, ab253508) was purchased from Abcam, Cambridge, UK. HLA-ABC monoclonal antibody W6/32 (IgG, 16-9983-85) and Human EGF Recombinant Protein (PHG0314) were from Thermo Fisher Scientific. HLA-A*03:01 RLRAEAQVK was provided by the NIH Tetramer Core Facility at Emory University, Georgia, USA. EcoRI (R0101M) was obtained from New England Biolabs. Nucleic Acid samples were mostly obtained from Integrated DNA Technologies Inc., Coralville, Iowa, with sequences shown in table S1. With the exception of the SAM-IV riboswitch sample, and reference DNA sample – synthesis of these is outlined in the following sections. All fluorescent dyes were obtained as lyophilized powder from ATTO-TEC GmbH, Siegen, Germany. The disordered proteins Starmaker-like (Stm-l), Prothymosin α (ProTα), were obtained as described in (3). A 13-mer proline-rich polypeptide was obtained by custom synthesis from Bio-Synthesis Inc., respectively. Details of the sequences of these proteins can be found in table S2. Samples were generally prepared using UltraPure DI water (Invitrogen) and PBS (pH 7.4, Gibco). Samples, except for antibodies were aliquoted, flash-frozen and stored at either -20 or -80°C as per manufacturer's instructions.

Protein labelling procedure

To protein stock solutions in DI water or PBS (pH 7.4), 1 M NaHCO$_3$ (Sigma-Aldrich) was added to a final concentration of 0.1 M in order to adjust the pH to 7.5 - 8.3. After addition of a molar excess of ATTO532 NHS-ester in DMSO or DI water, the reaction was incubated for 1 hour at room temperature, protected from light. Unreacted dye was removed by buffer exchanging twice into PBS (pH 7.4) using Zeba micro spin desalting columns with 7 K or 40 K MWCO (Thermo Fisher Scientific) following the manufacturer's instructions. Labelled insulin and dye were separated into PBS using size exclusion chromatography on a Superdex 75 10/300 GL column (Cytiva) on an AKTA pure protein purification system (Cytiva). Post-labelling concentrations and degree-of-labelling (DOL) were calculated from absorbance measurements at 280 nm and 532 nm on a Nanodrop spectrophotometer (Thermo Fisher Scientific), and are presented in table S3.

Electrospray mass spectrometry (ESI-TOF)

Reverse-phase chromatography on an Agilent 1100 HPLC system (Agilent Technologies, Palo Alto, CA, USA) was coupled in-line to a 1969 MSD-ToF electrospray ionisation orthogonal time-of-flight mass spectrometer (Agilent Technologies Inc, Palo Alto, CA, USA). Samples were diluted to a concentration of 0.02 mg/ml (for insulin) or 10 µM (for ATTO532 dyes) in 0.1% formic acid. 50 µL were injected on to a 2.1 mm x 12.5 mm *Zorbax* 5 µm 300SB-C3 guard column. The column oven was maintained at 40°C. Ultra-high purity water (Optima, LC-MS grade, Fisher Chemical) with 0.1% formic acid (Optima, LC/MS grade, Fisher Chemical) (solvent A) and methanol (Optima, LC/MS grade, Fisher Chemical) with 0.1% formic acid (solvent B) were used as mobile phases. Chromatography was started at 90% A and 10% B for 15 s, followed by a two-stage linear gradient from 10% B to 80% B over 45 s and from 80% B to 95% B over 3 s. Elution was performed isocratically at 95% B for 1 min 12 s followed by equilibration at 90% A and 10% B over 45 s. The flow rate was set to 1.0 mL/min. For mass spectrometry the instrument was configured with the standard ESI source and operated in positive ion mode. The capillary voltage was set to 4000 V and the nebulizer pressure to 60 psi. The source temperature was kept at 350°C with a drying gas flow rate of 12 L/min. The ion optic voltages were: fragmentor 250 V, skimmer 60 V and octopole RF 250 V. Data analysis was performed using Agilent MassHunter Qualitative Analysis B.07.00 software (Agilent Technologies, Palo Alto, CA, USA), see Fig. S1 and Fig. S2.

Native mass spectrometry method

HLA was desalted and buffer exchanged twice into 250 mM ammonium acetate using Zeba micro spin desalting columns with 7 K MWCO following the manufacturer's instructions and diluted to 10 µM. Spectra were acquired on a Q Exactive UHMR Orbitrap mass spectrometer (Thermo Fisher Scientific) with in-house prepared gold-plated capillaries using the following parameters: capillary voltage of 1.3 kV, capillary temperature of 100°C, in-source trapping of -10 V and HCD energy 1. Spectra were averaged and deconvolved using UniDec software (*65*), see Fig. S3.

DNA nanostructure synthesis

Two DNA nanostructures were synthesized as described in Ref. (*23*). The "square tile" consists of four terminally tethered 30 base-pair double-stranded (ds) DNA helices, and the "bundle" consists of two 60ds DNA helices tethered at two points along the length of the helix. Both structures consist of 240 nucleotides (nt), and are labelled with a single ATTO532 dye.

SAM-IV riboswitch and reference DNA synthesis

Single stranded (ss) RNA and DNA sequences, both 119 nt in length, were synthesized (sequence information in table S1) on an Applied Biosystems 394 automated DNA/RNA synthesizer using a standard phosphoramidite cycle of detritylation, coupling, capping, and oxidation on a 0.2 µmole scale. For DNA, TCA (3% in dichloromethane), 1*H*-tetrazole (0.45 M in acetonitrile), Cap A (10% acetic anhydride, 10% lutidine and 80% tetrahydrofuran) / Cap B (16% N-methylimidazole in tetrahydrofuran) and iodine (0.02 M in tetrahydrofuran, pyridine and water) were used. Pre-packed nucleoside SynBaseTM CPG 3000/110 (Link Technologies) resins were used and β-cyanoethyl protected phosphoramidites (dA-bz, dG-ib, dC-bz and dT where bz = benzoyl and ib = *iso*-butyryl, Sigma-Aldrich) were dissolved in anhydrous acetonitrile (0.1 M) immediately prior to use. The coupling time for dA, dC, dG and dT monomers was 45 s, and 600 s for the 5'-amino modifier phosphoramidite monomer. Stepwise coupling efficiencies were determined by automated trityl cation conductivity monitoring and were >98% in all cases.

For RNA, 2'-O-TC protected RNA phosphoramidites (A-bz, C-ac, G-ib, U, Sigma-Aldrich) monomers were used. Monomers were dissolved in anhydrous toluene: acetonitrile (1:1 v/v) (0.1 M) immediately prior to use. Capping and oxidation reagents were identical to those used in DNA synthesis while ethylthiotetrazole (ETT) (0.25 M in acetonitrile, Link Technologies) was used as a coupling reagent. The coupling time for all monomers during RNA synthesis was 3 min. Stepwise coupling efficiencies were determined by automated trityl cation conductivity monitoring and in all cases were >97%.

*Selective β-cyanoethyl removal*

RNA and DNA bearing the 5'-primary amines were treated on-column with diethylamine (20% in anhydrous acetonitrile) for 20 min at room temperature. The resin was then washed with acetonitrile (3 x 1 mL) and dried with argon.

*DNA deprotection*

DNA was cleaved from solid support and deprotected by exposure to a concentrated solution of aqueous ammonia in a sealed vial for 5 h at 55 °C. After drying *in vacuo*, oligonucleotides were dissolved in water and subject to further purification.

*RNA deprotection (2'-O-TC)*

The solid support was exposed to dry ethylenediamine: toluene (1:1 v/v) for 6 h at room temperature, washed with toluene (3 x 1 mL) then acetonitrile (3 x 1 mL), and dried using argon. The cleaved RNA was eluted with water and purified using RP-HPLC.

*RP-HPLC purification*

DNA and RNA and were purified using an Agilent system with Kinetex® C18 column (10 mm x 250 mm, pore size 100 Å, particle size 5 μm), a gradient of buffer A (0.1 M TEAA, pH 7.5) to buffer B (0.1 M TEAA, pH 7.5 50% v/v acetonitrile), and flow rate of 5 mL/min. A gradient of 0%-50% buffer B over 28 min at 55°C was used in all cases. Following HPLC, Amicon®Ultra-4 0.5 mL centrifugal filters (Merck, cat. no. UFC5x) were used to desalt and concentrate oligonucleotide samples.

*Post-synthetic oligonucleotide modification*

Freeze-dried oligonucleotide (20 nmol) was dissolved in $NaHCO_3$ buffer (0.5 M, pH 8.5, 30 μL) and mixed with ATTO NHS ester (200 nmol, 20 μL) dissolved in DMSO. The reaction was then left for 4 h at 25°C with 750 rpm shaking. After dilution with water, the samples were desalted using NAP™-10 Columns (Cytivia, cat. no. 17085402) according to the manufacturer's instructions. RNA or DNA oligonucleotides were purified using HPLC as described above.

*Oligonucleotide mass spectrometry*

All DNA and RNA were characterized by negative-mode electrospray using a UPLC-MS Waters XEVO G2-QTOF mass spectrometer and an Acquity UPLC system with a BEH C18 1.7 μm column (Waters). A gradient of methanol in triethylamine (TEA) and hexafluoroisopropanol (HFIP) was used (buffer A, 8.6 mM TEA, 200 mM HFIP in 5% methanol/water (v/v); buffer B, 20% v/v buffer A in methanol). Buffer B was increased from 0–70% over 7.5 min or 15–30% over 12.5 min for normal oligonucleotides and 50–100% over 7.5 min for hydrophobic oligonucleotides. The flow rate was set to 0.2 mL/min. Raw data were processed and deconvoluted using the deconvolution software MassLynx v4.1 and Unidec (*65*), see Fig. S4.

Insulin receptor Ectodomain (IR-ECD) purification and labelling

The Insulin Receptor ectodomain (IR-ECD) was purified as previously described by Gutmann et al. (*51*). For labeling fivefold molar excess of Alexa Fluor 532 NHS ester (A20001, Thermo Fisher Scientific) was used. The reaction was performed on ice for 1 hour and was quenched with a fivefold molar excess of buffered ethanolamine over Alexa Fluor 532 NHS. The labelled protein was separated from free dye on a Superdex 200 Increase 10/300 GL column equilibrated with 25 mM Hepes, 150 mM NaCl, pH 7.4. The protein was aliquoted and snap-frozen with liquid nitrogen.

ETs experimental procedure

*Fluorescence Microscopy*

Microfluidic devices were fabricated using procedures similar to those outlined in previous studies (*7, 24*), with the exception that in this study the silicon and glass substrates were thermally rather than anodically bonded. 10-100 µL of sample is loaded into one pair of loading reservoirs, while the other pair of reservoirs are sealed and connected to a vacuum pump (fig. S5). The pressure drop across the length of the channels loads the molecules into the trapping area under steady Poiseuille flow. Movies of fluorescent molecules diffusing through the trapping nanostructures are recorded either with the pressure drop applied (under flow), or without, depending on the modality of experiment. Movies are recorded for 1 to 10 minutes using an epi-fluorescent wide-field fluorescent microscope at a frame rate of 100 Hz, generally, with an exposure time of 5 ms. A detailed description of the setup can be found in previous work (*3*). Briefly, for the measurements reported in this study, a 3 W, 532 nm DPSS laser (Del Mar Photonics) is used to illuminate a ca. 150x150 µm region of the device through a water-immersion objective (UPlanSApo 60x W, 1.2 NA, Olympus). The emission is filtered with a 532 nm cut-off dichroic and long-pass filters (Chroma), and imaged on an sCMOS camera (Kinetix, Photometrix).

*Data analysis to determine the escape time of a molecular species*

Recorded movies of molecules in the trap landscape are analyzed as described in previous work (*3*). Durations of residence times, $\Delta t$, of molecules in traps are extracted from intensity vs. time traces, recorded for each trap locus. At present, a one-minute movie typically yields ca. $10^4$ escape events. The average escape time is the mean of the all the residence times collected, $t_{av} = \frac{1}{N} \sum \Delta t$, for a total number of residence times, *N*. The uncertainty on determining $t_{av}$ in such a measurement is given by the standard error of all recorded residence times $t_{av,e} \cong \frac{t_{av}}{\sqrt{N}}$. When fits of the data are performed, a histogram of residence times is fitted with a decaying single exponential function, $P_f(\Delta t)$ with $k = 1$ as shown in Eq. (2), e.g., to determine a fitted escape time, $t_{esc} = t_i$. Here $k$ is the number of components, $E_i$ is the amplitude of the component with escape timescale, $t_i$ and $c$ is a constant value capturing an experimental offset.

$$P_f(\Delta t) = \sum_{i=1}^{k} \left( \frac{E_i}{t_i} e^{-\frac{\Delta t}{t_i}} \right) + c \quad (2)$$

The fit error on $t_{esc}$ captures the standard error of the mean, and in general corresponds well to $t_{av,e}$ as expected.

## Acknowledgments

We gratefully acknowledge Andrew Turberfield for the kind gift of DNA nanostructures, the NIH Tetramer Core Facility (contract number 75N93020D00005) for providing HLA-A*03:01, and Robin W. Klemm for helpful discussions.

## Funding

European Research Council Consolidator Grant 724180 (XZ, TJDB, MK)

Biotechnology and Biological Sciences Research Council BB/W017415/1 (DS, XZ, MK)

Wellcome LEAP Delta Tissue Program (XZ, KCZ, JLPB, MK)

Royal Society Research Grants RGS\R2\242523 (AHES)

Royal Society of Chemistry Research grant R24-8767374007 (AHES)

Federal Ministry of Education and Research (BMBF) and the German Center for Diabetes Research (DZD e.V.) (UC)

## Author contributions

Conceptualization and model: MK
Methodology: XZ, TJDB, MK
Measurement: XZ, TJDB, DS
Data analysis and processing: TJDB
Protein synthesis/recombinant production, labelling and quality control: KCZ, MG, JLPB, UC
Nucleic acid design, synthesis and quality control: AHES
Visualization: TJDB, XZ, KCZ, DS, MK
Supervision: MK
Writing – original draft: MK
Manuscript – review & editing: All authors

## Competing interests

Novel aspects described in this work are covered by a patent application filed by MK to which XZ, TJDB, KZ and DS have contributed. MK is also founding a company aiming to commercialise aspects of this work.

## Data and materials availability

All data are available in the main text or the supplementary materials.

## Supplementary Materials
Supplementary Text
Figs. S1 to S24
Tables S1 to S5
References (*67–85*)
Movies S1 to S2

# Figures

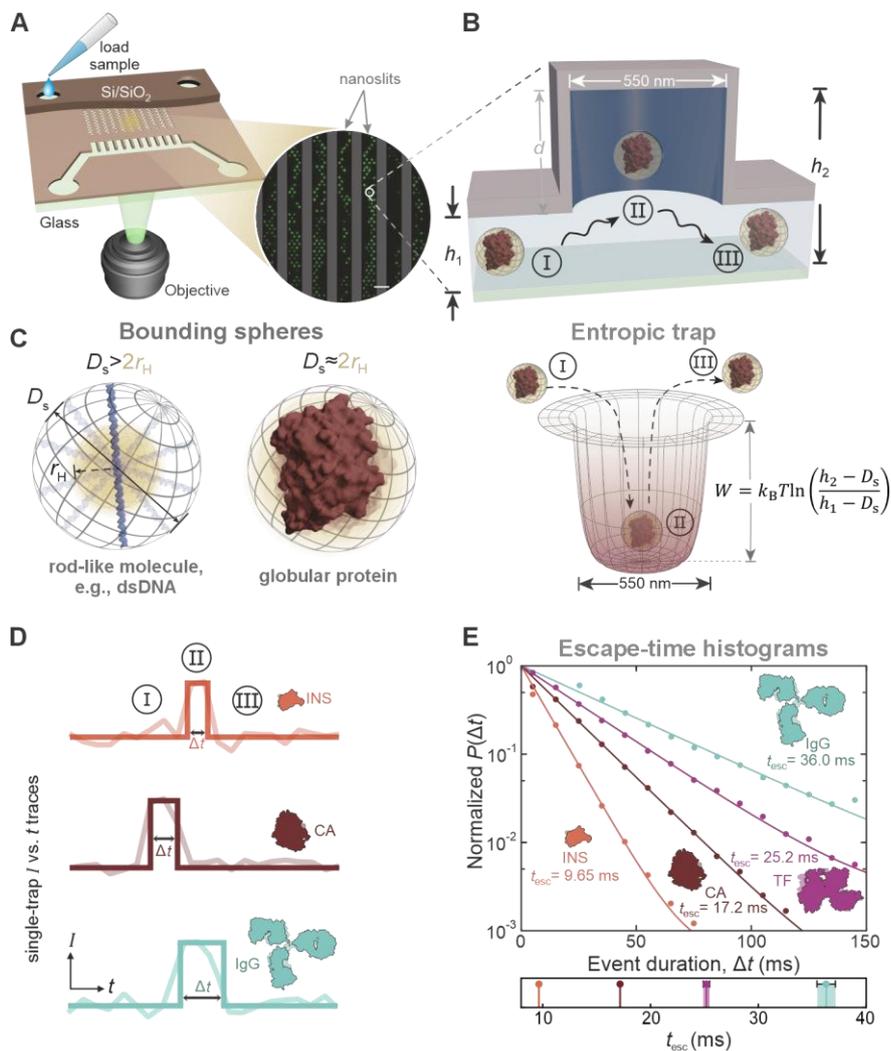

Fig. 1. **Principle and overview of escape-time stereometry (ETs) in solution.** (A) Schematic representation of a microfluidic chip carrying a series of fluid-filled parallel-plate nanoslits in which arrays of geometric indentations create local entropic traps for freely diffusing fluorescently labelled molecules in solution. Wide-field fluorescence microscopy image presents the central quarter of a typical field of view, time-averaged over 1 s, displaying ~25 molecules diffusing in the trap landscape. Scale bar is 5 μm. (B) Schematic depictions of a globular protein molecule entering (state-I), residing in (state-II) and leaving (state-III) a nanostructured indentation of diameter 550 nm, depth $d \approx 220 - 290$ nm and height $h_2 = h_1 + d$ nm located in a nanoslit of height $h_1 = 20 - 70$ nm (top). Bottom - corresponding entropic potential well of depth $W$ in which molecules reside for an average time given approximately by $t_{esc} \propto r_H \exp(W/k_B T)$ where $r_H$ is the hydrodynamic radius, $k_B$ is the Boltzmann constant and $T$ is the absolute temperature. (C) Graphic representation of bounding spheres (grey) of diameter $D_s$ enclosing a rod-like double-stranded (ds) DNA molecule (left) and a globular protein molecule (right). $r_H$ (yellow sphere) can be considerably smaller than $D_s/2$ for a non-spherical molecule, in contrast to a globular molecule. $D_s$ and $r_H$ values for a given species may be computed from a molecular structural model as described in the supplementary text S5.2. (D) Intensity vs. time trace of each trap locus in the lattice permits identification of molecular residence or escape events of duration $\Delta t$. Representative events detected using a step-finding algorithm (dark lines) superimposed on raw traces. Three typical escape-time events for insulin (INS), carbonic anhydrase (CA), and immunoglobulin G (IgG) illustrating that $\Delta t$ depends on the molecular weight of the species. (E) Probability density distributions $P(\Delta t)$ vs. $\Delta t$ for $N = 10^4 - 10^6$ recorded escape events fit to the form $P(\Delta t) \propto \exp(-\Delta t/t_{esc})$ to extract means $t_{esc}$ values. For visual comparison across samples, the fitted $t_{esc}$ values in each case are presented as vertical lines and symbols, with the measurement precision (given by the fit error) represented by a band on the reported mean value of thickness $\pm 1\sigma$ (bottom).

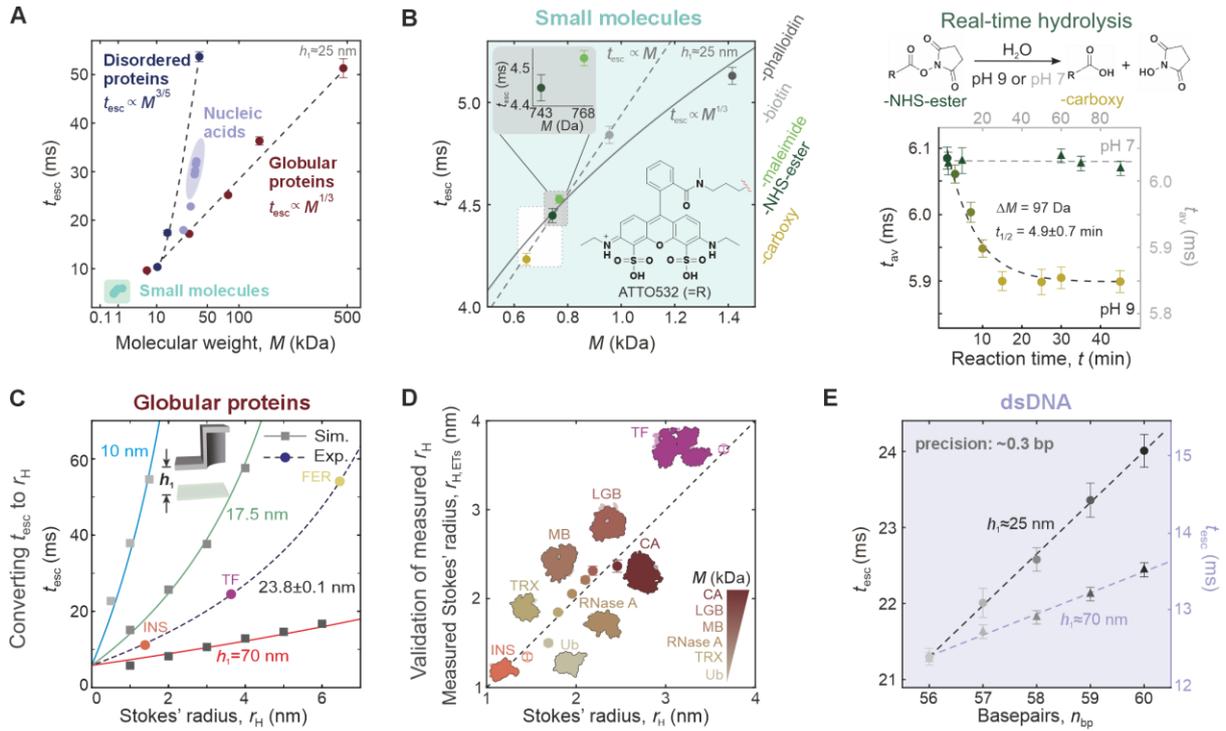

**Fig. 2. Sensitivity of measured escape-times to molecular weight.** (**A**) Measurements of escape time $t_{esc}$ vs. Molecular weight, $M$, presented on an abscissa scaled to the power of 1/3, and covering approximately 3 orders of magnitude, including small organic molecules (green symbols), globular proteins (brown symbols), disordered proteins (blue symbols) and dsDNA (purple symbols) in a device with $h_1 \approx 25$ nm. Globular proteins (insulin, carbonic anhydrase, transferrin, Anti-insulin IgG, and apoferritin) and disordered proteins (*3*) (Starmaker-like - Stm-l, Prothymosin α - ProTα, and a 13-mer proline-rich polypeptide sequence given in table S2) reveal good agreement with scaling behaviours $t_{esc} \propto M^{1/3}$ and $M^{3/5}$ respectively. (**B**) Plot of $t_{esc}$ vs. $M$ for small fluorescent organic molecules represented by different chemical derivatives of the fluorophore ATTO532 (left). The data display reasonable agreement with the $r_H \propto M^{1/3}$ expectation for small molecules (grey solid curve) (*66*). A linear trendline (dashed line) over a narrow range of $M$ illustrates how ca. 0.5% measurement precision on $t_{esc}$ implies the ability to detect differences of ca. 10 Da in the 1 kDa molecular weight regime. Right - Real-time tracking of the hydrolysis of the NHS ester derivative into the carboxyl form entails a change in molecular weight of 97 Da. The 3% overall change in $t_{av}$ over the course of the reaction is much larger than the measurement precision of ca. 0.5% in a minute-long measurement window permitting rate of hydrolysis to be measured at pH 9 (details in supplementary text S4.2). (**C**) Measured $t_{esc}$ values may be converted to $r_H$ and $D_s$ values describing the molecular species using BD simulations (as described in supplementary text S5.1). Simulated values for spheres of radius $r_H$ (square symbols) in a device with $d = 300$ nm illustrate the increasing non-linearity of the measured response with decreasing $h_1$, and are well fit by the calibration function Eq. (1) with $A \approx 0.3$ s/μm and $t_o = 5.83$ ms which is used for all protein experiments (red, green and blue curves). An example calibration measurement to determine the value of $h_1$ of the experimental device is performed using insulin (orange, INS), transferrin (purple, TF) and apoferritin (yellow, FER) whose Stokes' radii, $r_H$, are taken from molecular structures (as outlined in supplementary text S5.2). Measured $t_{esc}$ data are fitted with the calibration function yielding $h_1 = 23.8 \pm 0.1$ nm which agrees with the nominal design value of $h_1 \approx 25$ nm (dashed curve). (**D**) $r_H$ values measured for a test set of proteins in the range $M = 10 - 30$ kDa (filled symbols: ubiquitin - Ub, thioredoxin - TRX, RNase A, myoglobin - MB, β-lactoglobulin – LGB, carbonic anhydrase – CA; open symbols – INS and TF calibrator molecules) compare well with theoretical expectations, as well as with measurements using fluorescence correlation spectroscopy (FCS) (see fig. S10, table S4). (**E**) Single base-pair differences in dsDNA in the range of 56-60 bp are clearly distinguishable, with the magnitude of measured differences increasing strongly with decreasing $h_1$.

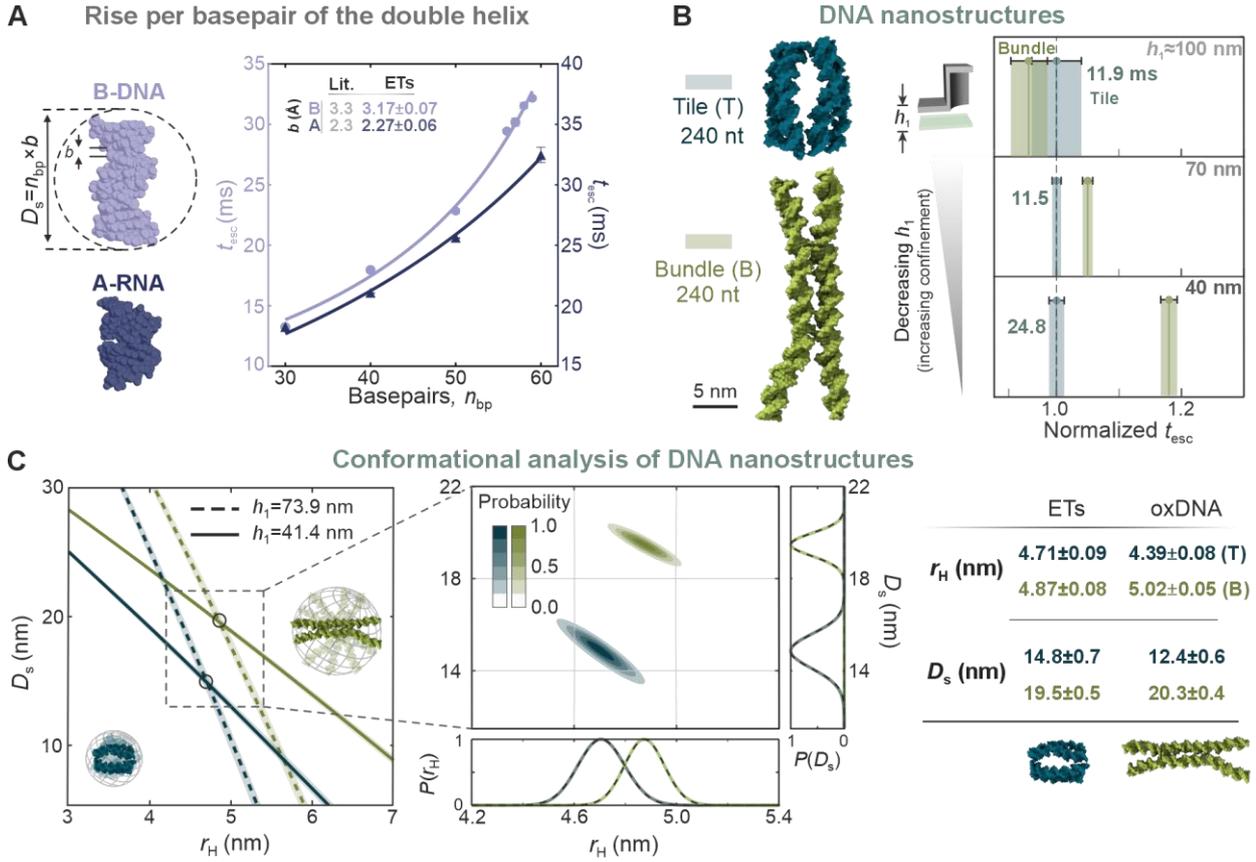

**Fig. 3. Sensitivity of the escape-time readout to molecular shape (3D conformation).** (**A**) Measurements of escape time $t_{esc}$ vs. number of basepairs, $n_{bp}$, for a series of dsDNA (B-form double helix) and dsRNA (A-form helix), fitted with Eq. (1) yielding $A = 0.35 \pm 0.01$ s/μm for RNA, and $0.18 \pm 0.01$ s/μm for DNA, and rise per basepair values, $b$, that are in good agreement with literature values from high resolution structural methods (*30*). (**B**) DNA nanostructures composed of an identical number of nucleotides (nt) but in different 3D conformations of a 'square tile' and 'bundle' display increasingly separated measured $t_{esc}$ values with decreasing slight height $h_1$. (**C**) Graphical representation of the solution of Eq. (1) using $t_{esc}$ measurements for two different mean calibrated values of $h_1$ (solid and dashed lines, see supplementary text S5.5 for details). Shaded bands around lines reflect the precision of the $t_{esc}$ measurement. Region of intersection of the two uncertainty broadened lines yields estimates of $D_s$ and $r_H$ for each nanostructure (left) and Gaussian probability density distributions of molecular physical properties $P(D_s, r_H)$, $P(r_H)$ and $P(D_s)$ (middle panel). Dashed superimposed curves demonstrate that the widths of the distributions reflect the statistical uncertainty on $t_{esc}$ measurement and do not stem from detected conformational flexibility of the nanostructures. Table comparing ETs measurements with inferences from coarse-grained structures simulated using oxDNA (right, see supplementary text S5.5 for details).

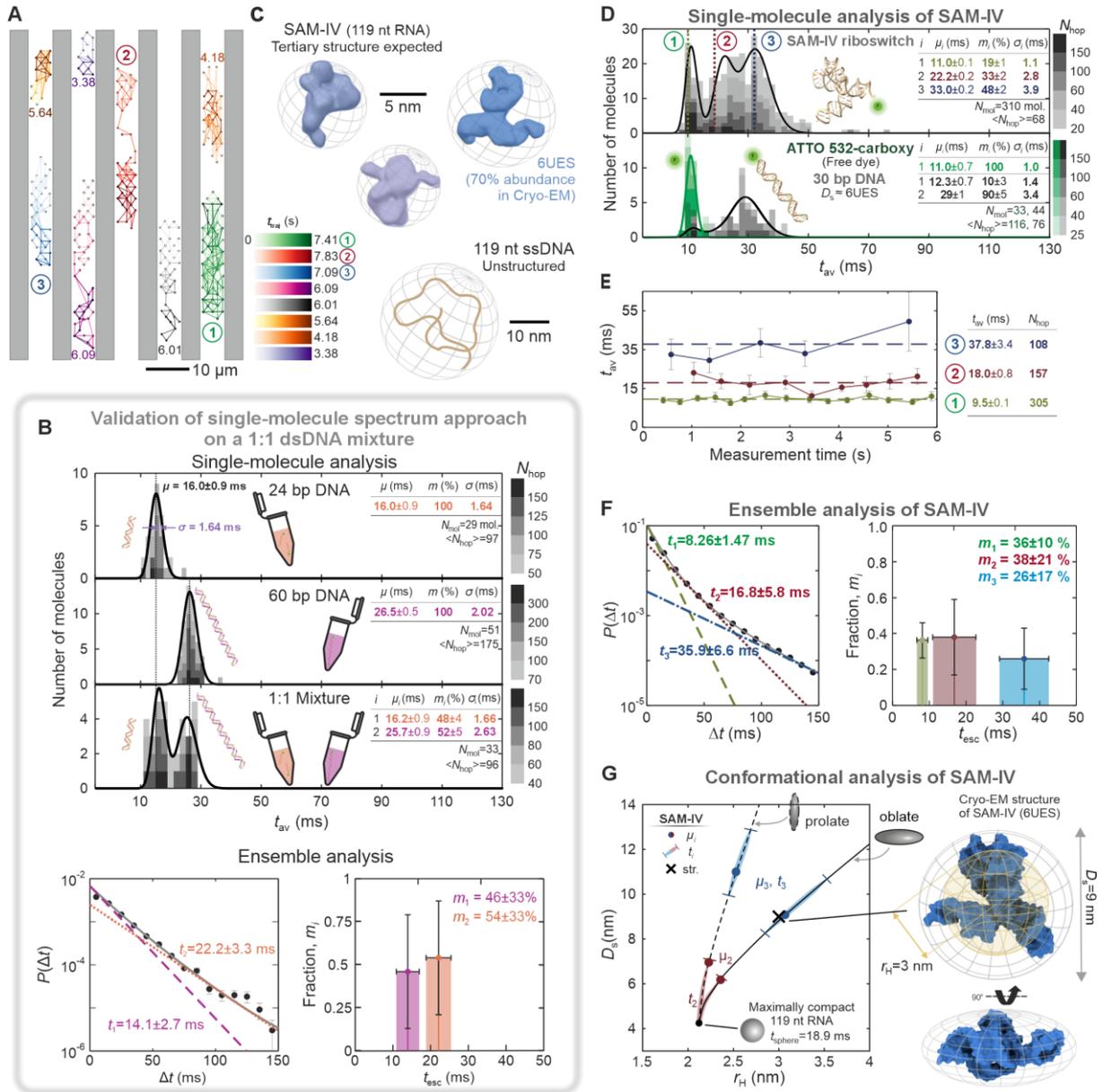

**Fig. 4. Single molecule spectra of SAM-IV riboswitch conformation and comparison with a structural model**. (**A**) Schematic depiction of measured individual molecular trajectories of the SAM-IV riboswitch diffusing in a landscape of entropic traps in slits of height $h_1 \approx 20$ nm. The time evolution of each trajectory is indicated by the color intensity which is scaled relative to the total trajectory duration $t_{traj}$, typically 2-10 s. (**B**) Validation of the single-molecule escape-time spectrum approach on a 1:1 mixture of 24 bp and 60 DNA (top). Each integer entry in the histogram corresponds to the average escape time, $t_{av}$, recorded for one molecular trajectory, with the grey level denoting the observed number of hops, $N_{hop}$. Pure 24 bp and 60 bp DNA samples show single peaks of average times, $\mu \approx 15$ and 26 ms respectively, whilst the mixture displays 2 components of approximately equal fractional weight, $m_i$. Fits of the spectra are performed according to Eq. (S24) (black curve, see supplementary text S6.1 for detail). The same single molecule trajectory data pooled and fit with a bi-exponential reveals two timescales and fractional weights that agree within error with the single molecule trajectory analysis (bottom). Measurement precisions on the characteristic timescales are well

estimated by the s.e.m and are captured by the fit errors on $\mu_i$ and $t_i$ in the single-molecule and ensemble approaches respectively. (**C**) Multiple possible 3D conformations reported for the SAM-IV riboswitch compared with a 119 nt unstructured ssDNA (see fig. S13). (**D**) Single-molecule spectra of the SAM-IV riboswitch (top), accompanied by superimposed spectra of ATTO532-carboxy dye and 30 bp DNA (bottom). Fitted timescales and component fractions $\mu_i$ and $m_i$ in each case permit assignment of species and molecular states to the 3 inferred peaks in the ETs spectrum of SAM-IV. State-1 maps on to free dye, also present in other nucleic acid samples ($\mu_1 \approx 12$ ms). Two distinct conformations attributable to SAM-IV (States 2 and 3) are much more compact than the unstructured 119 nt ssDNA control (see fig. S18), and display $\mu_i$ values similar to 30 dsDNA. (**E**) Recorded time traces of sequential escape events $\Delta t$, averaged over 16 consecutive escape events, for molecules 1, 2 and 3, display distinct and stable overall average values $t_{av}$ over the $> 6$ s measurement duration. (**F**) Ensemble measurements and corresponding multi-exponential fit of $N = 4.4 \times 10^4$ escape events of SAM-IV reveal 3 timescales, $t_i$, and weight fractions that capture the timescales and molecular fractions detected in the single molecule spectra. Comparing s.e.m. values on the measured average $t_i$ with those on the corresponding $\mu_i$ values in **D** illustrates how the measurement precisions delivered by single-molecule trajectory approach outstrip the ensemble approach despite the fact that the trajectory approach relies on half the number of total recorded events ($N = 2.1 \times 10^4$). (**G**) Theoretical $(D_s, r_H)$ plot for objects of fixed volume 40 nm$^3$ – corresponding to an RNA molecule with $M = 38.51$ kDa characterizing SAM-IV – presents limiting cases for oblate (solid line) and prolate ellipsoids (dashed line) which intersect at a point given by the most compact, spherical state corresponding to an escape time $t_{sphere} = 18.9$ ms (see supplementary text S5.6 for details). Measured escape times and associated uncertainties in (**D**) and (**F**) imply $(D_s, r_H)$ values as follows: values from single-molecule measurement $\mu_3$ (blue symbol) and the mean ensemble-fit timescale $t_3$ (blue shaded zone) indicate $(D_s, r_H)$ coordinates describing a structure centered on either the 'oblate' or 'prolate' curves. Inferred coordinates for an oblate ellipsoid lie very close to the point characterizing the cryo-EM structure (pdb code: 6UES – black cross). Timescale $\mu_2$ (red symbol) is consistent with a more compact state, on either the 'oblate' or 'prolate' curves, with red shaded regions corresponding to indications from ensemble-measurement value $t_2$.

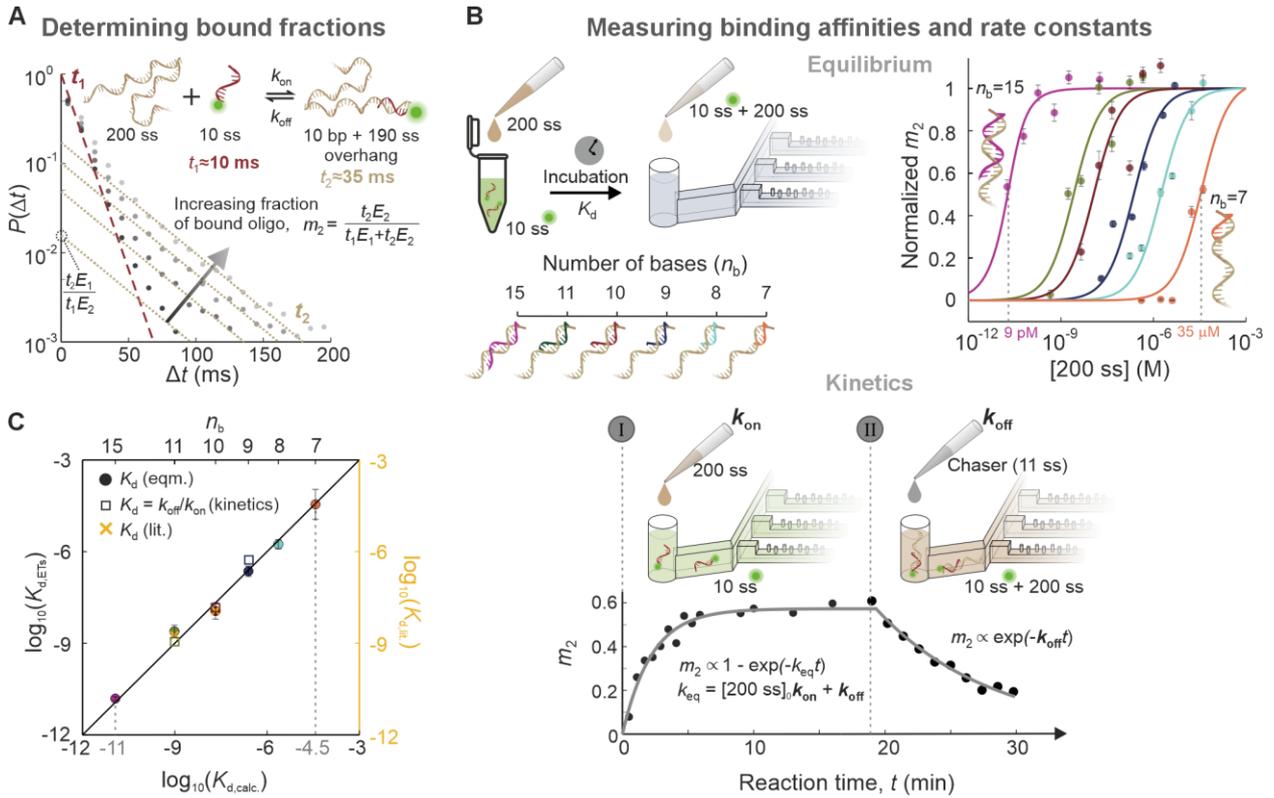

**Fig. 5. Measuring association and dissociation rates, and binding affinities in DNA hybridization**. (**A**) Illustration of the experimental approach to determine the fractions of free and bound molecules in a bi-molecular binding reaction. Here, a labelled 10 ssDNA oligo binds to an unlabelled 200 ssDNA fragment. Incubation with a high concentration of 200 ssDNA shifts the free oligo timescale from $t_1 \approx 12$ ms to the fully-bound timescale $t_2 \approx 21$ ms. At intermediate concentrations, escape time histograms obtained in 1-min long measurements are fit with a bi-exponential holding $t_1$ and $t_2$ fixed to yield amplitudes of two exponentials reflecting the bound fractions $m_1$ and $m_2$. $E_i$ denote event fractions which are distinct from fractional species abundances, $m_i$ (see supplementary text S7.1). (**B**) Measurements of $m_2$ following incubation of 0.02-1 nM of labelled oligo with 200 ssDNA of increasing concentration are fit with Eq. (S29) to obtain the affinity, $K_d$, of the interaction. Measurements for a range of oligos of length $n_b = 7$ to 15 bases demonstrates $K_d$ measurements in the range $10^{-11}$ M to $10^{-4}$ M (top right). Constructing $P(\Delta t)$ histograms of the reaction mixture in real time following mixing provides a measure of $m_2$ vs. reaction time, $t$, which is fit with Eq. (S30) to obtain the effective on-rate constant $k_{eq}$ (bottom left). Equilibrated, fully bound mixtures of oligo and 200 ssDNA are mixed with an excess of unlabelled oligo, and the decaying bound fraction, $m_2$, measured as a function of time to yield the dissociation rate constant $k_{off}$ using Eq. (S32) (bottom right). $k_{eq}$ and $k_{off}$ together determine the association rate constant, $k_{on}$, using Eq. (S31), from which $K_d$ can be determined independently using Eq. (S33). (**C**) Measurements of $K_d$ from both equilibrium (circles) and kinetic (squares) approaches compare well with theoretically estimated and literature values (crosses) over a range of nearly 7 orders of magnitude (see supplementary text S7.5 for further discussion).

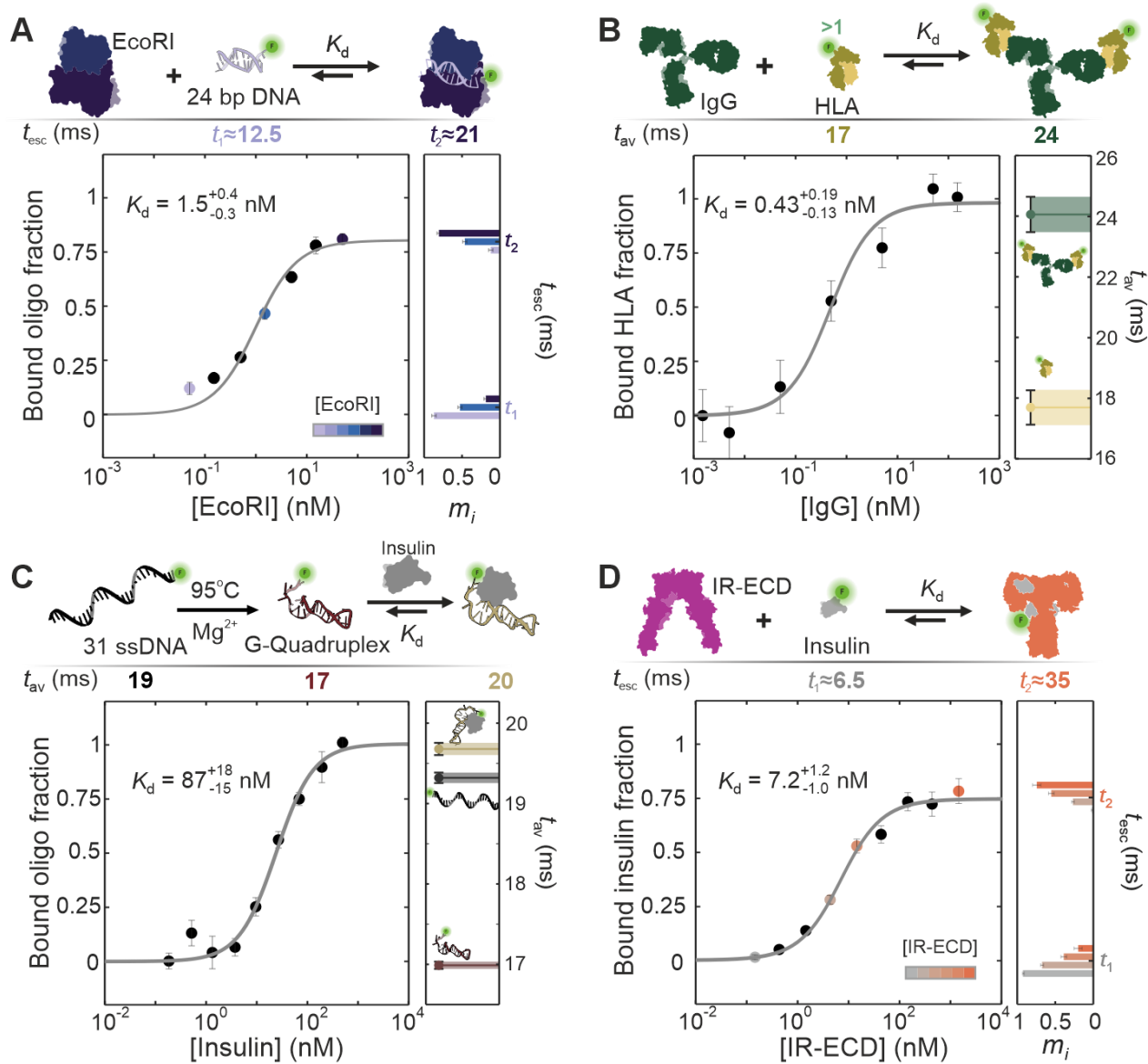

**Fig. 6. Measuring binding affinities in DNA-protein and protein-protein interactions.** When disparities in escape times characterizing the bound and free states, $t_1$ and $t_2$, are larger than a factor 2 the method in Fig. 5A can be used to determine $m_2$ and therefore $K_d$ values for the interaction of (**A**) the restriction enzyme EcoRI (MW ca. 28.5 kDa) with 250 pM of a labelled 24 bp DNA oligo (ca. 15 kDa) carrying the restriction sequence, (**D**) the insulin receptor ectodomain (IR-ECD – 280 kDa) with 2 nM of fluorescently labelled insulin (ca. 6 kDa). When the shift in timescale upon binding is much smaller (ca. 5-10%) we use the average escape time, $t_{av}$ as an indicator of $m_2$ (see fig. S21 and supplementary text S7 for more details). Using this method, we determine interaction affinities of (**B**) 100 pM of the fluorescently labelled HLA antigen (ca. 48 kDa) with the corresponding Immunoglobulin G (IgG – ca. 150 kDa). (**C**) Folding of the insulin-binding 31-base IGA3 aptamer results in a 10% reduction of $t_{av}$ that subsequently systematically increases upon binding of increasing concentrations of insulin added to 5 nM of the aptamer. In all cases, the measured $K_d$ values are in good agreement with the literature.

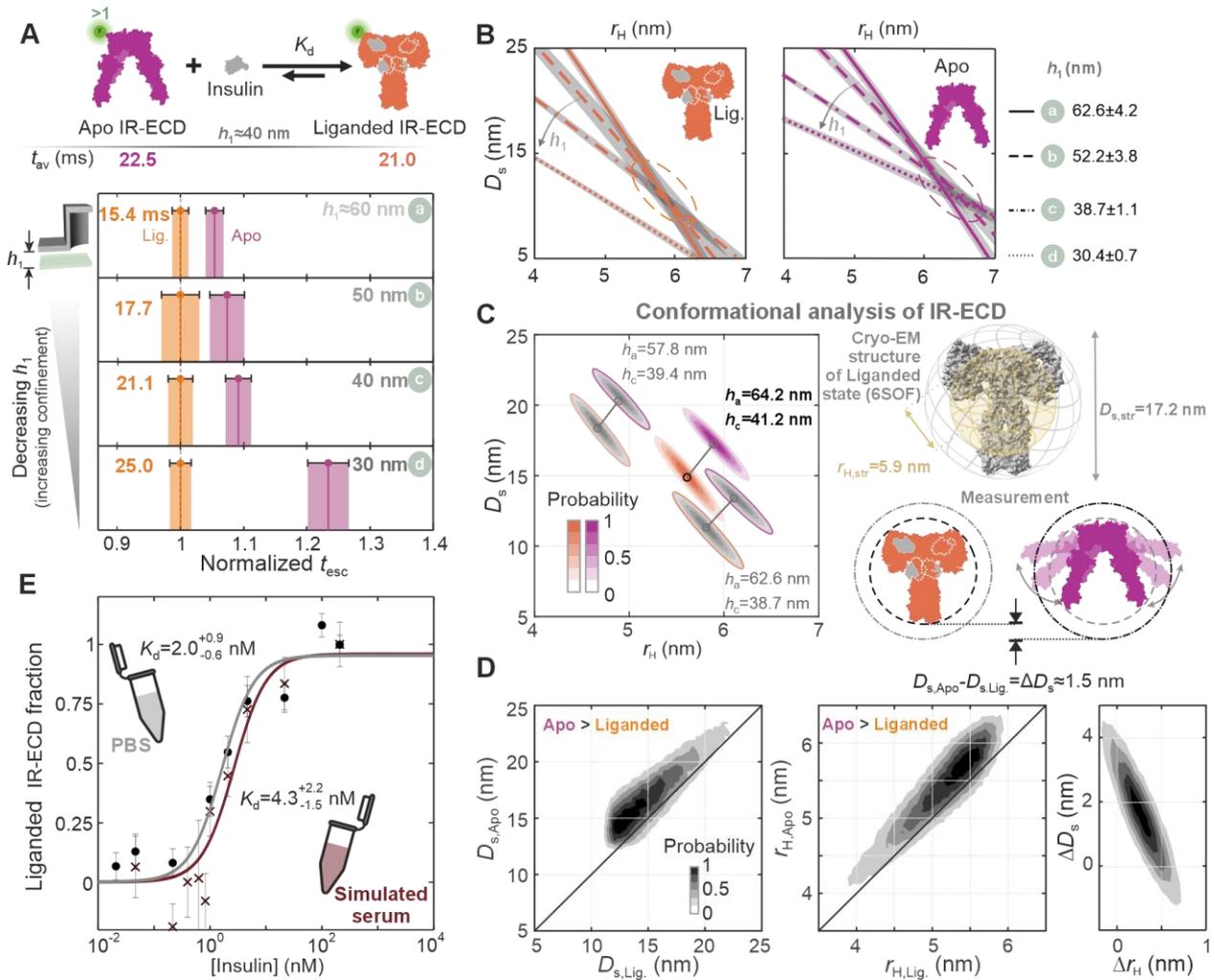

**Fig. 7. Detection and structural modelling of conformation changes of the Insulin Receptor Ectodomain (IR-ECD) upon insulin binding.** (**A**) Upon incubation with 100 nM insulin, well above the expected $K_d$ of the interaction reported in Fig. 6B, the fluorescently labelled IR-ECD displays a reduction in average escape time $t_{av}$ from $22.5 \pm 0.3$ ms to $21 \pm 0.1$ ms in a device where $h_1 \approx 40$ nm, used for $K_d$ measurements in (**E**). Examining the ligand-free 'apo' and the liganded states of the IR-ECD in slits of height ranging from $h_1 \approx 60$ to $30$ nm reveals progressively increasing fitted $t_{esc}$ values for both states, as well as a systematically increasing ratio of the two timescales, clearly indicating differences in 3D conformation between the two states (plots a-d). (**B**) Graphical presentation of Eq. (1) for the liganded state (left) and 'apo' state (right) using fitted $t_{esc}$ values in slits with quoted $h_1$ values determined by calibration (see fig. S14). Lines represent $(D_s, r_H)$ solutions expected for $h_1$ corresponding to the mean value obtained from. Shaded grey bands denote uncertainties given by the precision of the $t_{esc}$ measurement. Similar to Fig. 3, the intersection of any two bands gives the range of allowed values of $(D_s, r_H)$ characterizing the molecular state. (**C**) Three possible sets of solutions describing the 'apo' and liganded states obtained by assuming $h_1$ values that vary within one standard deviation of the mean values measured for devices 'a' and 'c' (left). Note that the true uncertainty in slit height (<ca. 1 nm) is expected to be much smaller than that obtained by the calibration procedure (*3*). Although absolute values for $D_s$ and $r_H$ are not determined with high accuracy at present, the difference between the two states is systematic and points to an increase of $\Delta D_s \approx 1.5$ nm and $\Delta r_H \approx 0.25$ nm for the 'apo' state compared to liganded case. Conceptual view of the inferred conformational difference between the 'apo' state (4ZXB) and the cryo-EM ligand saturated state (6SOF) as inferred from ETs measurements (right). (**D**) Plots of $D_{s,Apo}$ vs. $D_{s,Lig.}$ and of $r_{H,Apo}$ vs. $r_{H,Lig.}$ for several different pairs of $h_1$ values lying within the range inferred from calibration for the (a, c) device geometry pair. The plots confirm a systematic compaction in overall extent of the liganded state compared to the ligand-free 'apo' state (fig. S14). (**E**) Affinity measurement for the binding of the IR-ECD to insulin. The bound fraction is inferred from the reduction in average timescale $t_{av}$ of labelled IR-ECD due to the conformation change upon insulin-binding, as shown in (**A**) (see Fig. 6, fig. S24 and supplementary text S7.10 for details). Measurements in both PBS (black data) and in simulated human serum (red data) give similar $K_d$ values that are consistent with the literature.

# Supplementary Materials for

## Measurements of molecular size and shape on a chip

Xin Zhu[1]†, Timothy J. D. Bennett[1]†, Konstantin C. Zouboulis[1,2], Dimitrios Soulias[1], Michal Grzybek[3], Justin L. P. Benesch[1,2], Afaf H. El-Sagheer[4,5], Ünal Coskun[3,6-8], Madhavi Krishnan[1,2]*

[1] Physical and Theoretical Chemistry Laboratory, Department of Chemistry, University of Oxford; South Parks Road, Oxford OX1 3QZ, United Kingdom

[2] The Kavli Institute for Nanoscience Discovery, University of Oxford; Sherrington Road, Oxford OX1 3QU, United Kingdom

[3] Center of Membrane Biochemistry and Lipid Research, University Hospital and Faculty of Medicine Carl Gustav Carus, Technical University Dresden; 01307 Dresden, Germany

[4]School of Chemistry and Chemical Engineering, University of Southampton; Highfield, Southampton, SO17 1BJ, United Kingdom

[5]Institute for Life Sciences, University of Southampton; Highfield Campus, Southampton, SO17 1BJ, United Kingdom

[6]Paul Langerhans Institute Dresden (PLID) of the Helmholtz Center Munich at the University Hospital Carl Gustav Carus and Faculty of Medicine, Technical University Dresden; 01307 Dresden, Germany

[7]German Center for Diabetes Research (DZD), 85764 Neuherberg, Germany

[8]Max Planck Institute of Molecular Cell Biology and Genetics, 01307 Dresden, Germany

*Corresponding author. Email: madhavi.krishnan@chem.ox.ac.uk

† These authors contributed equally to this work

Table of Contents

List of Figures

# S1. Further experimental characterization and procedures

## S1.1. Examination of homogeneity of device properties, measurement reproducibility and measurement precision

In order to verify the reproducibility of ETs measurements, $t_{esc}$ was measured for 60 bp DNA (sequence shown in Table S1) at 0.5 nM concentration in PBS (pH 7.4), with samples prepared afresh for each measurement and run in the same device every day for a week, see Fig. S7B. We observed a standard deviation over the 7 consecutive measurements of ca. 0.5% of the $t_{esc}$ value. Although this is ca. 3 times the imprecision (which is itself close to the statistical limit) on any single measurement, we find this reproducibility to be indicative of highly consistent experimental conditions.

In order to examine the spatial uniformity of ETs devices within an illuminated field of view (FOV, ca. 150x150 μm), we split the FOV into 25 regions and analyzed escape events detected in each region separately, Fig. S7C. We observed a standard deviation over the 25 regions of ca. 0.9% of the $t_{esc}$ value, for measurements of 24 bp (sequence shown in Table S1) and 60 bp DNA at 0.5 nM concentration in PBS (pH7.4), see Fig. S7D. Although this is ca. 2 times the imprecision on any single measurement, we again find this consistency indicative of a highly uniform device geometry.

The measurement imprecision that we report in escape time measurements is evaluated as the fit error on $t_{esc}$. By varying the number of escape events, N, taken from the above measurement of 60 bp DNA, we verify that the fit error corresponds effectively exactly to the theoretical imprecision of $t_{esc}/\sqrt{N}$. This relation comes from the definition of the standard error on the mean value ($\sigma/\sqrt{N}$) for N events taken from an exponential distribution of escape times (characterized by mean, $t_{esc}$, and standard deviation, $\sigma = t_{esc}$) which describes the escape process. Simulated data drawn from an exponential distribution of escape times, with mean value equal to that of the 60 bp DNA measurement, was analyzed identically to the experiments and the fit error shows an identical trend as expected. Fig. S7E validates this feature of our measurement approach as discussed in detail in previous work (*3*).

Whilst the above measurements concern measurement stability and reproducibility in a single device, we do however see departures in average slit height, $h_1$, between nominally identical devices of up to 10 nm, although this is routinely at the level of ca. 2 nm for $h_1 = 25$ nm (see Fig. S5E). In order to account for this device-to-device variability we always calibrate our devices used for molecular stereometry, as outlined later in section S5.3.

## S1.2. Nanoslit surface passivation procedure

Prior to measurements of proteins, the devices are often, but not always, passivated to reduce non-specific binding of molecules to surfaces (sticking) following procedures described by Emilsson et al. (*67*). The device is first flushed with Piranha solution (3:1 Sulfuric acid and hydrogen peroxide, Sigma-Aldrich) for 20 min, followed by 20 min of 1 M KOH (Sigma-Aldrich). After a thorough clean with DI water, the device is then incubated with mPEG-silane (5 kDa MW, Sigma-Aldrich) prepared in a solution of 1 mL ethanol, 5 μL 10 mM acetic acid (both Sigma-Aldrich) and 25 μL water in a dark room for at least 48 hr. Then the device is cleaned with water for 5 min, and 0.5% Tween 20 in PBS (pH 7.4) for 20 min. To assess the change in slit height following the coating, the slit height was measured both before and after coating using the calibration procedure relying on four globular proteins, as outlined in section S5.3, data is shown in Fig. S5D. For example, for two devices of the same nominal fabricated height ($h_1 = 40$ nm) we found that $h_1 = 47.5 \pm 1.2$ nm without coating and $h_1 = 27.7 \pm 0.4$ nm with mPEG coating as shown in Fig. S5D. We consistently observe a ca. 20 nm reduction in $h_1$ upon coating as shown in see Fig. S5E, which is consistent with a 5 kDa mPEG brush thickness of ca. 10 nm as reported by Emilsson *et al.* (*67*). 0.03% Tween 20 was sometimes added to the measurement buffer to further facilitate the device passivation and minimize non-specific adhesion.

**S1.3.** High-throughput single-molecule counting down to femtomolar concentrations

In some instances, device passivation was not sufficient to completely reduce the sticking, leading to a lower than expected number of escape events detected in our experiments. Note that here we compare counts with those expected for highly charged DNA molecules that are known to display minimal adhesion to the slit surfaces. In general, we expect sticking to surfaces occurring in any part of the experimental protocol to affect the number of recorded counts. E.g., these differences in observed counts could also be due to sticking to preparation tubes, despite the use of LoBind tubes (Eppendorf) for sample preparation. To quantify this reduction in concentration a series of reference measurements of molecules which showed minimal signs of sticking were performed. The nominal concentration of 5 nt single-stranded DNA and insulin were measured before preparation at ca. 1 µM using a Nanodrop. Serial dilutions of each sample from 10 fM to 100 pM were prepared in PBS (pH 7.4) with 0.03% Tween 20 added. The samples were loaded into a freshly coated device with $h_1 \approx 70$ nm, and the counts were measured under flow for 1-20 minutes, depending on sample concentration. The number of escape events per unit time (event rate) were measured for each sample using an identical microscope field of view, and a flow rate of ca. 100 µm/s. The flow rate, is estimated from the volume flow rate, $Q$, using the Hagen-Poiseuille equation for a rectangular channel, Eq. (S1) (*68*), for a pressure drop $\Delta P$, across a channel of length, $L$, filled with fluid of viscosity, $\eta$, and assuming the channel width, $W'$ is much larger than the height.

$$Q \approx \frac{\Delta P \, h_1^3 \, W'}{12 \eta L} \left(1 - 0.63 \frac{h_1}{W'}\right) \tag{S1}$$

A linear fit of log(concentration) vs log(event rate) recovers the quantitative sensitivity of ETs in the current configuration to molecular concentration down to low femtomolar concentrations – a "Femtodrop" (Fig. S5F)! Increased throughput and lowering of this detection threshold by at least a factor 100 may be achieved through further optimization of device design. This approach provides an excellent measure of concentration at the ultra-low concentrations inaccessible to other low-sample volume laboratory measurement techniques such as the Nanodrop and is used to estimate actual measured concentrations of proteins in affinity measurements in section S7. The error bar on each data point in Fig. S5F is calculated as (event rate) $/\sqrt{N}$, for a total number of events, $N$.

**S1.4.** Negligible electrostatic contribution to measured escape times

Operating at high salt concentrations where both the magnitude of the surface electrical potential of silica, and the screening lengths are very small, we expect electrostatic effects to be negligible. Indeed, from our experience with electrometry a reasonable electrostatic contribution is achieved for $\kappa h < 6$. Here $\kappa^{-1}$, the Debye length is about 9 nm in an electrolyte containing monovalent salt at 1 mM concentration. For the ETs experiments, we have $\kappa^{-1} \approx 0.75$ nm at a salt concentration of 160 mM, putting our measurements in the regime $15 < \kappa h < 50$, where electrostatic interactions are expected to be at least 4 orders of magnitude weaker than thermal energy. Furthermore, were charge effects indeed present contrary to theoretical expectations, they would significantly contribute to measured escape times. As a result, $r_\mathrm{H}$ inferences for example would not agree well with expectations from structures and with independent experimental measurements, i.e., FCS. Finally, control measurements on DNA at different salt concentrations (e.g., 50 mM and 100 mM) show no difference, further confirming the absence of electrostatic effects in the large $\kappa h$ regime of ETs experiments as expected (Fig. S5C).

## S2. Data analysis and fitting
### S2.1. Fitting histograms of escape-time data

Molecular residence times, $\Delta t$, are identified from the intensity vs. time traces in each trap as described in previous work (3). The events are binned into a histogram with $n$ bins, where each bin is centered at $\Delta t_i = t_{\text{exp}} + (i-1)(t_{\text{exp}} + t_{\text{lag}})$ where $i$ represents the length of the event in number of frames, and $t_{\text{exp}}$ and $t_{\text{lag}}$ are the exposure time and lag time from the imaging. Then, event counts per bin ($N_i$) are normalized relative to the total number of events, $N$, as $\frac{N_i}{N(t_{\text{exp}}+t_{\text{lag}})}$, to give probability density distributions $P(\Delta t_i)$. The normalized error on each bin is defined as $\sigma_i = \frac{\sqrt{N_i}}{N(t_{\text{exp}}+t_{\text{lag}})}$, while errors on bins containing zero events are set equal to $[N(t_{\text{exp}} + t_{\text{lag}})]^{-1}$. Weighted non-linear regression is used for fitting such histograms with fit function $P_{\text{f}}(\Delta t)$ given by Eq. 2 in the main text, and repeated here for clarity, Eq. (S2). Here $k$ is the number of components, $E_i$ is the amplitude of the component with escape timescale, $t_i$ and $c$ is a constant value capturing an experimental offset,

$$P_{\text{f}}(\Delta t) = \sum_{i=1}^{k} \left(\frac{E_i}{t_i} e^{-\frac{\Delta t}{t_i}}\right) + c \quad (S2)$$

The first bin with $\Delta t_i = t_{\text{exp}}$ is always excluded from fitting to eliminate additional noise from both poor detection of short events arising from transiently trapped molecules, and from random spikes in background noise. Weights in the fitting process are defined as $w_i = 1/\sigma_i^2$ and errors on the fitted escape times are treated as the standard error of the mean.

### S2.2. Determining number of timescale components characterizing an escape-time histogram (multi-exponential fitting)

Since the escape-time histograms presented in this work have a large number of events per bin ($N_i$), the Poisson distribution in each bin population may be assumed to approach a normal (Gaussian) distribution. Consequently, the optimal number of exponents needed for fitting a histogram is determined by finding the maximum probability of the expected reduced $\chi$-squared being greater than the observed value, $P_{\text{DoF}}(\chi_r^2 > \chi_{r,o}^2)$. For a histogram of $n$ bins with measured $(\Delta t_i, P(\Delta t_i) \pm \sigma_i)$ and fit $(\Delta t_i, P_{\text{f}}(\Delta t_i))$ data points, the observed reduced $\chi$-squared is defined as $\chi_{r,o}^2 = \chi_o^2/DoF$ (69), where $\chi_o^2$ is the observed $\chi$-squared defined by Eq. (S3), $DoF = n - c_{\text{p}}$ is the degrees of freedom, where $c_{\text{p}}$ is the number of free parameters ($E_i, t_i, c_i$) in the fit function.

$$\chi_o^2 = \sum_{i=2}^{n} \left(\frac{P(\Delta t_i) - P_{\text{f}}(\Delta t_i)}{\sigma_i}\right)^2 \quad (S3)$$

To assess the reduced $\chi$-squared metric, a simulated multi-exponential distribution was generated using as "ground truth" values, the fit results inferred from the experimental histogram of SAM-IV riboswitch (see main text Fig. 4 and Fig. S6A, B). Fig. S6A illustrates the normalized probability density distributions $P_n(\Delta t)$ vs. $\Delta t$ (green curves) for $N \approx 4.4 \times 10^4$ recorded escape events of SAM-IV riboswitch (green points), also shown in Fig. 4F, after fitting with Eq. (S2) up to $k = 4$ components. Similarly, Fig. S6B shows the $P_n(\Delta t)$ vs. $\Delta t$ (grey curves) for an identical number of simulated escape events (black points) based on the fit escape times $t_i$ and event fractions $E_i$ obtained by fitting a 3-exponential decay to the experimental SAM-IV riboswitch (green points and $k = 3$ curve in Fig. S6A). For every fitted curve in Fig. S6A, B the $P_{\text{DoF}}(\chi_r^2 > \chi_{r,o}^2)$ was calculated as in Fig. S6C. The probability of a single exponential fit being in agreement with either the simulated (black) or experimental (green) data was less than 0.05%, while for the bi-exponential it increased to ~20% for both. Maximum agreement at ~85% for simulated and ~57% for experimental results was achieved when using a 3-exponential function. Adding a further component led to a decrease in $P_{\text{DoF}}(\chi_r^2 > \chi_{r,o}^2)$ for both the simulated and experimentally measured $P(\Delta t_i)$ distributions. As a result, the reduced $\chi$-squared test suggested that a tri-exponential function is the optimal for fitting both the simulated and experimental histogram for the SAM-IV riboswitch. We note that both the fitted escape times $t_i$, and component amplitudes $E_i$, from the simulated distribution (grey points) are in good agreement with the ground truth values (red points, Fig. S6D) used for simulations. The table lists the ground truth and fitted event fractions $E_i$ with their fit error, for the 3-exponent simulated distribution.

Although the analysis above focused on histograms with three species, larger $N$ values could enable the inference of a larger number of exponential components in the data. We illustrate this concept using a simulated case shown in Fig. S6E. Here we generate a histogram of $N \approx 1.5 \times 10^7$ events (black points) constructed based on five time- or molecular state-components: three of these components are given by escape times and event fractions obtained from the experimental ground truth values from the SAM-IV riboswitch case, but we include two additional time components that are described by longer escape times spaced with a similar fractional separation as the riboswitch timescales (ca. factor 2 between consecutive components). The six curves in Fig. S6E represent fits to the simulated data of the form of Eq. (S2), where $k = 1, …, 6$ and $P_{\text{DoF}}(\chi_r^2 > \chi_{r,o}^2)$ is determined for each of the 6 fits, as illustrated in Fig. S6F. The probability of single, double and triple exponential fits (i.e., $k = 1, 2, 3$ in Eq. S1) agreeing with the simulated data was less than 0.05%, while for the 4-exponential ($k = 4$) the probability increased to approximately ~18%. Best agreement at ~65% was achieved when using a 5-exponential function ($k = 5$). Adding a sixth exponent led to a decrease in $P_{\text{DoF}}(\chi_r^2 > \chi_{r,o}^2) \approx 55\%$, with the additional component notably duplicating the $t_i$ value from the fifth exponential, i.e., $t_5 \approx t_6$ for the $k = 6$ case. The analysis thus shows that the reduced $\chi$-squared test suggested a 5-exponential ($k = 5$) function provides an optimal fit to the simulated data capturing the known ground truth. Fig. S6G illustrates the good agreement between the fitted escape times $t_i$, and component amplitudes $E_i$, from the simulated distribution (grey symbols) and the ground truth values (red symbols) used in the simulation.

## S3. Theoretical model relating average escape time to molecular properties $r_H$ and $D_s$

The average residence time of molecules in traps, which we refer to as the escape time $t_{\text{esc}}$, may be written as follows:

$$t_{\text{esc}} = t_r \exp(W/k_B T) \tag{S4}$$

In the regime $W > 4\,k_\text{B}T$ where $W$ is the free energy difference between molecular states in the pocket and the slit, $T$ is the absolute temperature and $k_\text{B}$ is Boltzmann's constant. Furthermore, $t_\text{r} \propto \frac{D_\text{p}^2}{16 D_\text{t}}$ denotes an estimate of the residence time of the molecule in a pocket region of diameter $D_\text{p}$, in the absence of any modulation of slit height (i.e., $h_1 = h_2$). Here $D_\text{t}$ is the translational diffusion coefficient, which can be related to the Stokes's radius, $r_\text{H}$, of a particle in a fluid of viscosity $\eta$ by the Stokes-Einstein equation:

$$D_\text{t} = \frac{k_\text{B}T}{6\pi\eta r_\text{H}} \tag{S5}$$

where $T$ is the absolute temperature and $k_\text{B}$ is the Boltzmann constant. In contrast to previous studies using the electrostatic fluidic trap, in this work we operate exclusively at high salt concentrations (~150 mM), which ensures that the electrostatic contribution to $W$ is negligible and the trap is purely entropic (21). This was confirmed by measurements of a highly charged molecule (120 bp dsDNA) in both 50 mM and 100 mM NaCl in a typical device with $h_1 \approx 70$ nm, where we observe negligible difference in $t_\text{esc}$, see Fig. S5C. We therefore have $W \cong \Delta F_\text{ent} = -T(\Delta S_\text{trans} + \Delta S_\text{conf})$. Importantly, the translational entropy contribution for an idealized 'point object', with no spatial extent, may be given by $\Delta S_\text{trans} = -k_\text{B}\ln(h_2/h_1)$, whilst the second, configurational entropy term, $\Delta S_\text{conf}$, which may enter the picture for large non-spherical objects is discussed later.

In the case of low confinement of the escape process, such as for small hard spheres whose radius $R \approx r_\text{H} \ll h_1$, we may set $\Delta S_\text{conf} = 0$ in Eq. (S4) which thus reduces to Eq. (S6):

$$t_\text{esc} \propto r_\text{H}(h_2/h_1) \tag{S6}$$

In the case of greater nanoscale confinement such as for larger spheres, or shallower slits, where $h_1$ approaches the diameter of a sphere, $D$ (given by the regime $D < h_1 \ll h_2$) we expect a non-linear enhancement of escape times with increasing object size $D$, see Fig. S8. The reason behind the enhancement lies in the ratio of the effective number of translational states accessible to a finite-sized molecule in the pocket vs. in the slit. This quantity maybe expressed more accurately as a function of object diameter, i.e., by the ratio $\frac{h_2-D}{h_1-D}$, which now supersedes $h_2/h_1$ for a point object in Eq. (S6), and reveals how the diameter, $D$, of the object can play an even strong role in the experimental readout. Furthermore, a measure of the radius of the object continues to appear in $t_\text{r}$. Since we generally always have $D \ll h_2$, Eq. (S6) now reads

$$t_\text{esc} \propto r_\text{H}(h_2/h_1)\,(1-D/h_1)^{-1} = r_\text{H}(h_2/h_1)[1 + D/h_1 + (D/h_1)^2 + (D/h_1)^3 + \cdots] \tag{S7}$$

When $D$ is a substantial fraction of the slit height $h_1$, a large number of terms in Eq. (S7) contribute to $t_\text{esc}$, which therefore increases strongly with the diameter of the object, offering the prospect of greatly enhancing disparities in measured escape timescales for similar molecules (see Fig. 2C of main text and Fig. S8). For example, in the regime of strong confinement given by, e.g., $D/h_1 = 0.5$, the 4th order term contributes 3% to the measured value of $t_\text{esc}$ which is much larger than our typical measured precision of <1%. This showcases the power of $h_1$ as a tunable experimental parameter that directly impacts the sensitivity of the readout to small changes in molecular diameter. Throughout this work $260 < h_2 < 360$ nm, and note that because $h_2 \gg D$ the measured response is not very sensitive to small changes in $h_2$, in contrast to $h_1$.

$$t_{esc} \propto r_H \left( \frac{h_2 - D_s}{h_1 - D_s} \right) \tag{S8}$$

The response expected for $D < h_1$, is given by Eq. (S8) which is used to interpret all escape times in this study. Specifically we use Eq. (1) (same as Eq. (S21)), obtained from Brownian Dynamics simulations, to fit the experimental data, as described in detail later in Section S5.1. For globular molecules with hydrodynamic radius, $r_H$, we use a bounding sphere diameter, $D_s = D = 2r_H$ in Eq. (S8). For non-spherical molecules we use $D = D_s$ as discussed in the main text, which accounts for the rotational (orientational) contribution to the entropy difference. In the regime of extreme confinement given by $D \geq h_1$, further conformational entropy contributions may enter the picture, but this regime is not considered in the present study.

## S4. Escape-time measurements of biomolecules

Fluorophore-derivatives, proteins and nucleic acids (DNA) shown in Fig. 2A of the main text were all measured on the same device with $h_1 \approx 25$ nm.

### S4.1. Proteins

Insulin (INS), ubiquitin (Ub), thioredoxin (TRX), RNase A, myoglobin (MB), b-lactoglobulin (LBG), carbonic anhydrase (CA), apo-Transferrin (TF), human insulin detector antibody (IgG), apoferritin (FER), Starmaker-like (Stm-l), Prothymosin a (ProTa) and the proline-rich polypeptide (prepared, where applicable, as outlined in Materials and Methods section) are measured at 1-10 nM concentration in PBS (pH 7.4) with 0.03% Tween 20. The escape time histogram was fitted with a single exponential distribution for all the proteins apart from apoferritin and an escape time was determined. Apoferritin data was fitted with a bi-exponential distribution to accommodate a small fraction of ferritin monomer in the sample, as well as the expected 24-mer (78), see Fig. S9. The biexponential fit yielded $t_1 = 16.5$ ms, corresponding to a mass of 26 kDa which is consistent with the presence of some apoferritin monomeric subunits with $M \approx 20$ kDa. $t_2$ is the timescale plotted in Fig. S9A and corresponds to the 24-mer apoferritin complex. The globular proteins in main text Fig. 2A are fit with a line $t_{esc} = c_1 MW^{1/3} + c_2$. While the disordered proteins are fit with a line $t_{esc} = c_1 MW^{3/5} + c_2$, see Fig. S9B.

### S4.2. Dye derivatives

Five chemical derivatives of ATTO532 and ATTO542 are measured at 1 nM concentration in 100 mM NaCl, 0.1 mM Tris (pH 7, Carl Roth). Movies at a frame rate of 200 Hz, with $t_{exp} = 3$ ms, are recorded for each sample, and escape times determined from a single exponential fit to the escape time histogram. The ability of the measurement to distinguish between small differences in molecular weight is illustrated by the $t_{esc}$ difference measured between the NHS-ester and carboxy derivatives whose mass difference is just 25 Da – approximately two carbon atoms. Indeed, the ability of the measurement to distinguish small molecules can be estimated from a linear fit of $t_{esc}$ vs. molecular weight for the four smallest modifications ($M = 650 - 960$ Da), and evaluating the ratio of the uncertainty on the fitted escape times to the slope of the fitted line. This analysis suggests a measurement sensitivity of 8.8 Da which is less than a carbon atom. An additional dataset for ATTO542 and its derivatives is displayed in Fig. S9C. The mass sensitivity of the measurement is similar for chemical derivatives of both ATTO542 and ATTO532.

The short measurement times required (>10 sec) also enable tracking of reaction progress in real time. A sample of NHS-ester modified ATTO532 in 100 mM NaCl solution containing 0.1 mM Tris (pH 7) was measured intermittently over the course of 45 minutes. Movies of a duration of ca. 1 minute movie were recorded every 2 minutes and an average escape time, $t_{av}$, determined from a mean of all the residence time events of duration $\Delta t$. No significant change with time was observed. When the experiment was repeated in 100 mM NaCl containing 2 mM Tris (pH 9), a substantial time-dependent decrease in $t_{av}$ was observed. Finally, $t_{av}$ reached a plateau value consistent with that of the carboxylic-acid derivative, indicating hydrolysis of the NHS-ester. The half-life, $t_{1/2}$, of NHS ester or the inverse rate constant of the hydrolysis reaction was estimated by fitting the data to an exponential decay

$$t_{av} = t_{av,0} exp(-0.69t/t_{1/2}) + t_{av,f} \tag{S9}$$

Here $t_{av,0}$ denotes the average escape time measured at the start ($t = 0$) of the time-series and $t_{av,f}$ is an escape-time offset characterizing the reaction product.

### S4.3. DNA

A range of double-stranded DNA (dsDNA) fragments with lengths from 30 base pairs (bp) to 59 bp (sequences shown in Table S1) were measured at 1 nM concentration in PBS (pH 7.4). A single exponential fit to the escape time histogram is used to determine the escape time in two different height devices. The ability of the measurement to distinguish small differences in DNA length is estimated similarly to as described in S4.2 for the measurements in the shallower device with $h_1 \approx 25$ nm.

### S5. ETs measurements of biomolecular structure

In order to interpret structural parameters from escape time measurements main text Eq. (1) is used, and repeated here for clarity, Eq. (S10). To verify this relation and to determine the parameters $A$ and $t_o$ for our experimental design, we ran Brownian Dynamics (BD) simulations of spherical particles.

$$t_{esc} = A r_H \left(\frac{h_2 - D_s}{h_1 - D_s}\right) + t_o \tag{S10}$$

### S5.1. BD Simulations on spherical particles

BD simulations were performed in 3D for spherical particles diffusing in a nanostructured parallel-plate landscape. The trajectory of the particle is given by its instantaneous coordinates which were averaged over the exposure time $t_{exp}$ to generate an experimentally relevant trajectory from which escape times were determined as described previously *(3, 22, 28)*. We used translational diffusion coefficients, $D_t$, for a particle of radius, $R = r_H$ as given by Eq. (S5).

We simulated a series of spheres of radii in the range $0.5 < R < 10$ nm. Each series of spheres were simulated in a series of different trap geometries with $10 < h_1 < 70$ nm, and $d =$ $h_2 - h_1 = 300$ nm, reflecting the range of different device geometries used in experiments in this study. Trajectories consisting of 10,000 escape events were simulated for each particle in each device geometry, and escape times were fitted in the same way as experiments. Using Eq. (S11), and setting $2r_H = D_s$, a global fit of these escape times as a function of $r_H$ and $h_1$ yielded a functional form of $A$ as given by Eq. (S11), as shown in Fig. S8, as well as a value for $t_o$. We note that $t_o \approx t_{exp}$.

$$A = \alpha \left(\frac{h_1}{h_2}\right)^\beta + \gamma, \tag{S11}$$

We obtain fit values of $\alpha = 0.16 \pm 0.34$ s/nm, $\beta = -2.35 \pm 0.60$ and $\gamma = 0.265 \pm 0.062$ s/µm, which imply that the prefactor $A$ is only weakly dependent on $h_1$. For $h_1/h_2$ values in this study, $A$ varies between 0.27-0.33 s/µm. As mentioned in section S3, the product, $Ar_H$, can be thought of as the residence time, $t_r$, of a particle in a pocket region in the absence of any modulation of slit height, i.e., $h_1 = h_2$. This residence time can be estimated by the diffusion time of a particle to move a distance $s$ in a 2D plane, $t_r = \frac{s^2}{4D_t}$, where a naive estimate of $s$ may be given by the pocket radius and $D_t = k_B T/6\pi r_H$. For a typical value of $A = 0.3$ s/µm in this work, we have $s = 443$ nm, which is similar to the pocket radius of 275 nm. Indeed, we expect this value to exceed the radius, in line with the escape radius, $R_{esc}$, we impose on the simulated trajectories, as in Ref. (28). For this study we used an escape radius of the following form, $R_{esc}(\text{nm}) = 18718(r_H \times 10^9)^{-0.0037} - 18051 \approx 550 - 650$ nm for the range of interest in $r_H$ (1-5 nm). These parameters fully define the governing Eq. (S10) and are used to obtain molecular size information from escape-time measurements in this study.

**S5.2.** Independent estimates of biomolecular properties from 3D molecular models and FCS

In order to validate experimentally measured structural properties of molecules, a series of independent measurements, calculations and computational models were used to estimate these values.

Stokes' radius $r_H$ using HYDROPRO

Where 3D molecular structures were available, we used HYDROPRO to make a computational estimate of the translational hydrodynamic radius $r_H$ of the species of interest. The program generates a primary hydrodynamic particle model based on atomic coordinate files (e.g., PDB formatted file) by replacing amino acids (or nucleotides) with a collection of overlapping spheres (14). This is called the shell model in which the molecular surface is represented by a shell made up of the smaller overlapping spheres. Note that in this model only the surface exposed to the bulk solvent contributes to the frictional properties and thus the hydrodynamic behavior of the molecules. To compute the $r_H$ for each globular protein reported in main text Figs. 2C, D, we used the respective PDB formatted file ID and the relevant molecular weight, as displayed in Table S4. We assumed a specific volume of 0.73 cm³/g for all species. Replacing this value with values of multiple orders of smaller/larger magnitude did not alter the results reflecting the fact that within this view, frictional properties depend on the surface and not on the mass of the object.

Measured Stokes' radius $r_{H,FCS}$ using Dual-focus fluorescence correlation spectroscopy (2f-FCS)

Dual-focus fluorescence correlation spectroscopy (2f-FCS) was used as an independent measure of effective hydrodynamic radii, $r_{H,FCS}$, of the nucleic acids (dsDNA, dsRNA) presented in Table S5. As described previously a homebuilt dual-focus confocal epi-fluorescence microscope was used to record fluorescence signals from labeled molecules freely diffusing in solution (*18, 23*). Briefly, two identical 520 ± 15 nm laser beams with a 5 ps duration (NKT Photonics EXB-6 High Power Supercontinuum White Light Laser, in conjunction with a NKT Photonics SuperK Varia wavelength selector), pulsed alternatively at 20 MHz, are separated by a polarizing beam-splitter (Thorlabs), with a 25 ns temporal delay imposed on one arm via ~7.5 m of optical fiber. A Nomarski prism (Olympus) is used to alter the path of the two beams depending on their polarization. Following the objective (UPlanSApo 60x W, 1.2 NA, Olympus), two foci are formed which are separated by a small lateral gap ($w_0 = 335$ nm).

Once molecules are excited, the emitted fluorescence passes through the same objective, prism, dichroic mirror and is focused on a single pinhole. Then, fluorescence is divided by a 50/50 beam splitter and detected by two single photon avalanche diodes (Micro Photon Devices, band-passed using filters from Chroma, Thorlabs and Semrock). Single photons from both detectors are independently recorded in time-tagged time-resolved (TTTR) mode which allows the calculation of both the auto-correlation functions (ACF) for each focus separately and the cross-correlation function (CCF) for photons from the two foci. Cross-correlation yields a timescale for diffusive transport of a molecular species between the optical foci, and hence a characteristic diffusion coefficient, $D_t$, for the species, as outlined by (*70*). The hydrodynamic radius, $r_H$ is calculated from $D_t$ using the Stokes-Einstein Eq. (S5).

The diffusion coefficient is quantified by statistical bootstrapping which involves averaging over several fitting repetitions of the CCF, each with a different random subset of all recorded data. The temperature $T$ is determined within the precision of the lab thermometer (0.1°C). Each molecule listed in Table S5 was measured 3 times (freshly prepared sample per run) for 20 min. $r_{H,FCS}$ and the measurement error was calculated by taking the average and standard error of the mean, respectively, from these three repeats. All measurements were conducted on 30 μL of sample at 1-10 nM concentration of the labeled species in 100 mM NaCl and 0.2 mM Tris (pH 7.5). Note that 24x24 mm, #1.0 cover glasses (Menzel – Gläser) were used for all experiments and although the per measurement precision in 2f-FCS can be high (~2-5% imprecision), measurement accuracy as reflected in the standard deviation of fitted $r_H$ values from repeated measurements can be significantly poorer.

Theoretical estimates of $r_H$ for the nucleic acid double helix

In order to measure $D_s$ for NA helices from ETs measurements in a single slit height $h_1$, we require to independently determine $r_H$. Since the shape of double helical dsDNA and dsRNA is well defined, we can calculate this from a theoretical model. The double helix may be regarded as a cylinder of length $l$, and axial or aspect ratio $p = \frac{l}{w}$, where $w$ is the cylinder width. We then determine the hydrodynamic radius as follows.

For a cylinder, the translational diffusion coefficients along the long, principle $k$-axis, and symmetric $i, j$-axes are given by (*71, 72*):

$$D_{t,i} = D_{t,j} = \frac{k_B T}{4\pi\eta L}\left(\ln p + v_\parallel\right) \quad (S12)$$

$$D_{t,k} = \frac{k_B T}{2\pi\eta L}(\ln p + v_\perp) \quad (S13)$$

Where $v_\parallel$ and $v_\perp$ are end-effect corrections, and were determined empirically for $2 < p < 20$ by Tirado et al. (*72*) and interpolated for ease of use by De La Torre et al. (*73*):

$$v_\perp = 0.839 + \frac{0.185}{p} + \frac{0.233}{p^2} \quad (S14)$$

$$v_\parallel = -0.207 + \frac{0.980}{p} - \frac{0.133}{p^2} \quad (S15)$$

The spherically averaged macroscopic diffusion coefficient in a random direction is simply the average of the diffusion coefficients in each direction (*74*):

$$D_t = \frac{1}{3}(D_{t,i} + D_{t,j} + D_{t,k}) \quad (S16)$$

For cylinders, we thus have (*75*):

$$D_t = \frac{k_B T(\ln p + v)}{3\pi\eta L} = \frac{k_B T}{6\pi\eta r_H} \quad (S17)$$

Where the end-effect term, $v$ is given by (*74*):

$$v = \frac{1}{2}(v_\perp + v_\parallel) = 0.312 + \frac{0.565}{p} - \frac{0.100}{p^2} \quad (S18)$$

These equations are used to calculate $D_s$ values from escape time measurements of dsDNA and dsRNA outlined in Section S5.4.

Determination of $r_H$ and $D_s$ relation for the SAMI-IV riboswitch

In order to estimate $r_H$ and $D_s$ for the riboswitch, or indeed for any molecule, from an ETs measurement performed for single slit height $h_1$ we need to include some additional constraint in the analysis, e.g., an independent relationship connecting $r_H$ and $D_s$ if available. We therefore model the RNA as a spheroid of a fixed volume $V = 4\pi lw^2/3 = 40$ nm³ which provides a relation between $r_H$ and $D_s$ as outlined below. This volume was estimated based on the molecular weight of the SAM-IV riboswitch (38.51 kDa) and assuming a density of 1.6 g/cm³ typical for RNA (*76, 77*). Similar to the analysis for DNA in the previous section we define the unique axis to be the $k$-axis, of length $l$, the equatorial axes ($i,j$-axes) to be of length $w$, and the axial ratio is given by $p = l/w$. We adopt the Perrin S-factor notation which describes the diffusion coefficients about the different axes for spheroids as follows (*13, 78, 79*):

$$D_{t,k} = k_B T \left(\frac{4\pi}{3p^2 V}\right)^{\frac{1}{3}} \frac{(2p^2 - 1)S(p) - 2p^2}{16\pi\eta(p^2 - 1)} \quad (S19)$$

$$D_{t,i} = D_{t,j} = k_B T \left(\frac{4\pi}{3p^2 V}\right)^{\frac{1}{3}} \frac{(2p^2 - 3)S(p) + 2p^2}{32\pi\eta(p^2 - 1)} \quad (S20)$$

With $S(p)$ defined differently for prolate and oblate spheroids:

$$S(p) = \frac{2p}{\sqrt{p^2-1}} \ln\left(p + \sqrt{p^2-1}\right), \text{ for prolate spheroids } (p > 1) \tag{S21}$$

$$S(p) = \frac{p}{\sqrt{1-p^2}} \tan^{-1}\left(p^{-1}\sqrt{1-p^2}\right), \text{ for oblate spheroids } (p < 1) \tag{S22}$$

Using Eq. (S16) for the spherically averaged diffusion coefficient, we obtain a relationship between the long-axis $D_s = l$ and $r_H$ as follows:

$$D_t = \frac{S(p)k_B T}{\eta}\left(\frac{1}{1296 p^2 \pi^2 V}\right)^{\frac{1}{3}} = \frac{k_B T}{6\pi\eta r_H} \tag{S23}$$

By setting the bounding sphere diameter $D_s$ to the longest dimension of a prolate or oblate spheroid, $l$ or $w$, with fixed volume $V$, Eqs. (S21) and (S23) provide a relation between $r_H$ and $D_s$, which we use in section S5.6 and in Fig. 4G to model the SAM-IV riboswitch structure.

Determining $D_s$ from molecular 3D structures

A custom Python script was used to calculate $D_s$ for predicted structures of biomolecules. The "miniball" module (https://github.com/marmakoide/miniball?tab=readme-ov-file) was used to estimate the minimum diameter of a sphere which could fully enclose all the atoms of the predicted structure, as imported from the relevant PDB file. This package was vetted on the a synthetic tetrahedron structure of side $a$, and indeed returned a minimum sphere diameter of ca. $\sqrt{3/2}\,a$ as expected from the circumradius of a tetrahedron. Note that for a tetrahedron, we have $D_{\max} = a$ which is a quantity accessible in SAXS. However, in general $D_s$ and $D_{\max}$ may be expected to be quite similar, as illustrated by, e.g., short dsDNA fragments.

**S5.3.** Calibration of the measurement using globular proteins

INS, TF and Apoferritin were measured as outlined in section S4.1. $r_H$ for the proteins are calculated using HYDROPRO as outlined in section S5.2. Using Eq. (S10), $h_1$ was determined using α, $\beta$, $\gamma$ and $t_o$ parameters determined from simulations as described in section S5.1, and $h_2 - h_1 = d$ as measured using Atomic Force Microscopy (AFM) scans of the nanostructured surfaces prior to substrate bonding, see main text Fig. 2C. We refer to this procedure as device calibration. Using the calibrated $h_1$, six further proteins (Ub, TRX, RNase A, MB, CA, LGB) were measured, and their hydrodynamic radii inferred from Eq. (S10), assuming the proteins are globular and therefore that $2r_H = D_s$. We obtained remarkable agreement between measured $r_H$ values and the theoretical values for this test set of proteins, see main text Fig. 2D.

For further structural measurements in this study, devices were calibrated similarly using measurements on four globular proteins (INS, CA, TF, FER).

**S5.4.** Structural measurements on double-stranded nucleic acids

Average escape times for dsDNA and dsRNA of various lengths were measured as outlined in section S4.3 (sequences shown in Table S1). The device was calibrated using the four-point calibration method outlined in S5.3. The increased sensitivity of the escape time to non-spherical objects, compared to globular proteins with the same $r_H$, is illustrated by these measurements as shown in Fig. S11. For prolate objects we expect $D_s \gg 2r_H$ which considerably impacts $t_{esc}$ as shown in Eq. (S10).

We modelled DNA and RNA as rigid cylinders of radii 1.05 nm and 1.18 nm, and of lengths 0.34 nm/bp and 0.25 nm/bp respectively, and determine the $r_H$ in each case using the equations given in section S5.2. Note that we also measured the $r_{H,FCS}$ of our nucleic acids using 2f-FCS, and found good agreement between the experimental and calculated values, see Table S5. Using calculated $r_H$ values and assuming $D_s = bn_{bp}$ we fit Eq. (S10), to determine both $A$ and $b$ values for both DNA and RNA (Fig. S11), where $b$ represents the rise per base pair of the double helix. The fits yielded values for $b$ that are remarkably consistent with the rise per base pair values expected for the B and A-helices characteristic of dsDNA and dsRNA respectively (*18, 29*). We may attribute the differences in fitted $A$ values from that used for proteins to possible differences in the photophysical properties of the dye when conjugated to a DNA, RNA and amino acid.

For further measurements on DNA in this study (see section S6.4), devices were calibrated using a measurement of different lengths of DNA (15-60 bp, additional sequences shown in Table S1) and fitted using Eq. (S10) with $D_s = bn_{bp}$, to determine $h_1$ and $A$ for a given device.

### S5.5. Structural measurements on DNA nanostructures

Two DNA nanostructures were synthesized as described previously (*23*). The "square tile" consists of four terminally tethered 30 base-pair double-stranded (ds) DNA helices, and the "bundle" consists of two 60ds DNA helices tethered at two points along the length of the helix. Both structures consist of 240 nucleotides (nt), and are labelled with a single ATTO532 dye.

The nanostructures were measured using ETs at 0.1-1 nM concentration in 100 mM NaCl 0.2 mM Tris (pH 7.5) in three devices, each with a different geometry. 0.1 mM MgCl$_2$ was included for measurements of the bundle, but not for the square-tile. A single exponential fit to the escape time histogram is used to determine the average escape time $t_{esc}$. To measure the size and shape, $r_H$ and $D_s$, of the two species, we considered the two measurements in the shallower slits (devices 'a' and 'b') as smaller $h_1$ gives greater sensitivity to $D_s$. Each device was calibrated using a series of different lengths of dsDNA as outlined in section S5.4 and Fig. S12A. For devices with deeper slits, the value of $h_1$ in the governing Eq. (S10) was fixed to the value measured using AFM rather than from DNA-based calibration. This is because when $D_s \ll h_1$ the calibration procedure does not deliver accurate measurements of slit height determining the device geometry. According to Eq. (S10), a single measurement gives a curve of possible ($r_H$, $D_s$) values describing the molecule. By considering two independent measurements of the same molecule in devices with slits of different height $h_1$ we can simultaneously solve for $r_H$ and $D_s$. Since repeated measurements on $t_{esc}$ are normally distributed with a standard deviation given approximately by the fit error we obtain a corresponding map of a probability distribution of possible solutions in each case. An intersection of these distribution for two independent measurements yields a 2D distribution on a $D_s$ vs $r_H$ plot that may be fitted with a 2D Gaussian to infer $r_H$ and $D_s$ values, and corresponding uncertainties, see main text Fig. 3C. This approach ignores the potential inaccuracy of the calibration of $h_1$ for each device, and of $A$. Accounting for these inaccuracies we obtain the probability map of solutions Fig. S12B. This is generated by selecting a pair of heights, $h_{1,a}$ and $h_{1,b}$, for devices 'a' and 'b' respectively, from normal distributions following their respective uncertainties from calibration. We then select escape time values for $t_{esc,tile}$ and $t_{esc,bundle}$ drawing from normal distributions about the mean values from the measurements that follow their respective measurement uncertainties. The unique solutions to these pairs of values are calculated, $D_{s,tile}$, $r_{H,tile}$ and $D_{s,bundle}$, $r_{H,bundle}$, along with the differences between nanostructures. This process is repeated 10,000 times and a histogram of $D_s$-$r_H$ values for each species is produced, along with a histogram of the differences between nanostructures Fig. S12.

We note that for a given device, the same $h_1$ and $A$ is used for each species. The result is that although the range of possible $r_H$ and $D_s$ values for each species is rather large, as shown by the probability map Fig. S12B, given a specific $h_1$ and $A$ value, a highly precise measurement of *differences between the species* emerges as in main text Fig. 3C. We note that our definition of $r_H$ and $D_s$ requires $D_s > 2r_H$; this physical requirement is applied to all displayed probability maps in keeping with our interpretation of these parameters.

To validate our measured molecular dimensions, oxDNA (*32, 80, 81*) was used to simulate 2000 instances of the nanostructures, and $D_s$ was computed for each frame. A series of exemplary snapshots are included in Fig. S12C. The mean and standard error on the mean of $D_s$ for each nanostructure was calculated for comparison with experimental results. $r_H$ was calculated using HYDROPRO as outlined in section S5.2, for a series of representative frames to estimate a mean and standard deviation. These results are displayed in Fig. S12C.

$r_H$ for the structures was also measured using FCS, see Table S5. We note an offset between $r_{H,FCS}$ and the $r_H$ value measured using ETs, however since the latter is consistent with the $r_H$ predicted from oxDNA simulated structures (see Fig. 3C) we attribute this discrepancy to shortcomings in our FCS measurements. FCS is an ensemble measurement which averages signal from *all* fluorescent species in a sample. Whilst the ETs measurements presented for the DNA nanostructures are also an ensemble measurement, we only collect signal from fluorescent species which load into the slits. One possibility is that aggregated or larger fluorescent species in the sample which contribute solely to the FCS measurement, do not so to the same extent in ETs, since only objects smaller than the channel slit heights (41.4 and 73.9 nm respectively) will be loaded into the slits. The nucleic acid samples for which we report far better consistency between FCS and ETs-derived $r_H$ values in Table S5 are all commercially synthesized and purified by HPLC. By comparison, the two DNA nanostructure samples were synthesized as outlined in Materials and Methods section and were purified using gel electrophoresis, which could imply the presence of larger contaminant species in the sample.

Another possibility to consider is that the DNA nanostructures are 4-5 times larger in mass than the smaller nucleic acids molecules and may be regarded as deformable dielectric ellipsoids of ~15-20 nm length. It may be conceivable that a weak, transient confining effect of the DNA nanostructures within the laser focus influences the interpretation of diffusive transit times between the two foci, resulting in a slightly larger inferred hydrodynamic radius for the nanostructures.

### S5.6. Structural inferences on the SAM-IV riboswitch

The SAM-IV riboswitch (synthesized as outlined in the Materials and Methods section) was measured at 1 nM concentration in 5 mM MgCl$_2$ PBS (pH 7.4). The escape time histogram was fitted with a multi-exponential distribution, and yielded three distinct timescales $t_1$, $t_2$ and $t_3$, see main text Fig. 4F. In order to make structural inferences on molecular states implied by these different escape times, the riboswitch was modelled as a spheroid, constraining the $r_H$ and $D_s$ to values consistent with the expected volume of such an RNA molecule. Using a typical RNA density of 1.6 g/cm$^3$ (*76, 77*), and taking the mass of the SAM-VI riboswitch as 38.51 kDa, a range of 40 nm$^3$ isovolumetric spheroids were considered to constrain the range of possible states. We set the longest axis to $D_s$ since for a prolate spheroid, $l = D_s$. On the other hand, for an oblate spheroid, we have $w = D_s$. Given a fixed volume and using the relation between $r_H$ and $D_s$ outlined in section S5.2, ellipsoids of a given $D_s$ yield $r_H$ values that are unique to the prolate and oblate cases. Each of these permissible ($r_H$, $D_s$) combinations yields an expected escape time, $t_{esc}$, according to Eq. (S10), which may in turn be compared with measurements

The device was calibrated using the four globular proteins as outlined in S5.3, see Fig. S13A. Using Eq. (S10), the measured escape times were mapped onto possible ($r_H$, $D_s$) values. Note that the single molecule measurements yield slightly different values for the two conformational timescales and discussed in section S6.3.

To compare our measurement with 3D molecular structural estimates, the $r_H$ and $D_s$ of the published structure for the riboswitch (pdb:6UES, as inferred from cryoEM by Zhang et al. (*36*) was determined as outlined in section S5.2.

The SAM-IV riboswitch sample was additionally measured at 2 nM concentration in 5 mM $MgCl_2$ PBS (pH 7.4) with 10 μM SAM added. The escape time histogram revealed negligible difference in timescales and state fractions (Fig. S18).

**S5.7.** Structural measurements on IR-ECD

Labeled IR-ECD (prepared as outlined in Materials and Methods section), was prepared from frozen aliquots by thawing on ice, centrifuging at 10,000 rpm for 10 min at 4°C, and then running the supernatant through a Zeba micro spin desalting columns with 7 K MWCO (Thermo Fisher Scientific). The device was calibrated using four globular proteins as outlined in S5.3, see Fig. S14A. The sample is measured at a nominal concentration of 10 nM labeled IR-ECD in PBS (pH 7.4) and 0.03% Tween 20, both with and without 1 μM Insulin added, in devices with four different geometries. We note that nominal concentrations differ from true concentrations as described later (section S7.10). Prior to measurement the samples were incubated at room temperature for 1 hr. A single exponential fit to the escape time histogram is used to determine the escape time. To measure the size and shape, $r_H$ and $D_s$, of the Insulin-bound (Liganded), and free (Apo) samples, we consider two measurements in different height devices, as for the DNA nanostructures in section S5.5. We consider the intersection of $r_H$-$D_s$ solution curves from devices '*a*' and '*c*', (chosen since these geometries provide the largest difference in gradient between the two curves). Since the measured $t_{esc}$ is normally distributed with the standard deviation given by the standard error of the mean (fit error), we obtain a probability distribution of possible solutions for several different pairs of device heights, $h_{1,a}$ and $h_{1,c}$ drawn from distributions whose width is given by the calibration error on the slit height, see main text Fig. 7C.

In order to account for the uncertainties on $h_1$ values, we repeat this analysis for 10,000 pairs of $h_{1,a}$ and $h_{1,c}$ values and $t_{esc,Apo}$ and $t_{esc,Holo}$ values. We thus obtain a probability density map of pairs of $r_H$ and $D_s$ values for each conformation, along with a map of the difference in these two inferred parameters between the two molecular states (Apo and Lig.), see main text Fig. 7D. This analysis is conducted in a manner identical to the DNA nanostructures (outlined in section S5.5). We only consider physically realistic solutions in the domain $D_s > 2r_H$.

We repeat this analysis considering each pair of devices in turn as displayed in Fig. S14B. The most precise results are obtained for devices '*a*' and '*c*' as expected, on account of the largest different in slopes of Eq. (S10) in these two cases.

To compare with our measured values, the $r_H$ and $D_s$ was calculated for the published structures of the Liganded structure (PDB:6SOF, (*50*)), partially Ins-bound structure (PDB:7QID, (*50*)), and Apo structure (PDB:4ZXB, (*50*)), as outlined in section S5.2, see Fig. S14C.

## S6. Single-molecule ETs
### S6.1. Single-particle tracking and analysis

Ten-fold lower sample concentrations were used for ease of single particle tracking in single molecule experiments. Movies were analyzed using the same method as previously described to identify escape events. Single-particle trajectories were constructed from the escape events using single-particle tracking. We then obtain the average escape time for each individual molecule by determining an average event length, $t_{av} \approx t_{esc}$, of each trajectory consisting of $N_{hop}$ escape events. Single-molecule histograms of $t_{av}$ from trajectories with at least $N_{hop} > 20$ events were constructed using a bin width corresponding to the overall average uncertainty on $t_{av}$ (typically 1.5-5 ms). The histogram entries are grey-level coded according to the number of escape events per trajectory to emphasize regions of the plot where we have the lowest uncertainty on molecular-level escape times.

For a distribution of $t_{av}$ arising from trajectories each consisting of $N_{hop}$ escape events, $t_{av}$ follows an Erlang distribution, $f(t_{av}, k_i, \lambda_i)$, with shape parameter, $k_i = N_{hop}$, and scale parameter, $\lambda_i^{-1} = \tau_i/N_{hop}$, where $\tau_i$ is the average escape time of the $i$-th component, with molecule fraction $m_i$.

$$f(t_{av}, k_i, \lambda_i) = \sum_i m_i \frac{\lambda_i^{k_i} t_{av}^{k_i-1} e^{-\lambda_i t_{av}}}{(k_i - 1)!} \tag{S24}$$

For $N_{hop} > 20$, the distribution is well approximated by a normal distribution with mean, $\mu_i = k_i \lambda_i^{-1} = \tau_i$, and standard deviation, $\sigma_i = \sqrt{k_i} \lambda_i^{-1} = \tau_i/\sqrt{N_{hop}} = \mu_i/\sqrt{N_{hop}}$. Thus, although we fit Erlang distributions to all the data, for simplicity we quote $\mu_i$ and $\sigma_i$ values characterizing the corresponding normal distributions.

In an experiment, we do not measure $t_{av}$ from single-molecule trajectories consisting of identical numbers of escape events; we instead have a distribution of $N_{hop}$ values over different molecules. We expect that the $t_{av}$ distribution will be characterized by an Erlang distribution with shape parameter given by the average number of $N_{hop}$ of all the molecules in state $i$, $\langle N_{hop} \rangle_i$ which was verified by simulation (Fig. S15). In fitting data (experimental or simulated) we therefore initialize the parameters $k_i$ to $\langle N_{hop} \rangle$ which is an average over all molecules. Once an initial fit is obtained, we repeat the fit procedure, now initializing the $k_i$s to the $\langle N_{hop} \rangle_i$ values estimated for each component based on the preliminary fit. The values for $k_i = \langle N_{hop} \rangle_i$ returned at the end of the second fit are quoted in the tables of fit parameters.

The number and locations of state components $i$ fitted to a given measured distribution was initialized using a peak detection algorithm. Here we set the threshold amplitude for peak detection at 25% of the number of molecules in the bin reflecting the maximum in the distribution. The minimum separation distance between identified peaks was set to 2 ms. The abscissa values of the identified peaks were used to initialize $\mu_i$ in fits of the data to Eq. (S24). It is worth noting that the $\mu_i$ values returned by the fit can be highly robust to multiplicative factors of 3-5 on those dictated by the peak detection algorithm. For instance, for the data in main text Fig. 4D on the SAM-IV riboswitch, the fit value of $\mu_1 = 11$ ms is returned for initial peak guess lying between 2.2 and 55 ms. This suggests than the number of time-components is likely the only required input from the peak detection algorithm. This also highlights the robustness of the initialization process adopted for fitting multi-component Erlang distributions.

Precision and resolution of single-molecule measurements

The imprecision on a $t_{av}$ measurement (s.e.m.) on a single molecule is inversely proportional to the square root of the number of escape events in the trajectory, $N_{hop}$, as expected for an exponential distribution. This quantity also governs the *resolution* of a measurement consisting of two closely spaced yet distinct states, i.e., $\sigma_i = \mu_i/\sqrt{N_{hop,i}}$. In a measurement on many molecules in state $i$, the *imprecision* on the mean measured value of a given time component is given by $\sigma_i/\sqrt{N_{mol,i}}$, where $N_{mol,i} = m_i N_{mol}$. Here $N_{mol}$ is the total number of molecules measured (number of $t_{av}$ values in the distribution) and $m_i$ is the fractional abundance of molecules in state $i$. Since we have $N = N_{hop} N_{mol}$ for a sample characterized by a single time component, the s.e.m of an ensemble fit (characterized by the s.e.m. or the fit error on the time constant of the exponential) essentially exactly captures the error on the mean value of the corresponding Erlang/Gaussian distribution from a trajectory-based single molecule experiment.

Simulation study of multi-component fitting of single-molecule spectra

In order to validate the data analysis procedure relying on fitting of Erlang distributions of single-molecule $t_{av}$ values, we generated a single time-component distribution of simulated $t_{av}$ using $N = 10^4$ escape events grouped into $N_{mol} = 100$ trajectories. The number of hops per trajectory is $N_{hop} > 20$ and approach a normal distribution with a mean $\langle N_{hop} \rangle = 100$. The underlying exponential distribution from which the trajectories were drawn was characterized by a single escape-time component of mean value $\tau = 16.26$ ms $= \mu$. A histogram of $t_{av}$ was constructed and fit with Eq. (S24), see Fig. S15A. We note good agreement of the fitted distribution mean, $\mu = 16.23 \pm 0.17$ ms, with the ground truth value. We also note that the fit uncertainty on the mean is very similar to the theoretical statistically-limited imprecision of the s.e.m., evaluated as $\sigma/\sqrt{N_{mol}} \approx 0.16$ ms, where $\sigma = 1.62$ ms. We also note that the fitted $\langle N_{hop} \rangle$ value recovers the known average $\langle N_{hop} \rangle$ of the simulated data. Also shown in Fig. S15A (inset) is the ensemble fit of the same data pooled together. This simulation exercise reveals that for a single escape-time component the theoretical measurement imprecision using the single-molecule and ensemble methods are identical within statistical error.

In order to theoretically test the resolving power in a typical a single-molecule analysis, we turn once again to simulations. A distribution of simulated $t_{av}$ was generated from two sets of trajectories representing a mixture of two distinct molecular states at a ratio of 1.5:1 ($m_1 = 0.6$ and $m_2 = 0.4$). The number of molecules in each set of trajectories was $\langle N_{mol} \rangle_1 = 60$ and $\langle N_{mol} \rangle_2 = 40$, again characterized by normally distributed $N_{hop}$ values with mean values $\langle N_{hop} \rangle_1 = \langle N_{hop} \rangle_2 = 130$ and individual trajectories characterized by $N_{hop} > 20$. The underlying exponential distributions were characterized by escape times that differed by less than 15%, namely $\tau_1 = 16.3$ ms or $\tau_2 = 18.5$ ms. We note on the side, that the chosen $N_{hop}$ and $N_{mol}$ values are very reasonable for a typical experiment, and are in fact likely to greatly improve in future.

A histogram of $t_{av}$ was constructed and fit with Eq. (S24), see Fig. S15B. We again note good agreement of the fitted distribution means, $\mu_1 = 16.2 \pm 0.2$ ms and $\mu_2 = 18.4 \pm 0.6$ ms with the ground truth values. We note larger fit uncertainties on these values than the corresponding s.e.m.s which are 0.17 and 0.34 ms, for $i = 1$ and $i = 2$, respectively. We attribute this to binning of the data and fitting with a multi-parametric function. Importantly, we see that these uncertainties on the mean times are considerably smaller than the distribution standard deviations, $\sigma_i = \mu_i/\langle N_{mol} \rangle_i$, ($\sigma_1 = 1.42 > 0.17$ ms and $\sigma_2 = 1.61 > 0.34$ ms) which illustrates the distinction between the resolution and precision of the measurement.

Furthermore, although these time components were chosen to be separated by less than ~1.5$\sigma$, as evident from the histogram, yet the fit is able to successfully resolve the two states. We note in addition that the molecule fractions, $m_i$, of the two sets of trajectories are correctly recovered in the fitting process.

In sharp contrast, the ensemble fit of the same data yielded only a single degenerate timescale when fit with a bi-exponential function, thus illustrating the power of the single-molecule analysis over the ensemble approach. Overall, we find that the functional form fit in the single molecule approach (multi-Erlang/Gaussian) fosters extraction of much more information from the data (much smaller imprecisions on the relevant free parameters), reflecting the fact that there is indeed more information on the sample included in the analysis than in the ensemble approach.

**S6.2.** Validation of single molecule spectra experimental method using dsDNA

Three samples of DNA were measured at 0.01-0.1 nM concentration in 100 mM NaCl, 0.2 mM Tris (pH 7.5): pure 24 bp, pure 60 bp and a 1:1 mixture of these two samples (sequences available in Table S1). For the pure samples a single Erlang distribution was returned by the fitting function, while for the mixture we obtained a sum of two distributions with mean escape time values, $\mu_1$ and $\mu_2$, that were in good agreement with the respective pure samples. We note good agreement between the fitted $N_{\text{hop}}$ values and $\langle N_{\text{hop}} \rangle$, the average number of hops per trajectory. At present the number of hops collected is primarily limited by photobleaching of the molecule. Full analysis of the single molecule data is shown in Fig. S16.

The detected escape events for the mixture were also pooled and analyzed using the ensemble-fitting approach. A bi-exponential distribution was fitted to the escape time histogram, and $m_2$, the fraction of 60ds DNA in all the fluorescent molecules was determined using the method outlined in section S7.1. The fitted escape times and molecular fractions are displayed in Fig. 4B of the main text. We note good agreement between the escape times returned by the ensemble fitting, $t_1$ and $t_2$, and the single molecule fitting, $\mu_1$ and $\mu_2$, with the uncertainties on the mean ca. factor 3 smaller than for the single molecule fitting, capturing the simulation-based indications from the previous section. Similarly, the molecular fractions of each species, $m_1$ and $m_2$, obtained from the ensemble method, are in good agreement with the inferred single molecule component fractions: both approaches correctly recover the 1:1 input ratios of the two species, with the single molecule fit once again returning smaller uncertainties, by ca. factor 7, than the ensemble fit. Thus, we have validated the single molecule method as a quantitative way of dissecting mixtures experimentally.

**S6.3.** Single-molecule measurements on the SAM-IV riboswitch

The SAM-IV riboswitch sample was measured at 0.01-0.1 nM concentration in 5 mM MgCl$_2$ PBS (pH 7.4). Single-molecule trajectories were constructed, and a histogram of $t_{\text{av}}$ plotted and fit as described in section S6.1. As a control three additional samples were measured similarly, carboxyl-ATTO532, 30 dsDNA, 119 ssDNA (preparation details in Materials and methods section). Full analysis of the single molecule data is shown in Fig. S18.

Control measurements for the SAM-IV riboswitch

Measurements on sample of pure ATTO532 fluorophore returned a single time component with mean $\mu_{\text{dye}}$, permitting us to establish the identity of the fast first peak in the SAM IV measurements.

The 30 bp DNA returned a two-component fit with a low-amplitude component (10%) at $\mu_1 \approx \mu_{\text{dye}}$ in agreement with the 'dye control', and suggesting the presence of a small amount of free dye in the sample. Interestingly, the second component, $\mu_2 = 29$ ms, has a similar escape time to the peaks assigned to the SAM-IV riboswitch despite the fact that at 60 nt, the 30 dsDNA fragment is half the molecular weight of the riboswitch. This comparison provides direct qualitative evidence of the compactness the riboswitch.

The length and sequence of the 119 nt ssDNA were designed to offer a control measurement on a molecule with a similar molecular weight to the riboswitch, yet lacking significant secondary or tertiary structure (Fig. S13). Measurements on 119 nt ssDNA also returned two peaks, one of which was a low-amplitude component (5%) at $\mu_1 \approx \mu_{\text{dye}}$. However, the dominant second component at $\mu_2 = 81$ ms, reflects a far larger escape time than those observed for the SAM-IV riboswitch. Given that both species have approximately the same molecular weight, this large disparity in escape times directly indicates significant differences in 3D conformation. The single large timescale for 119 nt ssDNA points to an unfolded/disordered average conformation, which is illustrative of the lack of stable secondary and tertiary structure in the designed sequence, compared to ssRNA.

Measurements on the SAM-IV riboswitch

The structure of the $t_{\text{av}}$ distribution measured for the riboswitch clearly indicated a heterogeneous distribution of conformational states. Indeed, a fit to the data returned a three-component Erlang distribution with means $\mu_1$, $\mu_2$ and $\mu_3$. One distribution agreed with that of the dye-control, i.e., $\mu_1 \approx \mu_{\text{dye}}$.

The ensemble analysis of the riboswitch escape-time data also yielded three timescales $t_1$, $t_2$ and $t_3$ which are consistent with $\mu_1$, $\mu_2$ and $\mu_3$. We note that compared to the single molecule histogram the short timescales contribute a larger total fraction of events in the ensemble analysis and hence molecular fractions $m_1$ that are larger for the ensemble case than the single molecule analysis. This reflects a challenge that arises in constructing single particle trajectories for faster diffusing species which exhibit longer jumps between consecutive trapping events, leading to poorer recognition of complete, long trajectories and rejection from the analysis as a result. The analysis at present therefore has a slight "compositional bias" generally favoring slower diffusing molecules characterized by longer escape times. This was not evident in the validation mixture of 24 and 60 dsDNA since the timescales were more similar and there was no free dye in the mixture. Note also that the ensemble analysis is characterized by much larger fit parameter uncertainties on time-scales and molecular fractions than the single molecule data. This occurs despite the fact that the ensemble experiment contains more data – $N = 43840$ events compared to $N = 21480$ for the single molecule experiment.

Representative real-time single molecule traces were constructed by time-averaging the hops over either 4 or 16 consecutive hops as previously described (*3*). These traces confirm that the heterogeneity of the riboswitch sample arises from different conformational states that are stable over the observation window of 6 seconds. We did not obtain evidence of interconversion between different conformational states over this timescale for any of the 310 molecules measured, see Fig. S17.

**S7. Assessing the extent of molecular binding in a bi-molecular reaction**

To track the progress of intermolecular binding we are interested in the fractional abundance of the product, $m_2$. For a simple two reactant binding system, where A + B → AB, and the observable species are the labeled reactant $A$ and the product $AB$, we have $m_2 = [AB]/([AB] + [A])$.

**S7.1.** Determining the fractional abundance of the bound-state, $m_2$

In an ETs measurement, we track $m_2$ by labeling only one of the reactants, $A$ for example. Once the reaction has begun, we expect to have two fluorescent species $A$ and $AB$. We characterize our two fluorescent species by their respective escape times, $t_{\text{esc,A}} = t_1$, and $t_{\text{esc,AB}} = t_2$. For reasonable disparities between $t_1$ and $t_2$ (e.g., at least 50% fractional difference) and in a mixture of the two species, we measure a distribution of escape event lengths, $\Delta t$, that can be successfully fit by a biexponential function as in Eq. (S2). From this equation we note that our measurable in an experiment is the fractional number of events from each exponential distribution characterized by the timescale, $t_i$ - the 'event fraction', denoted as $E_i$. We can relate the event fraction to the true molecular fraction by Eq. (S25), which when applied to a two-reactant binding system gives Eq. (S26).

$$m_i = \frac{E_i t_i}{\sum_j E_j t_j} \tag{S25}$$

$$m_2 = \frac{\tau_2 E_2}{\tau_2 E_2 + \tau_1 E_1} \tag{S26}$$

This assumes that in an experiment we observe molecules from each species for the same amount of time, $\tau$, on average, such that we collect on average $\tau/t_i$ events from each species This assumption relies on each species bleaching at the same rate – a reasonable expectation the fluorophore is the same in both the free and bound-states. To determine $E_1$, $E_2$ and hence $m_2$ for our two-reactant binding system proceeding to equilibrium, the timescales $t_1$ and $t_2$ were fixed in Eq. (S2). In order to do so, $t_1$ was measured by species $A$ in absence of $B$, and $t_2$ was measured on an equilibrated sample containing $A$ in the presence of a large excess of $B$, ensuring $A$ is entirely in the bound state $AB$. We note that the overall average escape time of the mixture, $t_{\text{av}}$, is sufficient to assess the amount of bound-state present.

For a two-reactant binding system, we can relate $m_2$ to $t_{\text{av}}$, as follows:

$$t_{\text{av}} = \frac{\sum_i E_i t_i}{\sum_j E_j} \tag{S27}$$

$$m_2 = \frac{t_{\text{av}} - t_1}{t_{\text{av}}(1 - t_1/t_2)} \tag{S28}$$

The timescales $t_1$ and $t_2$ were assumed to be the minimal and maximal measured $t_{\text{av}}$ indicative of 'initial' and 'final' states of a time-dependent reaction, or alternatively, the free and 'fully' bound states in a binding equilibrium measurement. This assumes that at the highest concentration of unlabeled reactant in a binding reaction $t_{\text{av}} = t_2$ and therefore $m_2 = 1$. In instances where not all the reactant is bound at completion, such that $m_2 < 1$, the assumption $t_2 \approx t_{\text{av}}$ no longer holds. However, we note that the relative differences between $m_2$ values calculated using a nominal $t_2 \approx t_{\text{av}}$ are independent of $t_2$. As a result, $m_2$ can provide a measure of the extent of binding, which may be used to estimate $K_\text{d}$ values. Fig. S21 shows verification that $K_\text{d}$ values may be determined using $m_2$ values calculated using Eq. (S26) and (S28) interchangeably.

## S7.2. Inferring the equilibrium constant, $K_d$

The value of $m_2$ measured at equilibrium for a two-species binding reaction can be used to infer the binding affinity, $K_d$, of a reaction. By varying the concentration of unlabeled reactant $B$, $[B]_0$, added to a fixed concentration of labeled $A$, $[A]_0$, we can relate $m_2$ at equilibrium to the $K_d$ of the reaction as follows (*40*):

$$m_2 = m_2^{max} \frac{[A]_0}{2} \left[ [A]_0 + [B]_0 + K_d \sqrt{([A]_0 + [B]_0 + K_d)^2 - 4[A]_0[B]_0} \right] + m_0 \quad (S29)$$

$m_2^{max}$ denotes the $m_2$ value at which complex formation has saturated for the highest $[B]_0$ concentrations. In practice this is often observed to be less than 1 and varies due to the exact purity of reactant species (*40*).

## S7.3. Measurements of on- and off-rate constants in a binding reaction

The on-rate constant, $k_{on}$, of a reaction can be measured by tracking $m_2$ as a function of time after initiation of the reaction. The response of $m_2$ as a function of time is characterized by an equilibrium rate constant, $k_{eq}$, as shown in the following Eq.:

$$m_2 = m_2^{max}(1 - e^{-k_{eq}t}) \quad (S30)$$

When $[B]_0$ is in large excess of $[A]_0$, we can express $k_{eq}$ in terms of $k_{on}$ and $k_{off}$, the reaction off-rate constant, as follows (*40*):

$$k_{eq} = k_{on}[B]_0 + k_{off} \quad (S31)$$

$k_{off}$ can be measured independently by tracking $m_2$ as a function of time after binding is complete. This can be achieved by addition of a high concentration of a labeled 'chaser' species $C$ to the fully equilibrated binding reaction. $C$ is designed to bind strongly and essentially irreversibly to free $B$ in solution. The result of this is that once an $AB$ complex spontaneously dissociates the likelihood of $A$ rebinding with another $B$ molecule is essentially zero, resulting in a gradual reduction in the amount of $AB$ in solution. The rate of this decrease in $m_2$ thus yields an off-rate constant $k_{off}$ which may be given by (*51*):

$$m_2 = m_2^{max} e^{-k_{off}t} \quad (S32)$$

The binding affinity can be determined from the measured on and off rate constants according to:

$$K_d = \frac{k_{off}}{k_{on}} \quad (S33)$$

## S7.4. Estimating the overall uncertainty in measurements of $m_2$

The general formula for the total error $f_e$ on quantity $f$ arising from errors in variables $x$ and $y$ propagation on which $f$ depends may be written as follows (*89*):

$$(f_e^2) = \left(\frac{\delta f}{\delta x}\right)^2 x_e^2 + \left(\frac{\delta f}{\delta y}\right)^2 y_e^2 + 2\left(\frac{\delta f}{\delta x}\frac{\delta f}{\delta y}\right) \text{Cov}(x, y) \quad (S34)$$

where $x_e$ and $y_e$ denote uncertainties on $x$ and $y$ respectively.

Using Eq. (S26) and (S34), ignoring uncertainties on $t_1$ and $t_2$, the uncertainty on $m_2$, $m_{2,e}$, can be expressed as follows:

$$(m_{2,e}^2) = \left[\frac{m_2}{E_2}(1-m_2)\right]^2 E_{2,e}^2 + \left[\frac{m_2^2 \tau_1}{\tau_2 E_2}\right]^2 E_{1,e}^2 - \frac{m_2^4}{(\tau_2 E_2)^2}\left(\frac{2\tau_1 E_1}{\tau_2 E_2}\text{Cov}(\tau_2 E_2, \tau_1 E_1)\right) \quad \text{(S35)}$$

Because $E_1$ and $E_2$ are not independent ($E_1 = c - E_2$, with an unknown $c$), the covariance is non zero. We note that theoretically $c = 1$, but because of spurious events in the first bin of the escape time histogram not fitted as mentioned in section S2, $E_1 + E_2 \neq 1$. $\text{Cov}(\tau_2 E_2, \tau_1 E_1) = -\tau_1 \tau_2 E_{2,e}^2$ or $\text{Cov}(\tau_2 E_2, \tau_1 E_1) = -\tau_1 \tau_2 E_{1,e}^2$. We note that theoretically these should be identical, but in practice we get different uncertainties for $E_1$ and $E_2$. We therefore use the larger of the two errors to estimate the overall uncertainty, the final expression given as follows:

$$(m_{2,e}^2) = \left[\frac{m_2}{E_2}(1-m_2)\right]^2 E_{2,e}^2 + \left[\frac{m_2^2 \tau_1}{\tau_2 E_2}\right]^2 E_{1,e}^2 + \frac{m_2^4}{(\tau_2 E_2)^2}\left(\frac{2\tau_1^2 E_1}{E_2}\max\{E_{1,e}^2, E_{2,e}^2\}\right) \quad \text{(S36)}$$

When using $t_{av}$ to determine $m_2$, using Eq. (S28) and (S34), and ignoring uncertainties in $t_1$ and $t_2$, the uncertainty on $m_2$ is given as:

$$(m_{2,e}) = \frac{t_1}{t_{av}^2(1 - t_1/t_2)} t_{av,e} \quad \text{(S37)}$$

**S7.5.** DNA hybridisation

DNA hybridization provides a carefully controllable and tunable molecular binding system to quantify the ability of ETs to measure intermolecular binding. We focused on a sequence of ssDNA 11 nt long reported on by Palau et al. (*39*). We obtained a series of fluorescently labeled oligos from 7-11 bases in length that represent progressive truncations of the parent 11 nt sequence. We also included a longer fluorescently labeled 15 nt ssDNA sequence. Two unlabeled 200 nt sequences were designed with complementary 5'ends to the 11 ssDNA and 15 ssDNA respectively. The sequences of the 200 nt oligos were optimized to have as few stable secondary structures as possible using the Mfold web server (*82*). The sequences for all DNA used in this study are presented in Table S1.

Equilibrium measurements of affinity constants

To measure the binding affinity, $K_d$, of each short DNA oligo with the complementary 200 ssDNA, a fixed concentration (0.02-1 nM) of oligo, $[A]_0$, was incubated with varying concentrations of complementary 200 ssDNA, $[B]_0$, in 50 mM NaCl Tris-EDTA buffer (pH 8, Sigma). Where possible, $[A]_0$ was chosen to be well below the $K_d$ of interest for each oligo. Incubation times, and range of 200ss concentrations added, varied depending on the oligo length. After equilibrium was reached, the fraction of oligo hybridized, $m_2$, was determined for each reaction by fitting the escape time histogram with a biexponential distribution as outlined in section S2. The $K_d$ value was determined by fitting $m_2$ as a function of $[B]_0$, according to Eq. (S29).

Typically, $m_2^{max} \neq 1$, which we attribute to always having a non-zero fraction of reactant oligo in an inactive form. For ease of comparison between $K_d$ values of different length oligos, a normalized $m_2$ is used such that the $m_{2,fit} = 1$ at maximal $[B]_0$.

For the slowest reacting mixtures (i.e., with smallest $[B]_0$) equilibrium was deemed to have been reached by an invariance of measured $K_d$ with further incubation time, see Fig. S19. For higher $[B]_0$ values we incubated the sample for periods much longer than the theoretically expected equilibration time. For several samples the $m_2$ was also determined using $t_{av}$, and compared with those determined from the bi-exponential fitting of the escape time histogram, see Fig. S21. The $K_d$ values extracted were the same within error, validating that $t_{av}$ can be used as a quantitative measure of the $K_d$.

We note that the time required for equilibration is much longer than expected from literature estimates of diffusion limited on-rates expected for DNA hybridization (*40*). We attribute this to the possible influence of non-specific association of biomolecules to the walls of the reaction tubes during incubation. Such issues can be addressed by measurements using the reaction kinetics approach to independently determine $K_d$, described next.

<u>Measurements of reaction kinetics</u>

To measure the on-rate constant, $k_{on}$, of each short, labeled ssDNA oligo at a fixed concentration $[A]_0 = 0.1 - 1$ nM, in 50 mM NaCl Tris-EDTA buffer (Sigma-Aldrich) was loaded into the device. A small volume of complementary 200 ssDNA, at varying concentrations $[B]_0$, was added into the device and sequential measurements were recorded over a duration of 5-180 mins depending on the $[B]_0$. The fraction of oligo hybridized, $m_2$, was determined as outlined in section S7.3. Eq. (S30) was used to determine $k_{eq}$ for each $[B]_0$. Data for the $k_{eq}$ measurements of 10 ssDNA are shown in Fig. S20A. Because $k_{off}$ is generally very small, a fit of the measured $k_{eq}$ data to Eq (S31) is rather insensitive to this quantity. $k_{off}$ was therefore determined independent in order to infer $k_{on}$ (Eq (S31)).

To measure the off-rate constant, $k_{off}$, of each short DNA oligo, after the $m_2$ had plateaued in the $k_{eq}$ measurements, a small volume of high concentration (0.5-5 µM) unlabeled 11 ss, the chaser oligo $C$, was added to the device and sequential measurements were recorded until $m_2$ had decayed to close to 0. Eq. (S32) was used to determine $k_{off}$ for each $[B]_0$. Data for the $k_{off}$ measurements of 9, 10 and 11 nt ssDNA are shown in Fig. S20A. An average value was used to determine $k_{on}$ in conjunction with the $k_{eq}$ measurements using Eq. (S31), inset of Fig. S20A.

To explore whether the chaser was facilitating the unbinding of $AB$ and thus influencing the $k_{off}$ measurement, $k_{off}$ was measured for a series of different reaction mixtures, all incubated with constant $[A]_0$ and $[B]_0$ but with varying chaser concentrations $[C]$. $k_{off}$ was determined to be the same within error for the whole range of ratios of $[C]/[B]_0$ measured in this study, thus indicating minimal cooperative unbinding in our system over that concentration range. For higher ratios of $[C]/[B]_0$, a factor two increase in $k_{off}$ was indeed observed, indicative of facilitated unbinding, but this regime is not relevant to our study, see Fig. S20B.

We note that in occasional cases there was a short time delay after loading of the B before any binding was observed. Then after this the expected inverse exponential increase in bound fraction occurred. In these instances, this offset was subtracted from the measurement times to give a more accurate measure of the reaction on rate. Since the response curves are exponential, analysis of any part of the response should yield the same measured rate. We attribute this offset to inconsistent loading times through the microfluidic channels. A similar effect was occasionally observed after addition of the chaser and the offset similarly accounted for.

Calculation of theoretically expected binding affinities

Theoretical $K_d$ values were computed for each oligo using the nearest-neighbor Watson Crick parameters as outlined by SantaLucia et al. (*41*).

**S7.6.** 24bp + EcoRI binding affinity measurement

Labeled 24 bp DNA was measured at 250 pM in PBS (pH 7.4) with varying concentrations of EcoRI added (details of samples in Materials and Methods and Table S1). Prior to measurement, the sample was incubated at room temperature for 2 hours. The fraction of oligo bound to the EcoRI, $m_2$, was determined from fitting the escape time histogram with a biexponential function as outlined in section S7.1. $K_d$ for the system was determined using Eq. (S29), with oligo concentration, $[A]_0$, and EcoRI concentration, $[B]_0$. Bar plots in main text Fig. 6A illustrate the relative fractions of unbound and bound oligo for varying concentrations of EcoRI. Repetitions of the affinity measurement are shown in Fig. S22. The $m_2$ was also determined using $t_{av}$ as outlined in section S7.1, and compared with those determined from the bi-exponential fitting of the escape time histogram, see Fig. S21. The $K_d$ values determined were the same within error, further confirming that $t_{av}$ can be used as a quantitative measure of $K_d$.

**S7.7.** HLA + IgG binding affinity measurement

Labeled HLA (prepared as outlined in Materials and Methods section) was measured at 1 nM in PBS (pH 7.4) with varying concentrations of added unlabeled IgG. Prior to the measurement the samples were incubated at room temperature for 1 hour. The fraction of HLA bound to the IgG, $m_2$, was determined from $t_{av}$ as outlined in section S7.1. $t_{av}$ was used to track the binding since the HLA is a complex known to dissociate giving a complicated mixture of timescales especially at low nanomolar concentrations (*83, 84*). $t_1$ and $t_2$ were determined by the $t_{av}$ measured for the sample with the lowest and highest concentrations of added IgG, respectively. A considerably lower density of molecules was observed in the device than the nominal concentration of sample, measured using the Nanodrop on sample at concentrations that are several orders of magnitude higher. A corrected concentration of 0.1 nM HLA was measured with reference to a 60 bp dsDNA sample as outlined in section S1.3, and was used as $[A]_0$ in Eq. (S29) with the corresponding IgG concentration, $[B]_0$, in order to estimate $K_d$ for this system. Repetitions of the affinity measurement are shown in Fig. S22.

**S7.8.** Insulin aptamer + Insulin binding affinity measurement

A labeled insulin aptamer sequence as reported by Yoshida et al. (*46*) and listed in Table S1, was prepared via annealing in 10 mM MgCl$_2$ (Sigma-Aldrich) PBS (pH 7.4) for 10 min at 95°C, and gradual cooling down at 2°C/min. The folded aptamer was measured at 5 nM concentration in the same buffer with varying concentrations of added insulin. The fraction of aptamer bound to insulin, $m_2$, was determined from $t_{av}$ as outlined in section S7.1. $t_{av}$ was used to track the binding since the timescales $t_1$, and $t_2$ for the unbound and bound aptamer respectively, were within 10% of one another, and therefore the fitting of a biexponential distribution to the escape time histogram was unreliable. $K_d$ for the system was determined using Eq. (S29). Repetitions of the affinity measurement are shown in Fig. S23B.

We also note the difference in timescale of the aptamer measured in PBS (pH 7.4) compared to that measured after annealing in 10 mM MgCl$_2$ PBS (pH 7.4). This is consistent with the formation of a G-quadruplex structure, as reported previously (*46, 85*), which is expected to cause an increase in compaction of the DNA and thus a reduction in escape time. This was also observed in FCS measurements of the unfolded, folded and insulin-bound states, shown in Fig. S23A.

**S7.9.** Insulin + IR-ECD binding affinity measurement

Labeled insulin (prepared as outlined in Materials and Methods section) was measured at 2 nM concentration in HEPES Buffered Saline (HBS, pH 7.2, Merck) with varying concentrations of unlabeled IR-ECD (prepared as outlined in Materials and Methods section). Prior to measurements the samples were incubated overnight at 4°C. The fraction of insulin bound to the IR-ECD, $m_2$, was determined from fitting the escape time histogram with a bi-exponential distribution as outlined in section S7.1. $K_d$ for the system was determined using Eq. (S29) with insulin concentration, $[A]_0$, and IR-ECD concentration, $[B]_0$. We assume that for this measurement of $K_d$ the Insulin-IR-ECD is approximately a monovalent binding system, since, despite the four possible insulin binding sites on the IR-ECD, at the rather large concentration ratio of 15:1 [IR-ECD]:[Insulin] corresponding to the mean $K_d$ value measured, the likelihood of multivalent Insulin-IR-ECD binding is very low. Repetitions of the affinity measurement are shown in Fig. S22.

**S7.10.** IR-ECD + Insulin binding affinity measurement

Labeled IR-ECD (prepared as outlined in Materials and Methods section and S5.7) was measured at a calibrated concentration of 0.08 nM (calibrated using 60 dsDNA as a reference molecule as outlined in section S1.3) with varying concentrations of unlabeled insulin. This corresponded to a nominal concentration of 10 nM as measured using Nanodrop before dilution. Prior to measurement, the sample was incubated at room temperature for 1 hr. The fraction of IR-ECD bound to insulin, $m_2$, was determined from $t_{av}$ as outlined in section S7.1. $t_{av}$ was used to track the binding since the timescales $t_1$, and $t_2$ for the unbound and bound aptamer respectively, were within 10% of one another, and therefore the fitting of a biexponential distribution to the escape time histogram was unreliable. The $K_d$ for the interaction based on conformation change was calculated using Eq. (S29), see main text Fig. 7E. It has been reported by Gutmann et al. (*50*) that at least 3 insulins were required to bind to the IR-ECD before a significant conformation change occurs. The impact and relative contributions of the underlying processes on the $K_D$ estimated from a conformation change is not clear.

This measurement was repeated in simulated human serum (Biochemazone, BZ278) with 0.03% Tween 20 in solution. This commercially obtained simulated serum contains a large number of non-specific components that are present in real serum and thus provides a good mimic of the background present in a diagnostic measurement. We determined the $K_d$ in simulated serum similarly to in PBS using a calibrated IR-ECD concentration of $[A]_0 = 0.1$ nM, see main text Fig. 7E. These measurements were also repeated with longer incubation at 4°C overnight. We obtained larger $K_d$ values in this case, which we attribute to non-specific binding of insulin to surfaces of the preparation tubes used for incubation causing a reduction in effective concentration in solution, $[B]_0$.

The measurement was also repeated in human serum (Sigma, H4522), and the $K_d$ was determined using a calibrated IR-ECD concentration of $[A]_0 = 0.05$ nM, see Fig. S24B. In contrast to the measurements in PBS and simulated serum where the $K_d$ values of the interaction were consistent; in the human serum we report a $K_d$ value an order of magnitude lower. The literature has previously reported lower affinity values for structurally stabilized IR-ECD. Our measurement in serum taken at face value could suggest that the lower $K_d$ measured stems from structural stabilization due to binding by serum components. If so, this measurement provides the first indication that ETs is able to detect free insulin in serum at physiologically meaningful concentrations that lie in the 10 pM range.

As a control, IR-ECD was measured under identical conditions in PBS but with varying concentrations of Epidermal Growth Factor (EGF) instead of insulin. Prior to measurement the sample was incubated at room temperature for 1 hr. No change in escape time was observed upon addition of EGF, see Fig. S24A, and hence no conformation change of the IR-ECD was observed, indicating the specificity of the IR-ECD to insulin.

**S7.11.** Sources of inaccuracy in molecular binding measurements

Accurate measurements of binding affinities are inherently challenging due to inaccuracies in measurement of pico-nanomolar concentrations of the reactants. In general, absorption measurements at low micromolar concentration were used before dilution for measurement. We found considerable variability in these absorption measurements and our measurements therefore must contend with a potential factor 2 uncertainty in quoted concentrations for a given sample. Where a range of concentrations were measured, for example $K_d$ curves with varying $[B]_0$, sequential dilution was used to ensure accurate relative measurement across a $K_d$ curve – here the factor 2 uncertainty in the original stock manifests as a shift in the *x*-axis for all data points. Where a single concentration is measured, as for $[A]_0$, we note that the $K_d$ is relatively insensitive to $[A]_0$ concentration as long as it is well below the $K_d$. We therefore aimed for this regime for all measurements.

A perhaps more concerning source of concentration inaccuracy is nonspecific binding (sticking) of the sample to preparation tubes and also within the device. Non-stick Eppendorf tubes and device passivation were used to mitigate these issues but in multiple cases we observed considerably lower concentrations of protein than nominally measured before dilution. For fluorescent species we use quantitative molecule counting to calibrate the concentration of freely diffusing molecules.

A final source of inaccuracy in concentration measurements is in determining the concentration of the "active" binding fraction in a sample. We consistently saw reactions complete with <100% binding. We attribute this to a non-negligible 'inactive' fraction of the sample that is unable to bind. This has been widely reported in the literature (*40*), and means that the active $[A]_0$ concentration, $[A]_{0,\text{active}} < [A]_0$. This fraction $[A]_{0,\text{active}}/[A]_0$ can be estimated from the saturation $m_2$ value at completion of the reaction and is often ~30%. Since the $K_d$ is by design insensitive to $[A]_0$, the $[A]_{0,\text{active}}/[A]_0$ fraction is unimportant, however if generally true, then we may expect a similar fraction of $[B]$ that is similarly inactive, thus suggestive of ~30% inaccuracy in $[B]_0$ and hence in the inferred $K_d$ value.

## S8. Supplementary Figures

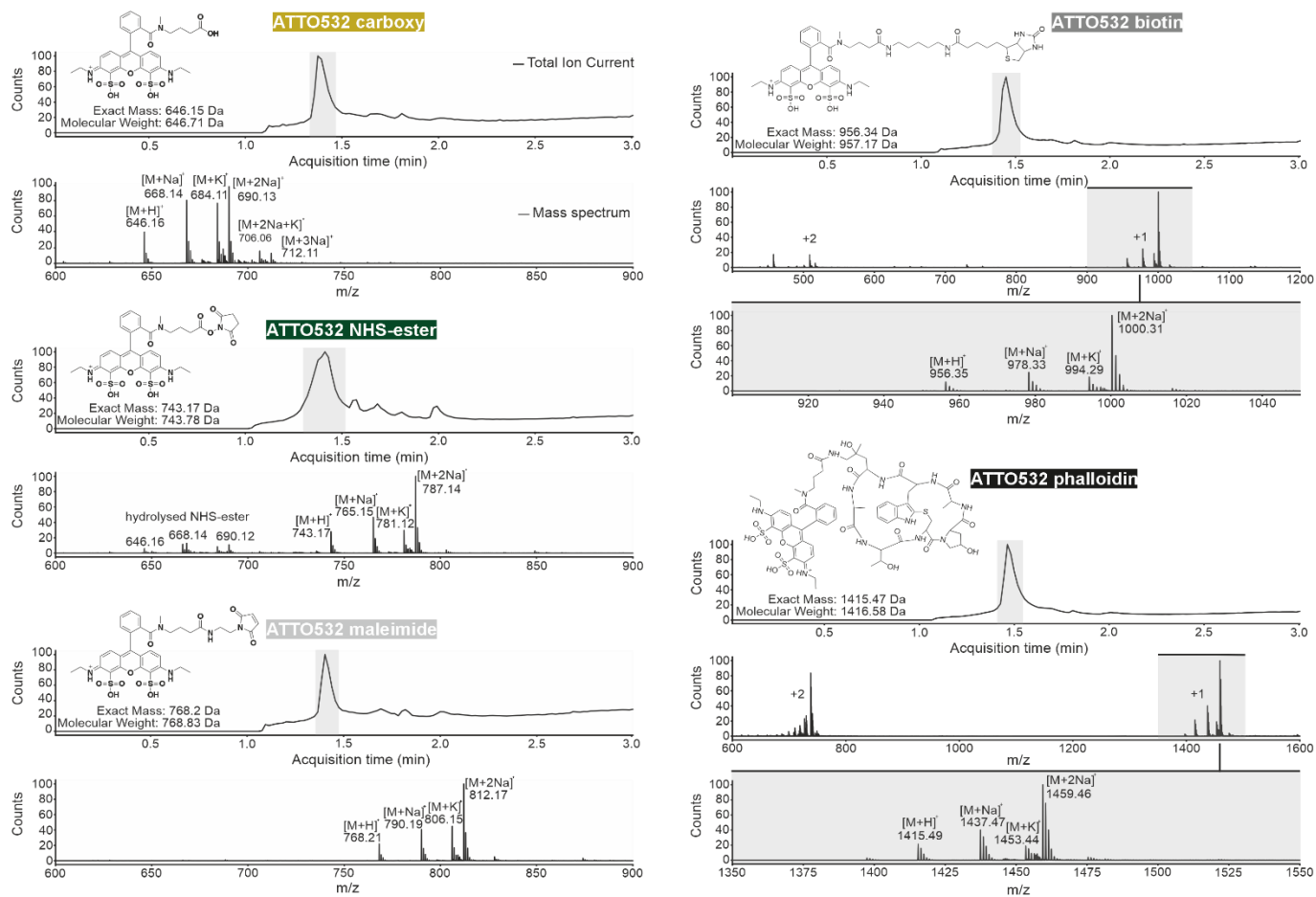

**Fig. S1. LC-MS spectra of ATTO532 dye derivatives.** Total ion chromatogram (top) and mass spectrum (bottom) are presented in each case.

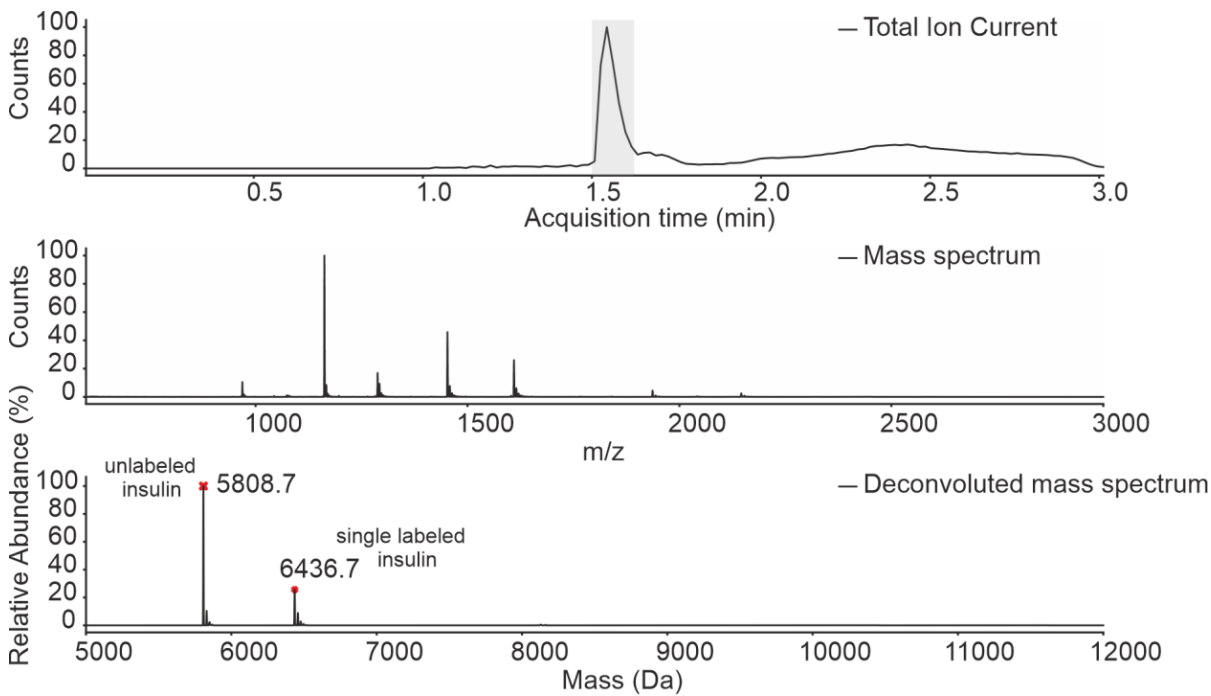

**Fig. S2. Intact mass spectrometry analysis of ATTO532-labeled insulin**. Total ion chromatogram, mass spectrum and deconvoluted mass spectrum are shown. Only unlabeled insulin and insulin carrying a single ATTO532 label were observed.

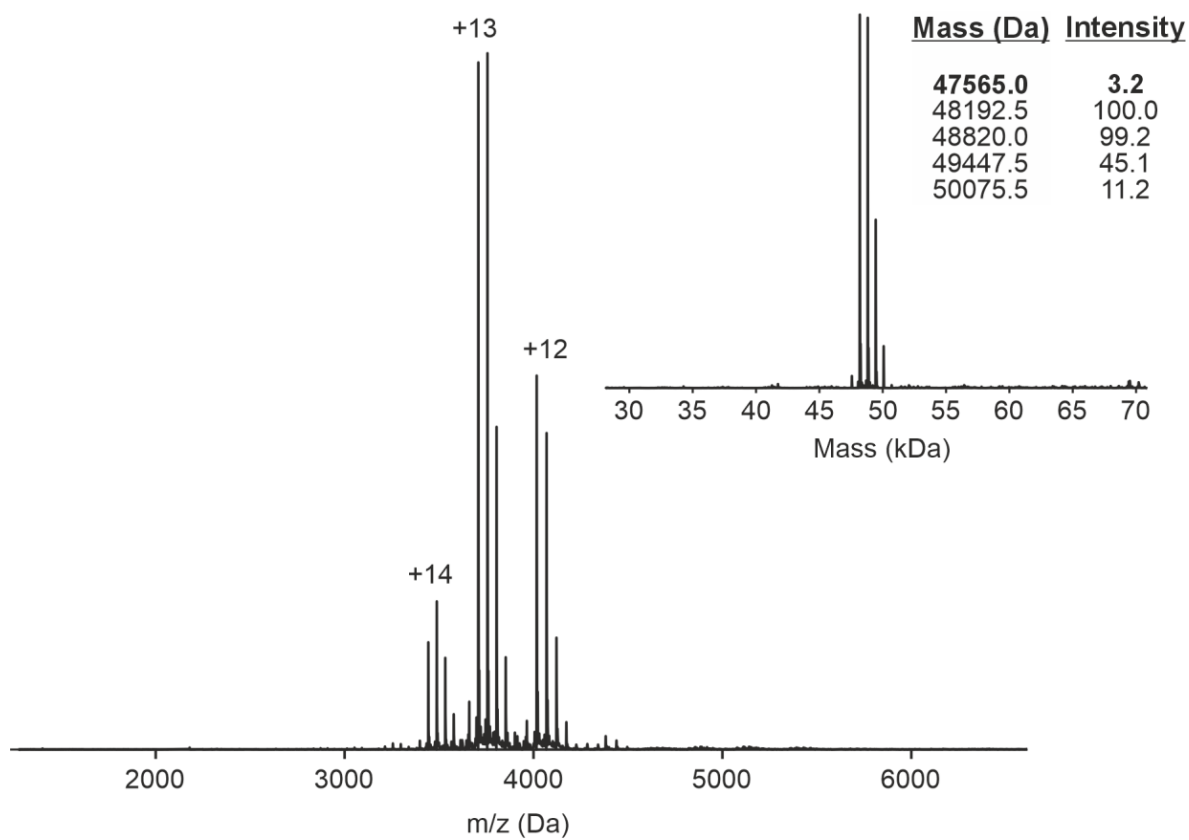

**Fig. S3**. **Native mass spectrometry analysis of ATTO532-labelled HLA-A*03:01.** The table lists all measured masses. We expect unlabelled HLA to have a nominal mass of 47492 Da which is close to the measured 47565 Da (first table row). Adduct masses of 627.5 Da correspond to a single ATTO532 molecule. The majority of HLA molecules carry one or two labels, but up to four labels were observed on a small fraction (~11%) of the labelled HLA complex.

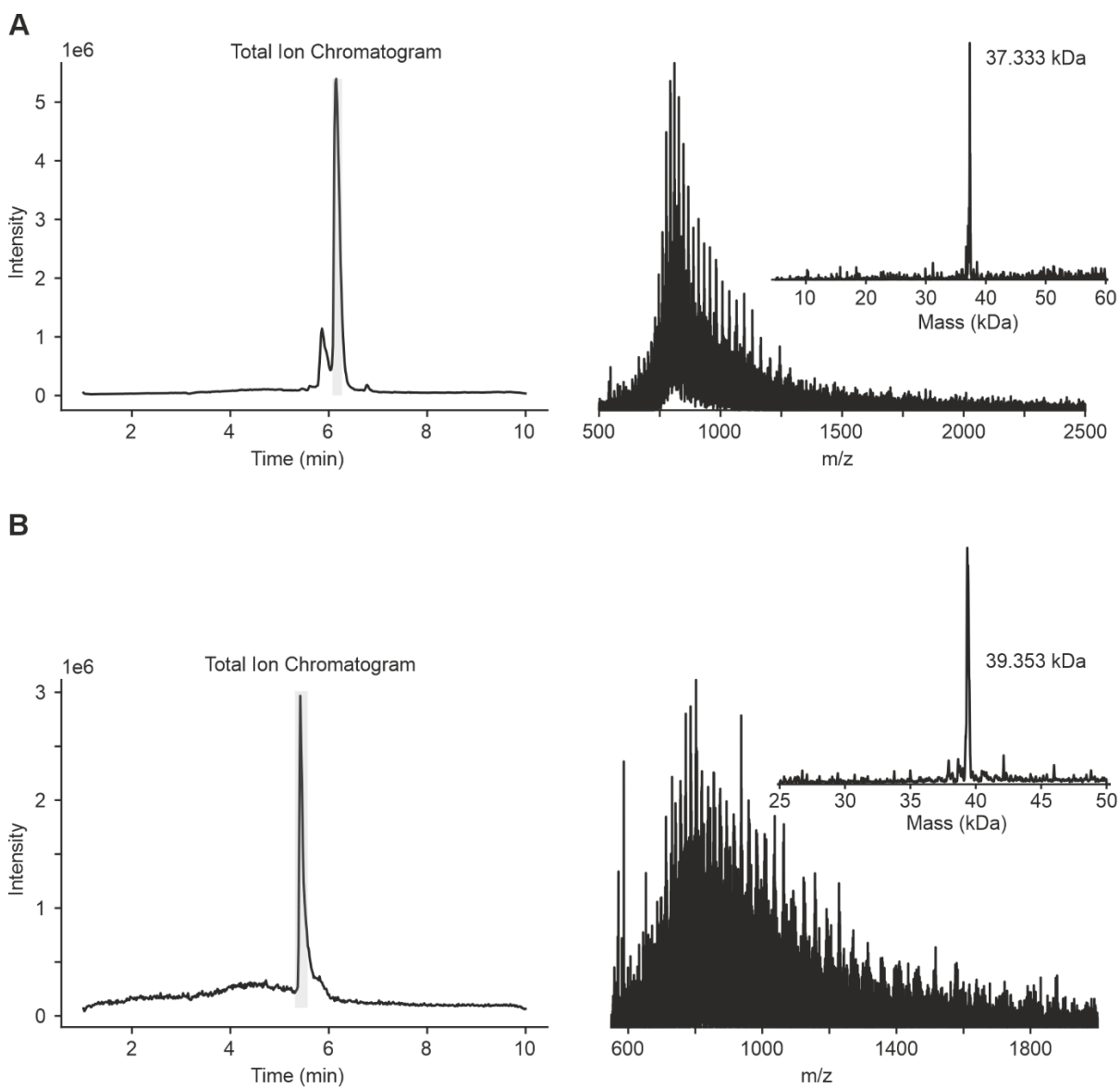

**Fig. S4. LC-MS spectra of (A) 119 nt reference ssDNA and (B) SAM-IV riboswitch.** Total ion chromatogram and mass spectrum are shown in each case. Measured masses agree well with expectations from the sequences: for DNA, $M \approx 37$ kDa, for RNA: $M \approx 39$ kDa. Sequences are shown in S9.Table S1.

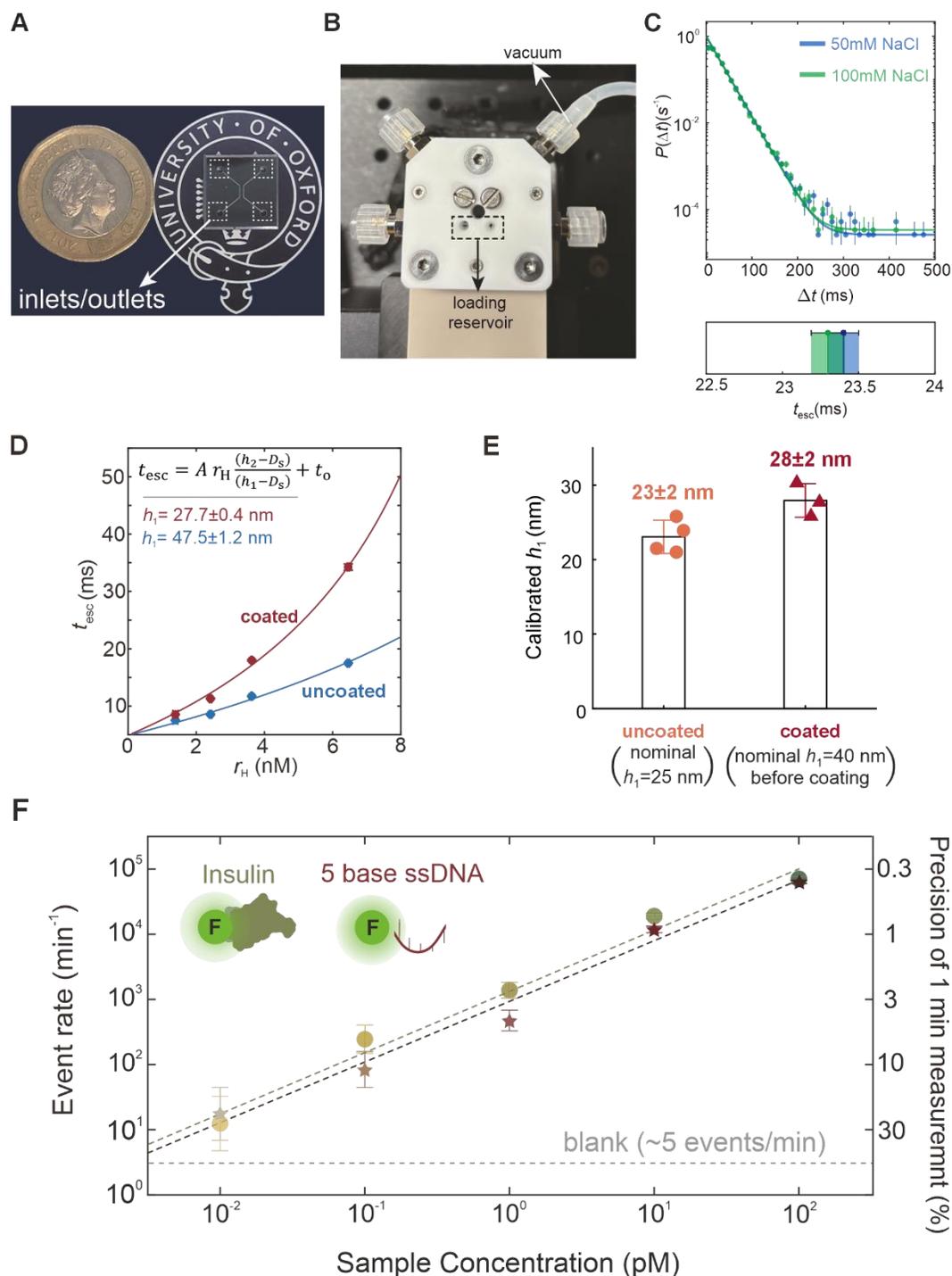

**Fig. S5. Diagram of ETs device, evidence of reproducible passivation of slits with m-PEG-silane and demonstration of low copy-number molecular counting.** (**A**) Microfluidic chip used in this work, next to 1£ coin for scale. (**B**) Teflon chuck housing the chip (not visible) allows for easy mounting on the widefield fluorescent microscope stage, and application of vacuum. Loading reservoirs are open to air while the outlet is sealed and connected to vacuum (ca. 50 mbar) to facilitate loading of sample into the chip. (**C**) Escape time histograms for two measurements of 120 bp dsDNA under different salt concentrations in a device with $h_1 \approx 70$ nm at pH 8. The measured escape times are indistinguishable (lower panel) thus confirming that the electrostatic contribution to $W$ in Eq. (S4) is negligible under typical ETs experimental conditions. (**D**) Calibration curves for two nominally identical devices, one with m-PEG coating and one without, were constructed using a series of globular proteins (INS, CA, TF, FER) with different $r_H$ values, as calculated using HYDROPRO. Measured escape times were fitted with Eq. (S2) with $D_s = 2r_H$, $t_o = 5.83$ ms and $d = h_2 - h_1 = 260$ nm as measured by AFM. (**E**) $h_1$ in devices with nominally the same slit geometries typically vary by less than 2 nm both before and after the coating. (**F**) Passivation of slits with m-PEG silane supports quantitative molecular counting down to 10 fM sample concentration. Samples of labelled 5 ssDNA and INS measured over 5 orders of magnitude in concentration. The event rate measured scales linearly with concentration as expected, showing that ETs devices can be used as precise and sensitive concentration sensors.

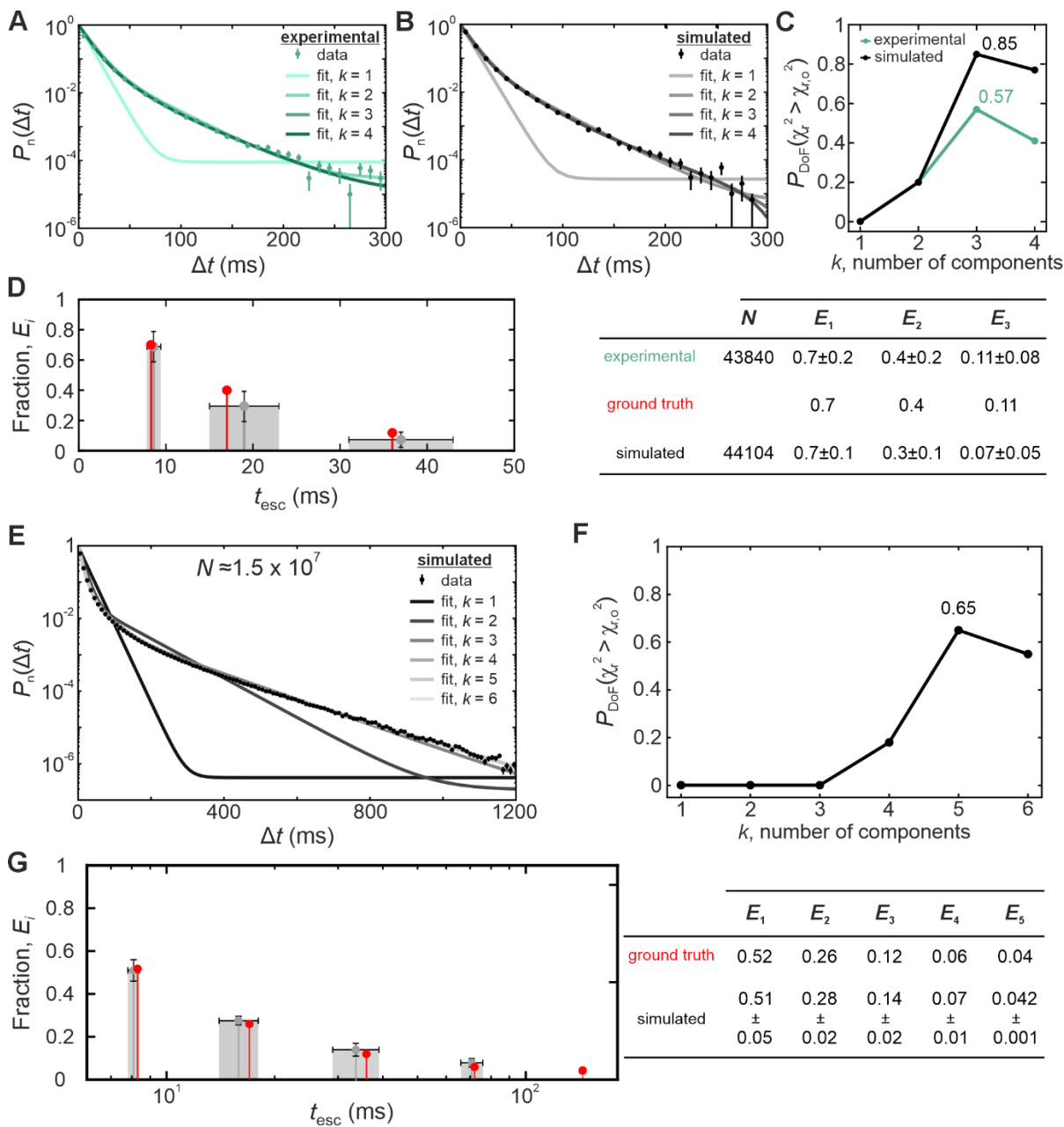

**Fig. S6. Multiexponential histogram fitting and ground-truth validation.** (**A**) Measured normalized probability density distributions $P_n(\Delta t)$ vs. $\Delta t$ (green curves) for $N \approx 4.4 \times 10^4$ recorded escape events of SAM-IV riboswitch (green data and lines), also shown in Fig. 4F, were fitted with Eq. (S2) up to $k = 4$ components. (**B**) Fits of $P_n(\Delta t)$ vs. $\Delta t$ (grey curves) for $N \approx 4.4 \times 10^4$ simulated escape events (black data) using as simulation inputs the fit escape times ($t_i$) and event fractions ($E_i$) obtained by fitting a 3-exponential decay to the experimental SAM-IV riboswitch in (**A**). (**C**) Probability of the expected reduced $\chi$-squared being greater than the observed $P_{\text{DoF}}(\chi_r^2 > \chi_{r,o}^2)$ for the experiment in (**A**) (green) and simulation in (**B**) (grey), confirms that both experimental and simulated data are best fit with Eq. (S2) when the maximum number of components is $k = 3$. (**D**) Ground truth values (3 vertical dashed red lines), based on the experimental fit results (green values), used to simulate 3-component histograms with $N \approx 4.4 \times 10^4$. Each grey normal distribution represents the fit escape time ($t_i$) and fit error for each component from a simulated histogram, illustrating good agreement with the given ground truth values. Table lists the event fractions ($E_i$) for each case for experiments and simulations. (**E**) $P_n(\Delta t)$ vs. $\Delta t$ (grey curves) for $N \approx 1.5 \times 10^7$ simulated escape events (black points) based on a ground truth given representing 5 states corresponding to the 3-exponent fit experimental escape times ($t_i$) and event fractions ($E_i$) from **A** (ground truth) and including two additional components. The data were fitted with Eq. (S2) for up to a maximum of $k = 6$ components. (**F**) $P_{\text{DoF}}(\chi_r^2 > \chi_{r,o}^2)$ for the fit curves in (**E**) shows that the generated simulated data are best fit with Eq. (S2) when $k = 5$ (~0.65). (**G**) Ground truth timescales displayed as 5 red dotted lines. Each grey band represents the fit escape time ($t_i$) and corresponding fit error, illustrating good agreement with the ground truth values.

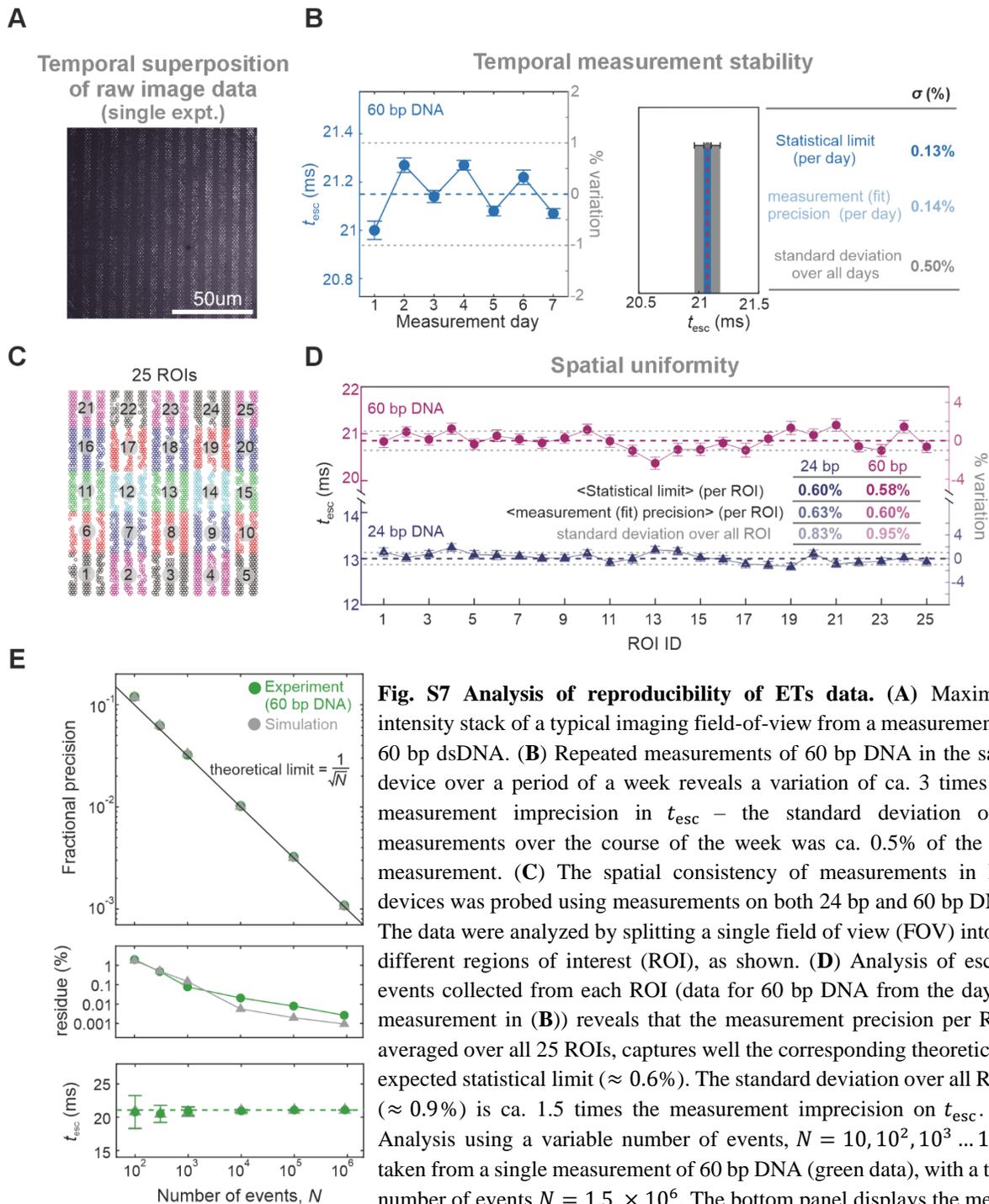

**Fig. S7 Analysis of reproducibility of ETs data.** (**A**) Maximum intensity stack of a typical imaging field-of-view from a measurement of 60 bp dsDNA. (**B**) Repeated measurements of 60 bp DNA in the same device over a period of a week reveals a variation of ca. 3 times the measurement imprecision in $t_{esc}$ – the standard deviation of 7 measurements over the course of the week was ca. 0.5% of the $t_{esc}$ measurement. (**C**) The spatial consistency of measurements in ETs devices was probed using measurements on both 24 bp and 60 bp DNA. The data were analyzed by splitting a single field of view (FOV) into 25 different regions of interest (ROI), as shown. (**D**) Analysis of escape events collected from each ROI (data for 60 bp DNA from the day #5 measurement in (**B**)) reveals that the measurement precision per ROI, averaged over all 25 ROIs, captures well the corresponding theoretically expected statistical limit ($\approx 0.6\%$). The standard deviation over all ROIs ($\approx 0.9\%$) is ca. 1.5 times the measurement imprecision on $t_{esc}$. (**E**) Analysis using a variable number of events, $N = 10, 10^2, 10^3 \ldots 10^6$, taken from a single measurement of 60 bp DNA (green data), with a total number of events $N = 1.5 \times 10^6$. The bottom panel displays the mean of fitted $t_{esc}$ values for 100 samples of size $N$. The error bars represent averages of the fit errors over the 100 realizations. Top panel - Each symbol in the plot represents the mean fractional uncertainty for the corresponding case in the bottom panel. This plot shows that the fractional uncertainty in experimentally measured $t_{esc}$ follows exactly the theoretically expected value of $1/\sqrt{N}$ (solid black line). Simulated data with the same mean as the experimental data also reveal the same behaviour when analyzed in the same way (grey data). The residue plot represents the percentage departure of this fractional uncertainty from the theoretical $1/\sqrt{N}$ expectation (middle).

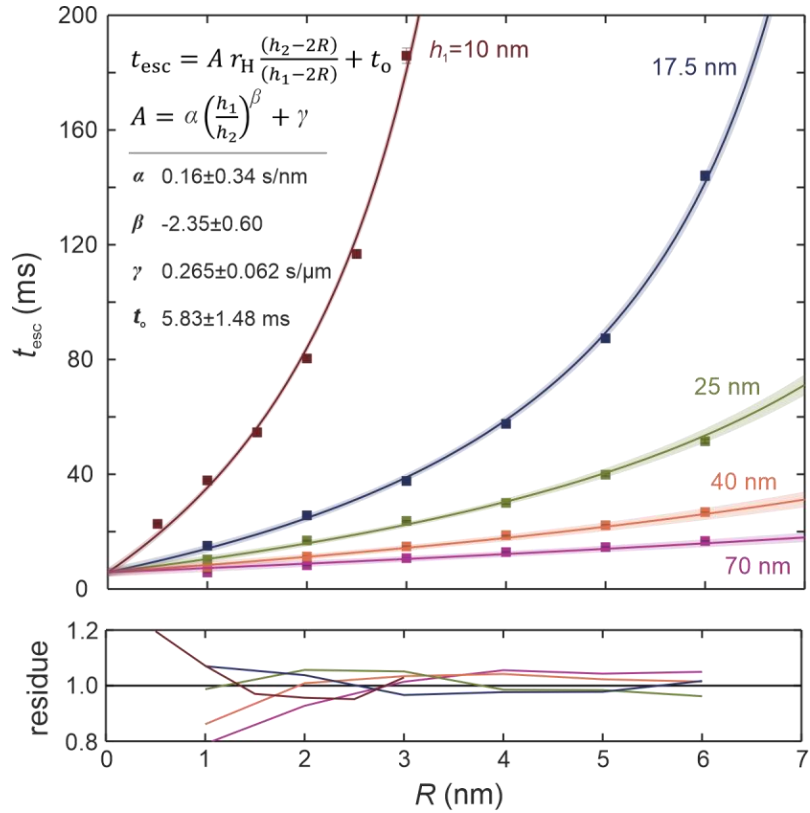

**Fig. S8. Brownian Dynamics (BD) simulations of spherical particles diffusing in a nanostructured landscape.** BD simulations of spheres of different radii, $R = r_H$, diffusing in a landscape of traps are analyzed to determine $t_{esc}$. $N = 10000$ escape events were simulated for a range of different geometries given by $10 < h_1 < 70$ nm and $d = h_2 - h_1 = 300$ nm. A global fit (lines, with shaded region displaying 1 standard deviation of uncertainty on the overall fit), using Eq. (S10) with $D_s = 2R$, to all five datasets yielded good agreement with the theoretical equation; the average r.m.s. of the residuals is less than 3% (bottom panel). Fit parameters were obtained as follows: $\alpha = 0.16 \pm 0.34$ s/nm, $\beta = -2.35 \pm 0.60$ and $\gamma = 0.265 \pm 0.062$ s/μm, $t_o = 5.83 \pm 1.48$ ms. For further analysis, only the mean parameter values were used to specify Eq. (S10) in any given device geometry.

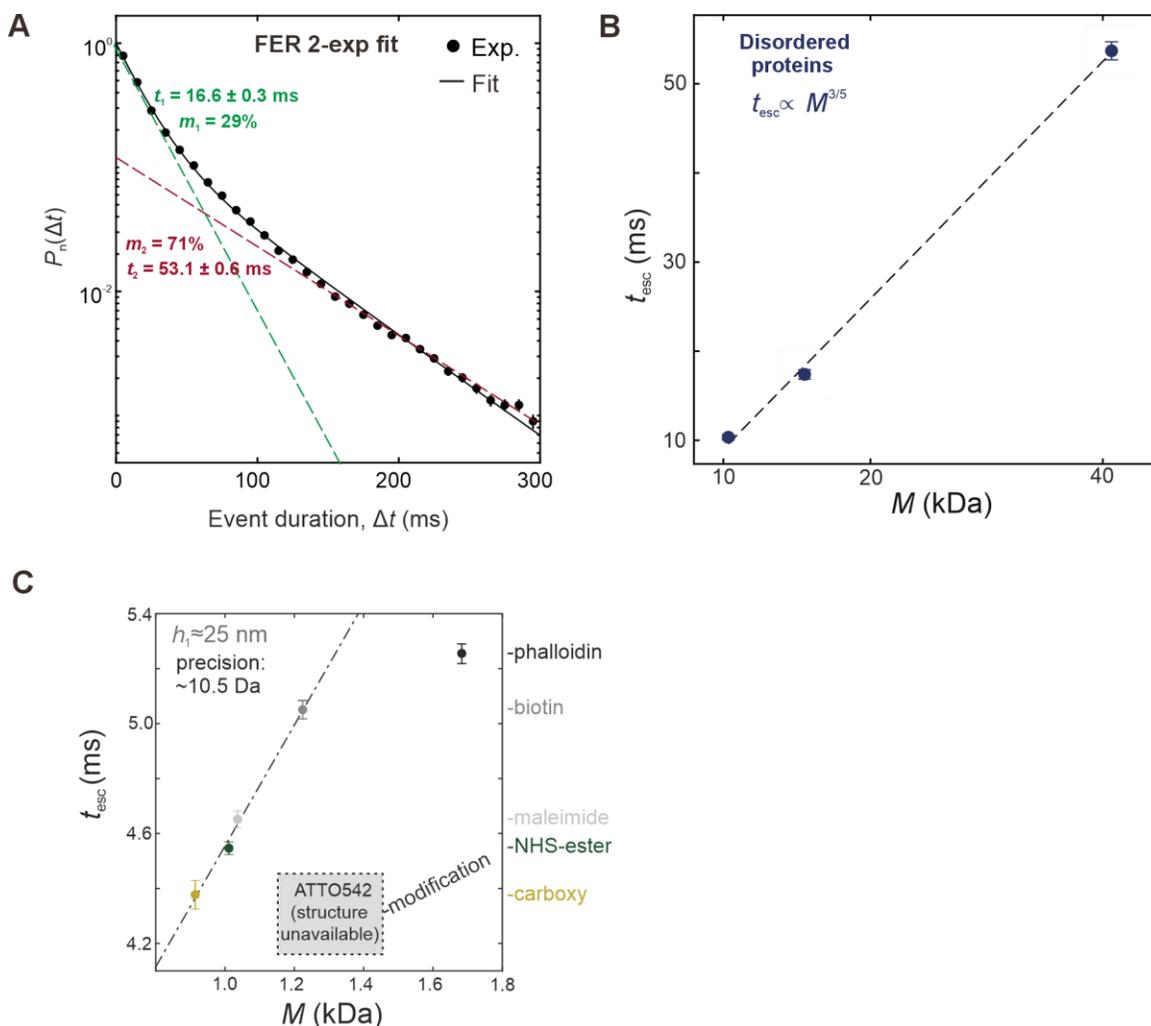

**Fig. S9. Measurements on Apoferritin (FER), disordered proteins and ATTO542 dye derivatives**. **(A)** Bi-exponential fit (black curve) of Apo-ferritin escape time histogram (data points). $t_1$ (green) is as expected for an Apo-ferritin monomer ($M \approx 20$ kDa), while $t_2$ (red) is taken to correspond to the 24-mer complex ($M \approx 465$ kDa). **(B)** Escape times for disordered proteins (13-mer proline rich polypeptide, ProTa and Stml-1) measured at 5-20 nM concentration in PBS. A linear fit (dashed line) of $t_{esc}$ vs. $M^{\frac{3}{5}}$ agrees with the data. A movie of duration ca. 5 min was recorded for each sample. **(C)** Plot of $t_{esc}$ vs. $M$ for small fluorescent organic molecules represented by various chemical derivatives of the fluorophore ATTO542. Similar to Fig. 2B from the main text, ~0.5% measurement precision on $t_{esc}$ implies the ability to resolve differences ca. 10 Da in the $M = 1$ kDa molecular weight regime. Once again, the 25 Da difference between the NHS and maleimide esters is well resolved.

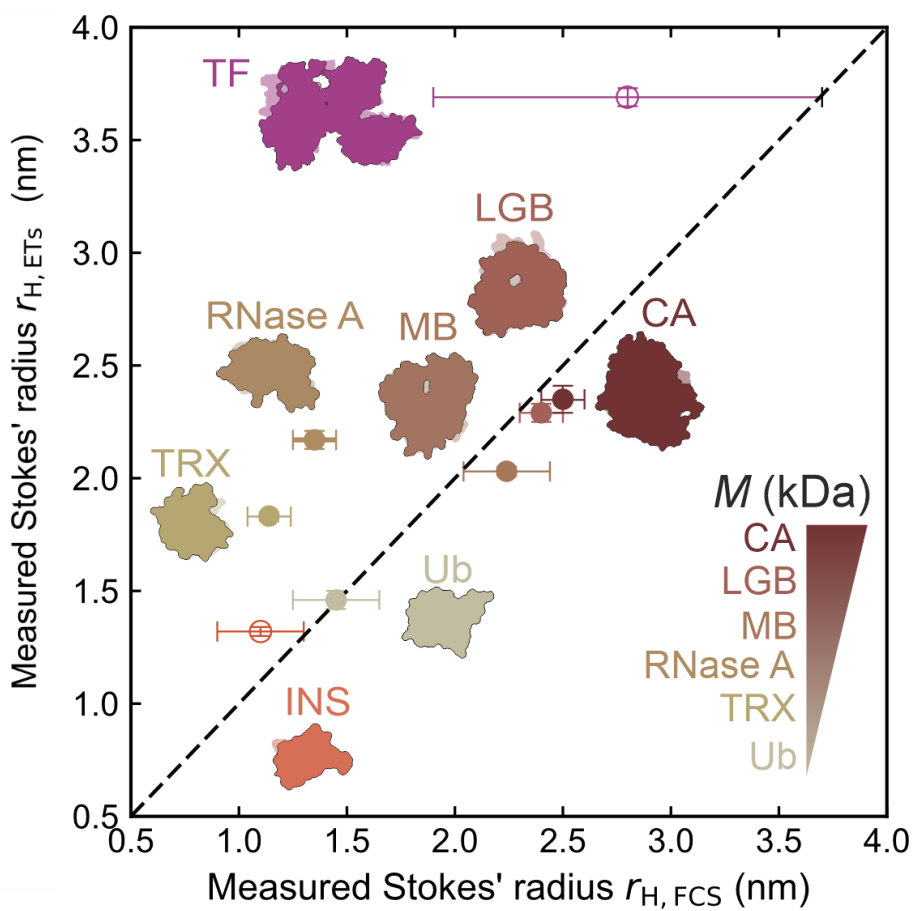

**Fig. S10. Comparison of Stokes' radii measured using ETs and 2f-FCS for a set of globular proteins (Fig. 2D, main text).** $r_{H,ETs}$ for the test set of globular proteins in the range $M = 10 - 30$ kDa (INS, Ub, TRX, RNase A, MB, LGB, CA, TF, see main text Fig. 2D) compare reasonably with measurements using 2f-FCS, $r_{H,FCS}$. Open symbols represent molecules used for device calibration. 2f-FCS measurements are in general prone to greater inaccuracies on account of very long measurement times needed, and the presence of free dye and multimers in solution – all of which are readily addressed and handled using ETs. In comparison, Fig. 2D shows that $r_{H,ETs}$ measurements agree very well with values computed for the molecular 3D structures.

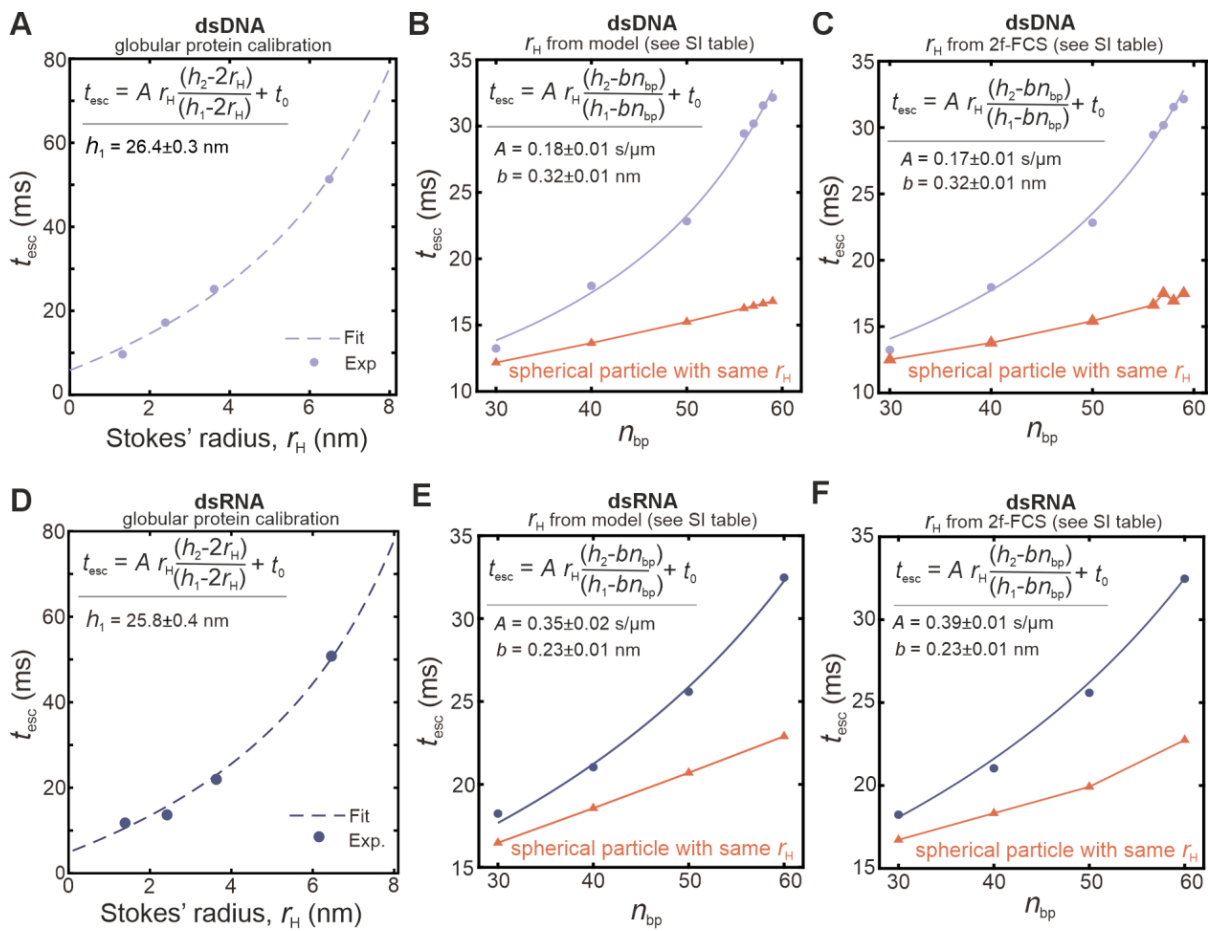

**Fig. S11. Measurements of the rise per basepair, $b$, in dsDNA and dsRNA.** Calibration curves (dashed lines) for two devices used for $b$ measurements of dsDNA **(A)** and dsRNA **(D)**: a series of globular proteins (INS, CA, TF, FER) with different $r_H$ values, as determined from molecular structures using HYDROPRO, were measured and escape times fitted with Eq. (S10) to determine $h_1$, using $D_s = 2r_H$, $t_o = 5.83$ ms and $d = h_2 - h_1 = 286$ nm and 250 nm, respectively, as measured by AFM. **(B, C, E, F)** Measurements of escape time $t_{esc}$ vs. number of basepairs, $n_{bp}$, for a series of dsDNA **(B, C)** and dsRNA **(E, F)**. Data is fitted to determine $A$ and $b$ using Eq. (S10), with $t_o = 5.83$ ms, $D_s = bn_{bp}$, $h_1$ as noted in calibration plots (panels **A** and **D**), and using $r_H$ values either as calculated using the theoretical equation (Eq. (S17), panels **B** and **E**) or as measured using 2f-FCS (Table S5, panels **C** and **F**). Whilst the fit value $A$ changes slightly depending on the source of $r_H$, the inferred $b$ value is robust to this choice. In all cases, orange data illustrates the approximately linear response expected from Eq. (S10) for "equivalent spheres" ($D_s = 2r_H$) with the same $r_H$ as for the corresponding length of the nucleic acid. This weaker response function for equivalent spheres illustrates the fact that $r_H \approx 3 - 5$ nm, over this range, is far smaller than the slit height $h_1 \approx 25$ nm.

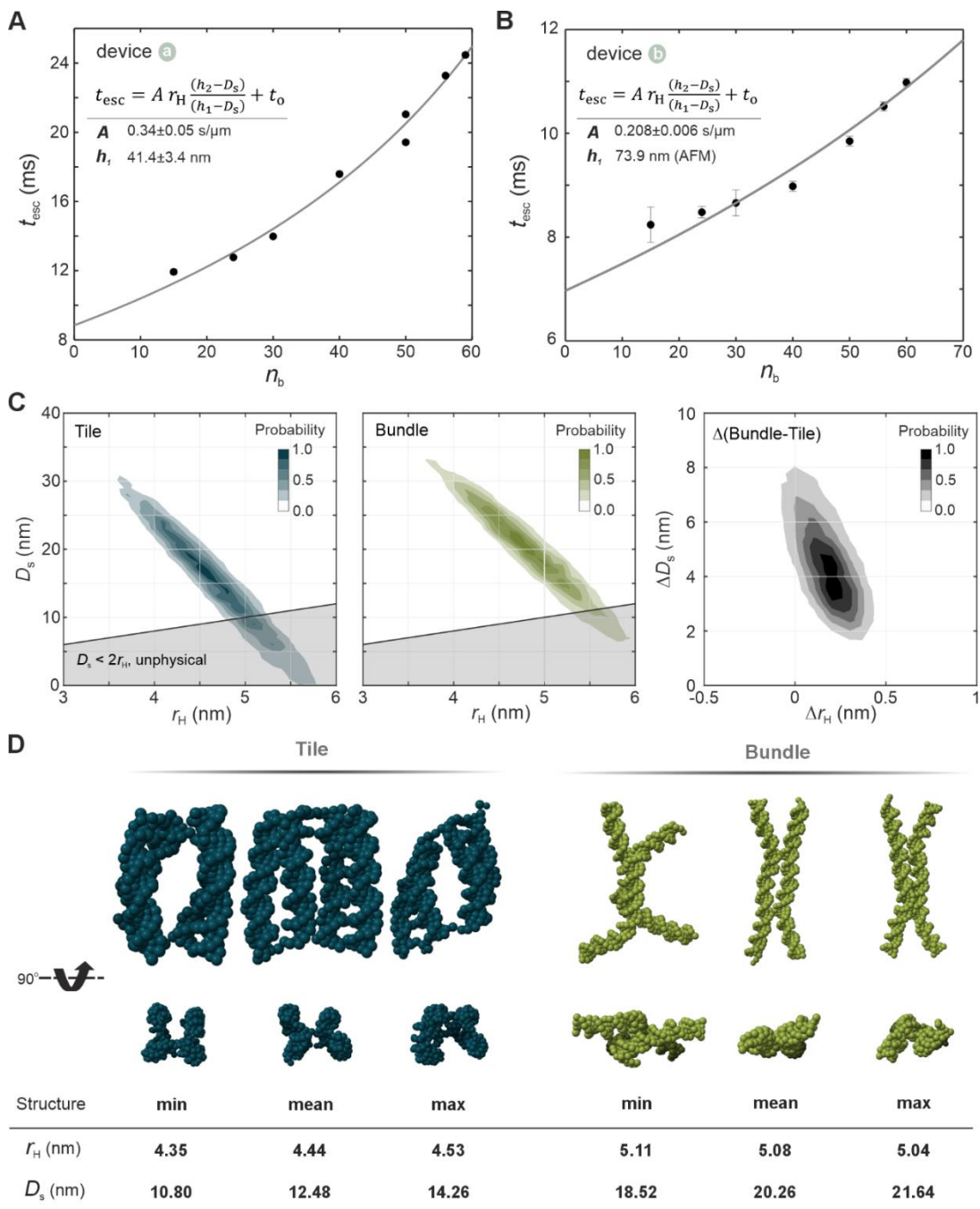

**Fig. S12. Measurement of $D_s$ and $r_H$ for DNA nanostructures.** Calibration curves for two devices (a, b) used for DNA-nanostructure measurements: a series of dsDNA with different number of base pairs, $n_{bp}$, was measured and escape times fitted with Eq. (S10) where $D_s = 0.34 n_{bp}$ $t_o = 5.83$ ms, and $d = h_2 − h_1 = 286$ nm. (**A**) For device (a), $A$ and $h_1$ were determined to be 0.34±0.05 s/μm and 41.4±3.4 nm, respectively. (**B**) For device (b), $A$ was determined to be 0.208±0.006 s/μm and $h_1$ was set to the nominal value as determined by AFM, 73.9 nm. (**C**) Full range of $(D_s, r_H)$ values obtained for the nanostructures accounting for the calibration uncertainties. To do this, we inferred a number of $(D_s, r_H)$ values based on 10000 sets of $h_{1,a}, h_{1,b}$ and $A_a, A_b$ pairs drawn from normal distributions with means and uncertainties following from the calibration shown in (**A, B**). For each set of device parameters, $(D_s, r_H)$ implied by the measured $t_{esc}$ values were determined for both DNA nanostructures Then a histogram was constructed to create the probability density maps in the left and middle panels. The difference between $D_s$ and $r_H$ values for the two species was also calculated and similarly plotted in the right panel. Our definition of $D_s$ requires that for non-spherical objects $D_s > 2r_H$. This excludes $(D_s, r_H)$ solutions in the parameter space indicated by the grey shaded region. (**D**) A series of snapshots from oxDNA simulations of the nanostructures presented in order of decreasing compaction (based on $D_s$ value). $r_H$ values were calculated for each structure using HYDROPRO, while $D_s$ was evaluated using a custom Python script ("miniball").

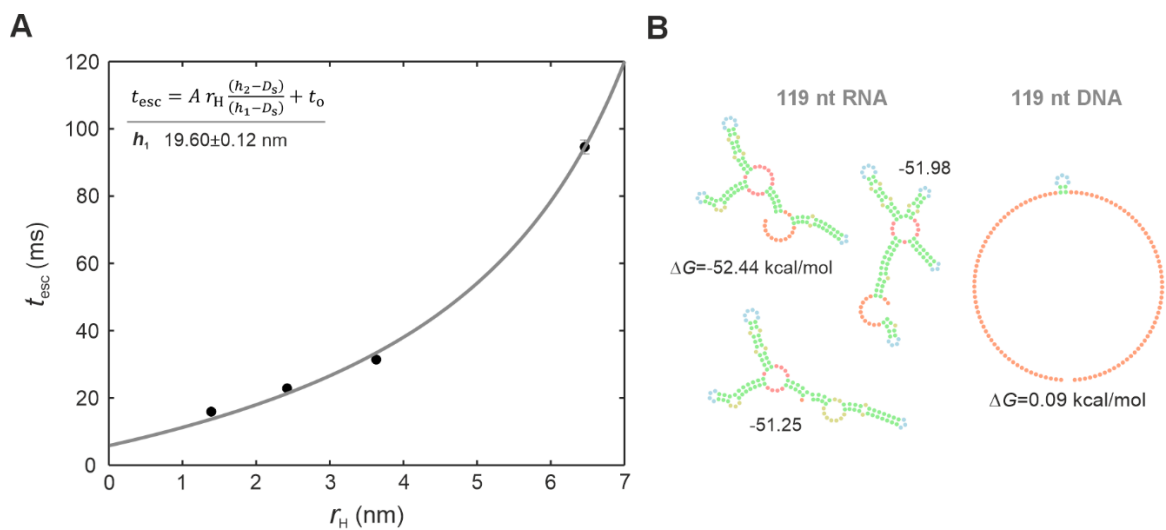

**Fig. S13. Further data for the SAM-IV riboswitch experiment**. (**A**) Calibration curve for the device used to measure the SAM-IV riboswitch: a series of globular proteins (INS, CA, TF, FER) with different $r_H$ values, as determined using HYDROPRO, were measured and escape times fitted with Eq. (S10) where $D_s = 2r_H$, $t_o = 5.83$ ms and $d = h_2 - h_1 = 259$ nm as measured by AFM. $h_1$ was determined to be $19.60 \pm 0.12$ nm. (**B**) The most stable secondary structures predicted by Mfold for the SAM-IV riboswitch and reference sample of 119 nt DNA, and their respective thermodynamic stabilities, $\Delta G$, for 1 M [Na$^+$] at 37°C. Multiple stable secondary structures were observed for the RNA, compared to essentially no secondary structures obtained for the selected DNA sequence. Note that in solution RNA folds into tertiary structures.

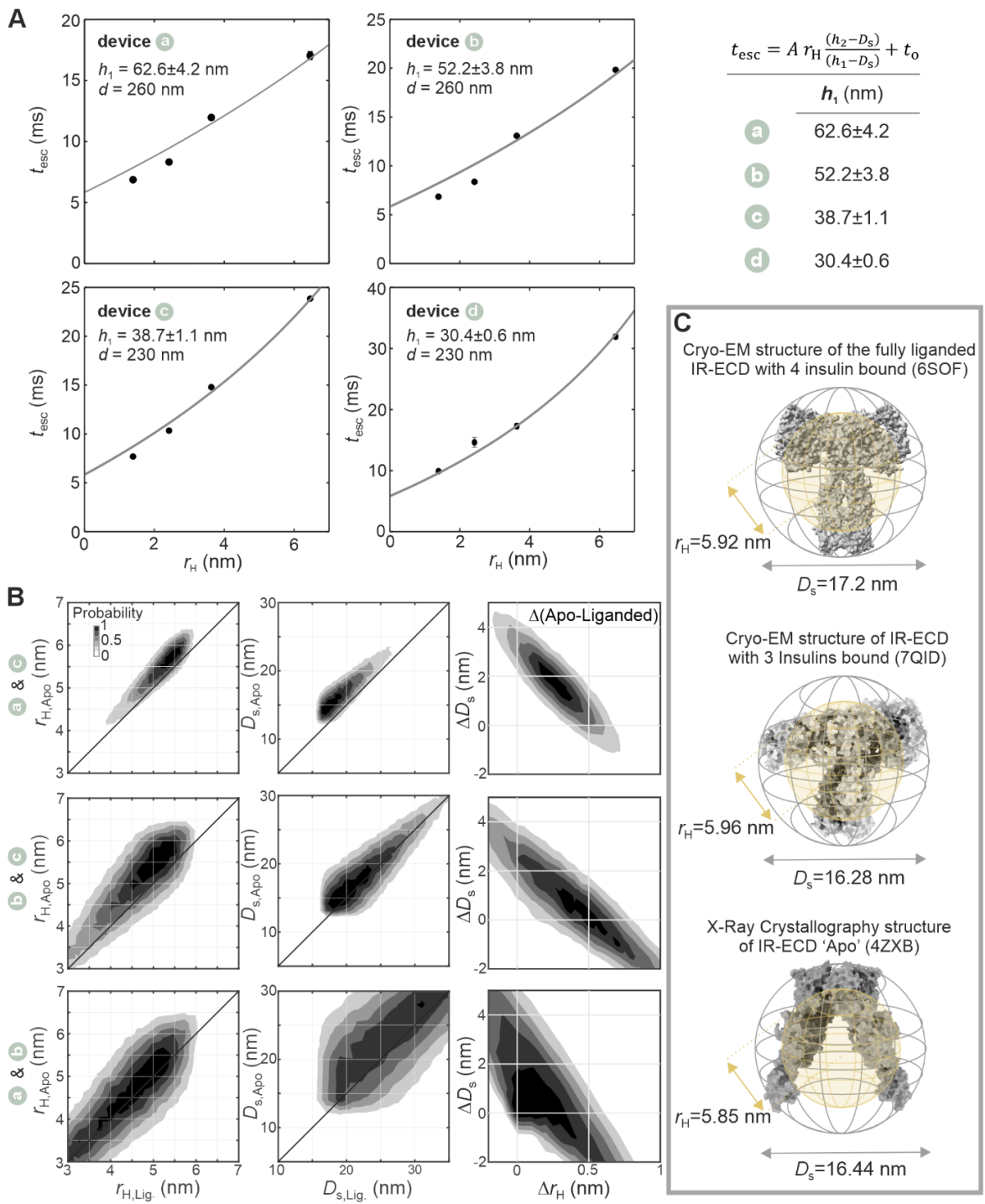

**Fig. S14. Measurement of $D_s$ and $r_H$ of the IR-ECD.** (**A**) Calibration curves for the devices used to measure the IR-ECD: a series of globular proteins (INS, CA, TF, FER) with different $r_H$ values, as calculated using HYDROPRO, were measured and escape times fitted with Eq. (S10) with $D_s = 2r_H$, $t_o = 5.83$ ms and $d = h_2 - h_1$ as measured by AFM. Fitted $h_1$ values are displayed in the figure legends. (**B**) Pairwise determination of $(D_s, r_H)$ values obtained for the IR-ECD for each device pair (a, c), (b, c) and (a, b) accounting for the calibration uncertainties. For each pair of devices $(i, j)$, 10,000 sets of $h_{1,i}, h_{1,j}$ pairs were selected from normal distributions following their respective uncertainties from calibration in (**A**). For each set, $(D_s, r_H)$ was calculated for molecular state (Apo. and Lig.) based on measured $t_{esc}$ values similarly selected from normal distributions following the measurement uncertainties. A histogram was constructed to create the probability density maps in the left and middle panels. The difference between the $D_s$ and $r_H$ values for the two states was also calculated and similarly plotted in the right panel. Note that we only include data which satisfies our physical constraint $D_s > 2r_H$. Regardless of the pair of device geometries selected, we consistently note $D_{s,Apo} - D_{s,Lig.} \approx 1.5$ nm and $r_{H,Apo} - r_{H,Lig.} \approx 0.25$ nm. The pair of geometries (a) and (c) yields the most precise readout of the three geometry-pairs examined as explained in the text (top row). (**C**) Calculated $D_s$ and $r_H$ values for structures reported in literature. From top to bottom: fully liganded (PDB: 6SOF) (*50*), partially liganded (PDB:7QID) (*50*), Apo (PDB:4ZXB) (*54*).

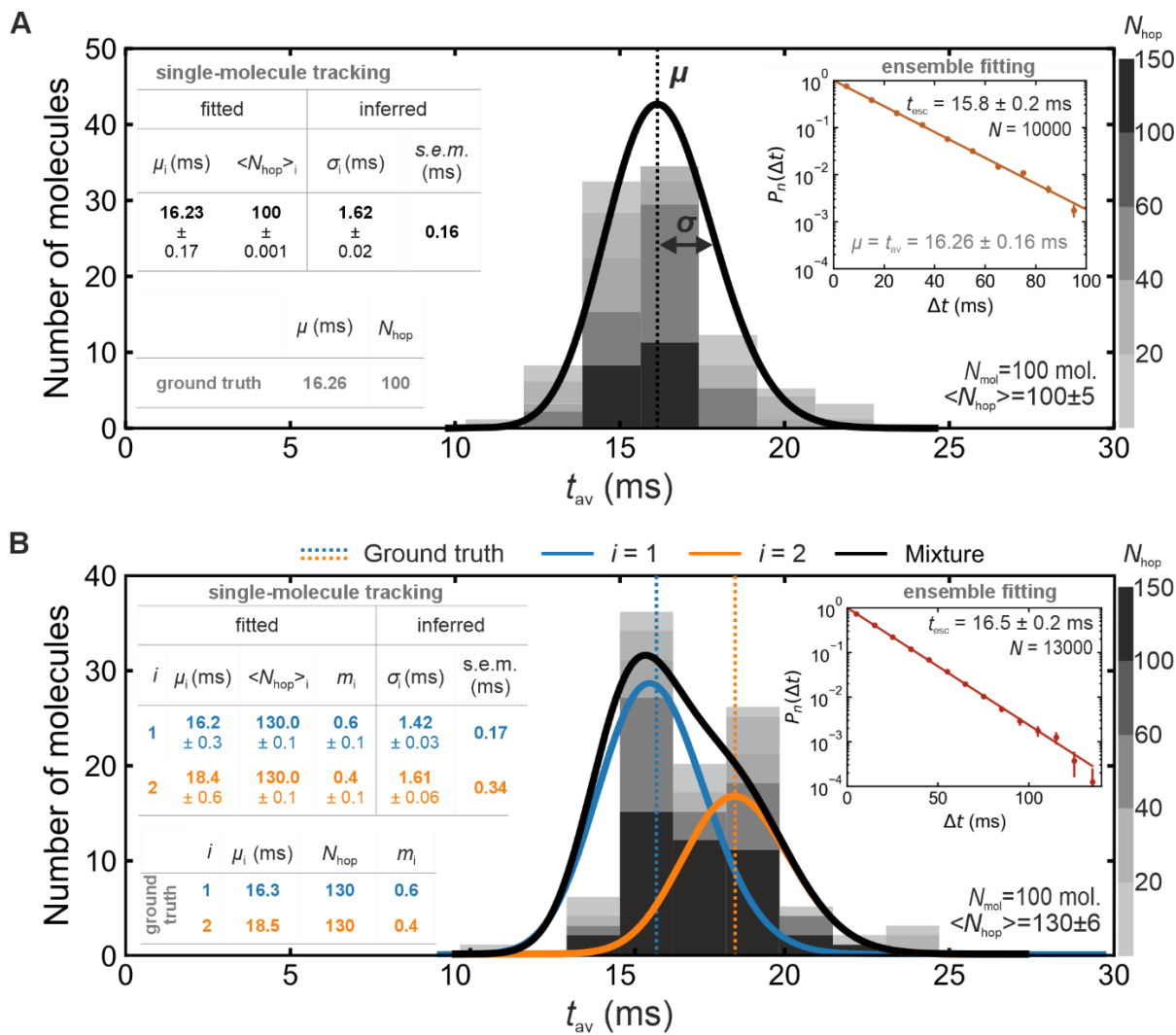

**Fig. S15. (A)** A simulated $t_{av}$ distribution characterized by $N_{mol} = 100$ trajectories consisting of normally distributed number of hops characterized by, $\langle N_{hop} \rangle = 100$, taken from an exponential distribution with mean $\mu$, is histogrammed (grey bars) and fitted with an Erlang distribution (Eq. (S24), black curve). We note consistency in fitted $\mu_1$ and $\langle N_{hop} \rangle_1$ values compared to the ground truth. The fit imprecision on $\mu_1$ is identical to the theoretical limit, i.e., $\sigma_1/\sqrt{N_{mol}}$, and is a factor 10 smaller than the distribution standard deviation, $\sigma_1$ (which governs measurement resolution when the data contain more than one state). There is good agreement between the fitted $\mu_1$ and the average value returned from the ensemble fit. **(B)** A simulated $t_{av}$ distribution characterized by two sets of trajectories (describing a mixture of two species) consisting of normally distributed numbers of hops, $\langle N_{hop} \rangle_i = 100$, taken from an exponential distribution with mean $\mu_1$ (blue curve) and $\mu_2$ (orange curve). The mean escape time per trajectory $t_{av}$, binned into a histogram as before (grey bars) and fitted with an Erlang distribution (Eq. (S24), black curve) with two components as suggested by the peak detection algorithm. The fitted $\mu_i$, $\langle N_{hop} \rangle_i$ and $m_i$ values compare well with the ground truth. Again, the fit uncertainty on the $\mu_i$ compare well with the estimated statistical limit denoted as s.e.m., and is approximately 10 times smaller than the resolution $\sigma_i$. A bi-exponential fit of all the escape events pooled together (ensemble fit – inset) returns a single redundant timescale, thus illustrating the power of the single-molecule analysis.

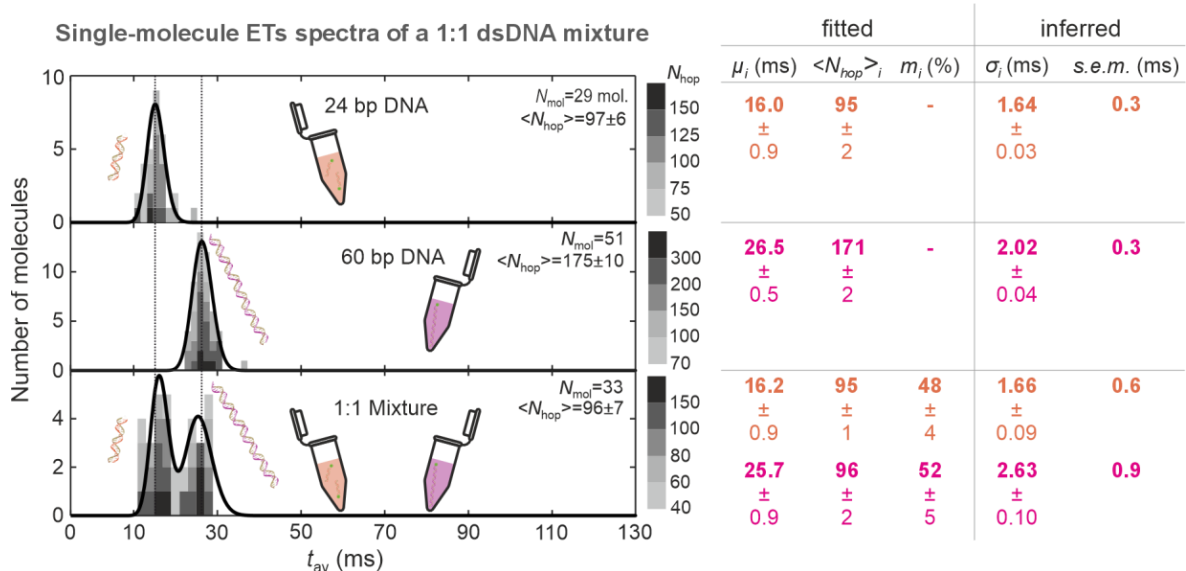

**Fig. S16. Validation of single-molecule spectra measurements using dsDNA.** Single-molecule spectra of $t_{av}$ measurements on individual molecules are constructed for measurements of 24 bp DNA, 60 bp DNA, and a 1:1 mixture of the two. These spectra are fitted with a sum of Erlang distributions (Eq. (S24), black curves), which are in fact well approximated by Gaussian distributions for $N_{hop} \geq 50$ achieved in this study. Fitted parameters for the means, $\mu_i$, average number of events per trajectory, $\langle N_{hop} \rangle_i$, and molecule fractions, $m_i$, are displayed in the table on the right.

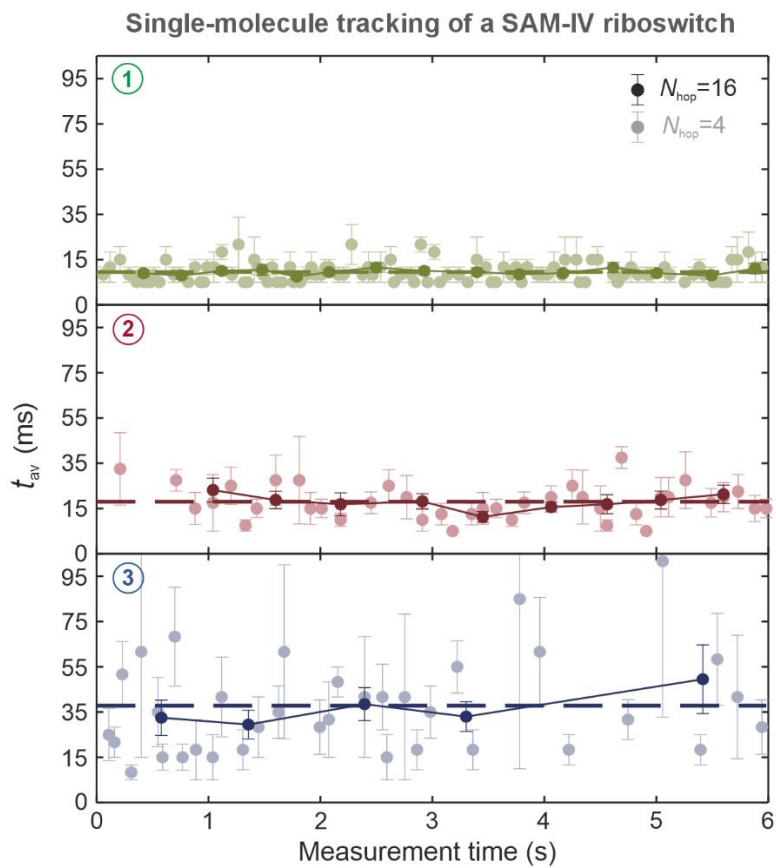

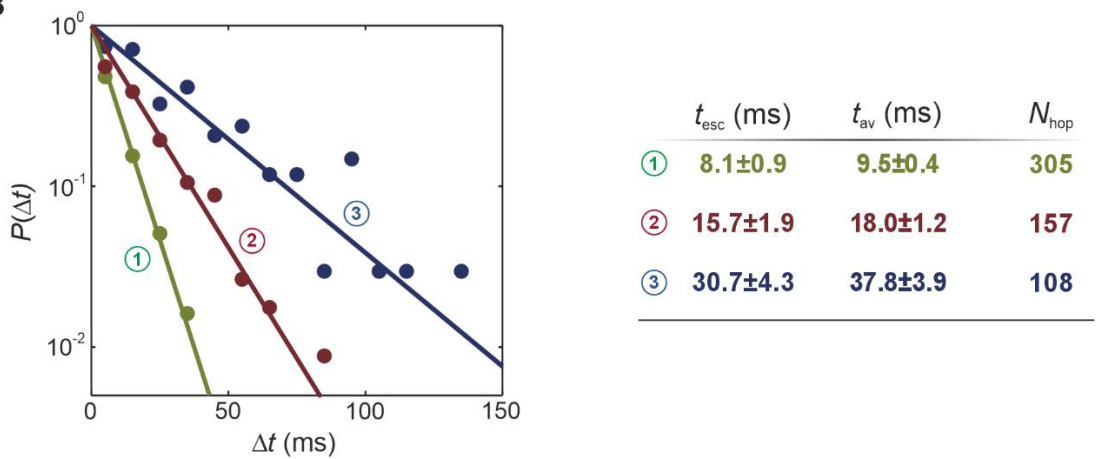

**Fig. S17. Single-molecule tracking of the SAM-IV riboswitch.** (**A**) Recorded time traces of sequential escape events $\Delta t$, averaged over 4 (light symbols) and 16 (dark symbols) consecutive escape events, for three individual molecules in the SAM-IV riboswitch sample, measured in PBS containing 5 mM $MgCl_2$. The temporal fluctuations in average escape times are consistent with statistical expectations over the 6 s of observation time suggesting no detectable conformational fluctuation, or interconversion between the different molecular states. (**B**) Escape time histograms are also constructed from the same single molecule data and fit with a single exponential distribution (left panel), yielding $t_{esc} \approx t_{av}$ for each molecule as expected (right panel).

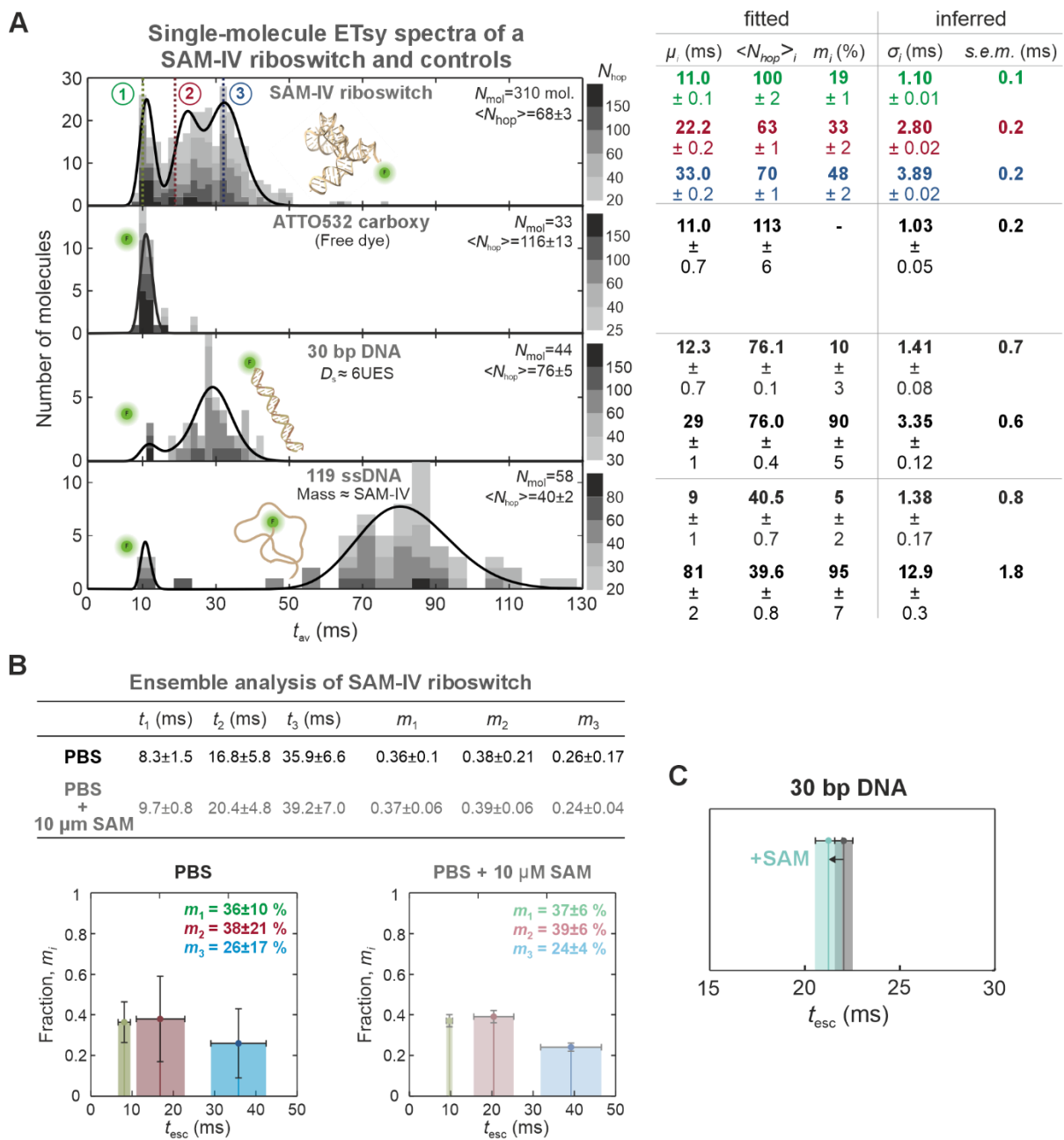

**Fig. S18. Single-molecule measurement of the SAM-IV riboswitch and binding of S-adenosyl methionine (SAM) to the riboswitch.** (**A**) Single molecule spectra of $t_{av}$ are constructed for measurements of a SAM-IV riboswitch, ATTO532 carboxy, 30 bp DNA and 119 ssDNA. These spectra are fitted with a sum of Erlang distributions (Eq. (S24)), which are in fact well approximated by Gaussian distributions for the large $N_{hop}$ values in this study. Fitted parameters for the means, $\mu_i$, average number of events per trajectory, $\langle N_{hop} \rangle_i$, and molecule fractions, $m_i$, are displayed in the table on the right. (**B**) Ensemble analysis of measurements of SAM-IV riboswitch incubated in PBS with and without 10 μM SAM. Escape time histograms were fitted with tri-exponential distributions (fit values in table, top panel) and the obtained timescales and molecule fractions are plotted. We do not observe significant shifts in the timescales $t_2$ and $t_3$, characteristic of the riboswitch, or their respective molecule fractions upon the addition of SAM. (**C**) As a control experiment, 30 bp DNA was also compared in the different buffers, displaying, as expected, a negligible change in $t_{esc}$ with the addition of SAM.

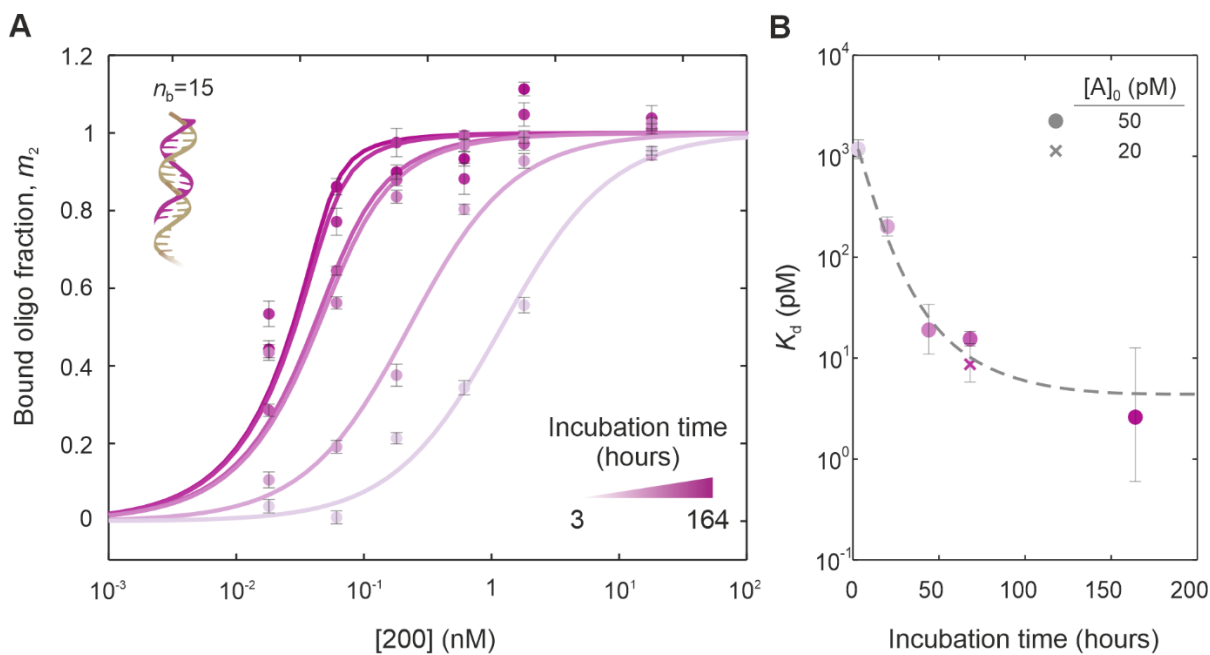

**Fig. S19. Assessment of the attainment of equilibrium in the measurement of $K_d$ for a 15 base ssDNA oligo binding to 200 base ssDNA.** (**A**) Titration curves bound-fraction for a labelled sample of 15 base ssDNA incubated with complementary 200 base ssDNA for varying lengths of time at room temperature before measurement. In each case, the $m_2$ values were determined from bi-exponential fitting and fit with Eq. (S29) to determine the $K_d$ (panel **B**). We approach saturation of the measured $K_d$ after ~1 week. However, since in this case, the final inferred $K_d \approx 2.5$ pM is not accurate as it is much smaller than $[A]_0 = 50$ pM, the initial concentration of 15 ssDNA added. In Fig. 5 from the main text, we report the value obtained after 68 hours of incubation with $[A]_0 = 20$ pM (cross symbol). Dashed line in panel **B** is a guide to the eye.

**Fig. S20. Measurement of on and off-rates and assessment of the role of cooperative unbinding in $k_{\text{off}}$ measurements.** (**A**) Left - Measurements of $m_2$ vs. reaction time, $t$, performed with 0.5 nM of labelled 10 ssDNA oligo with increasing concentrations of complementary 200 ssDNA are fit with Eq. (S30) to obtain the equilibrium rate constant $k_{\text{eq}}$. Inset shows that $k_{\text{eq}}$ increases proportionately with concentration of 200 ssDNA. Right – Equilibrated, fully bound mixtures of 0.5-2 nM 9-11 ss oligo and 200 ssDNAs at various concentrations are mixed with 0.5-5 µM of unlabelled 11 ssDNA oligo, and the decaying bound fraction, $m_2$, measured as a function of time to yield the cooperative rate constant of dissociation, $k_{\text{off}}$, from a fit to the data with Eq. (S32). $k_{\text{off}}$ can vary by more than an order of magnitude for single-base differences. $k_{\text{eq}}$ and $k_{\text{off}}$ together determine the rate constant of association, $k_{\text{on}}$ (left inset), from which $K_d$ can be determined independently. (**B**) $k_{\text{off}}$ was measured for 10 ssDNA equilibrated with complementary 200 ssDNA using a variety of different chaser (unlabelled 11 ssDNA) concentrations. Grey datapoints show the data as shown in (**A**), with mean value and standard error on the mean displayed by the dashed line and shaded region. Here, $[A]_0 = 0.5$ nM, $[200\text{ss}] = 10 - 100$ nM and $[\text{chaser}] = 500$ nM. Red and blue datapoints show $k_{\text{off}}$ measurements with $[A]_0 = 1$ nM, $[200\text{ss}] = 20$ and 200 nM, respectively, and $[\text{chaser}] = 80 - 24{,}000$ nM. Varying the ratio of $[\text{chaser}]/[200\text{ss}]$ has negligible effect on $k_{\text{off}}$, until $[\text{chaser}]/[200\text{ss}] \geq 100$, which is outside the range of the data used to determine the $k_{\text{off}}$ in this work.

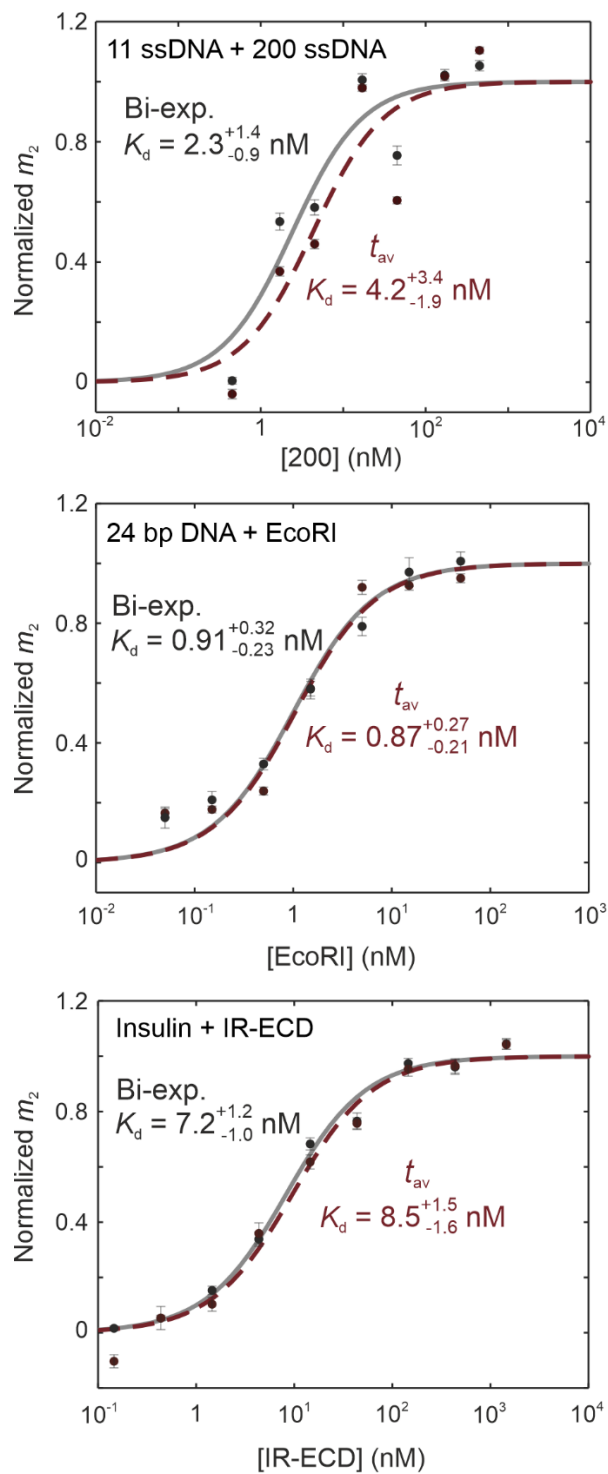

**Fig. S21. Comparison of $K_d$ measurements using bi-exponential and $t_{av}$ fitting methods to determine $m_2$.** Identical datasets of affinity measurements for three different systems (top: 11 ssDNA + 200 ssDNA, middle: 24 bp DNA + EcoRI, bottom: Insulin + IR-ECD) were analyzed using both methods to extract $m_2$. When fitted with Eq. (S29), we obtain $K_d$ values that are consistent within error when extracting $m_2$ using bi-exponential fitting (grey, solid curve), and $t_{av}$ (red, dashed curve), for all three systems. This validates the method for use on systems where bi-exponential fitting is challenging.

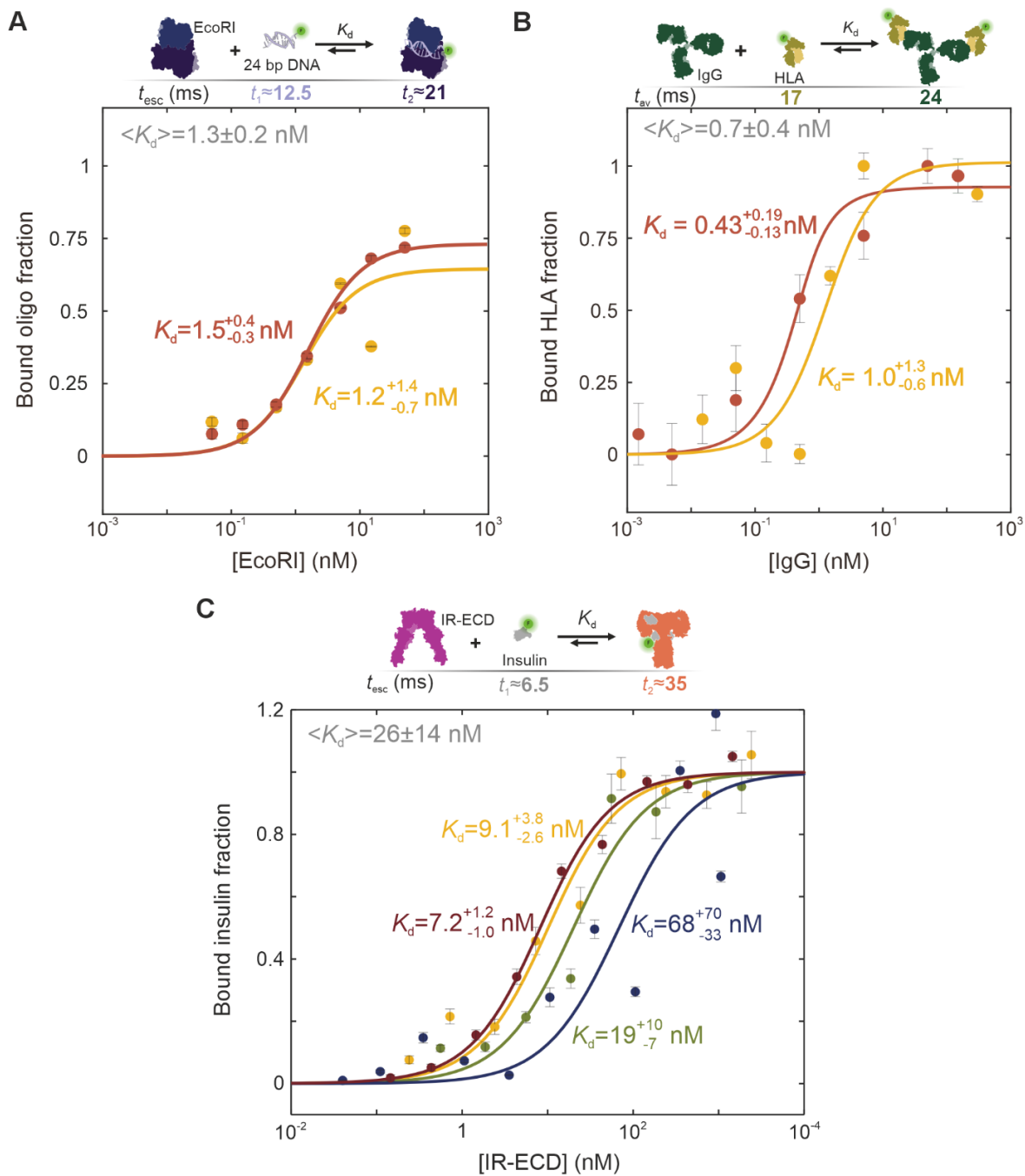

**Fig. S22. Measuring $K_d$ for the interactions of EcoRI, HLA-IgG and IR-ECD and their binding partners.** Data for 2-4 repetitions of affinity measurements for a labelled sample of (**A**) 24 bp DNA incubated with EcoRI, (**B**) HLA antigen incubated with corresponding IgG and (**C**) insulin incubated with the IR-ECD are shown. In each repetition, 0.25 nM DNA was incubated with varying concentrations of EcoRI in PBS at room temperature for 2 hours before measuring, 0.1 nM HLA was incubated with varying concentrations of IgG at room temperature for 1 hour before measuring, and 2 nM insulin was incubated with varying concentrations of IR-ECD in HBS at 4°C overnight before measuring. (**A, C**) The $m_2$ values extracted from bi-exponential fitting were fit with Eq. (S29), and a mean and standard error on the mean of 2 and 4 repeats gives a $K_d = 1.3 \pm 0.2$ nM and $K_d = 26 \pm 14$ nM, respectively. In the main text Fig. 6, we report the lowest $K_d$ values obtained, 1.2 nM and 7.2 nM, respectively. (**B**) The $m_2$ values extracted from $t_{av}$ were fit with Eq. (S29), and a mean and standard error on the mean of both repeats gives a $K_d = 0.7 \pm 0.4$ nM. In the main text Fig. 6, we report the lowest $K_d$ value obtained, 0.43 nM.

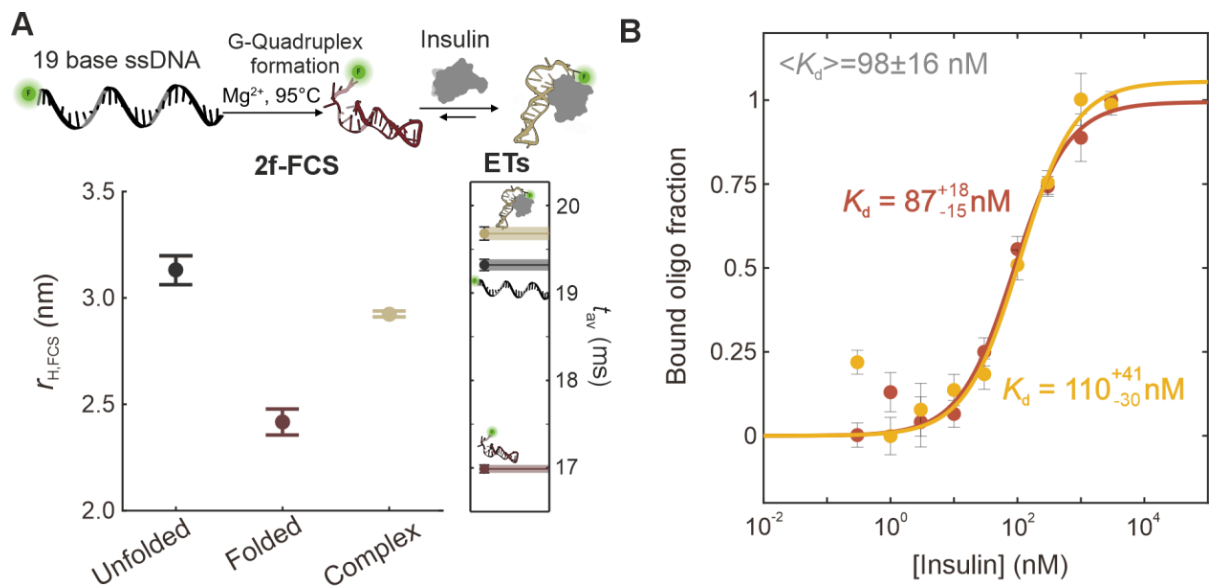

**Fig. S23. Measuring $K_d$ for the binding of insulin to an Insulin aptamer.** **(A)** 2f-FCS measurements of the insulin aptamer in unfolded, folded and insulin-bound states displaying a similar general trend as that measured in ETs (right panel). **(B)** Two repetitions of affinity measurements for a labelled sample of folded aptamer incubated with insulin are shown. In each case, 1 nM of fluorescently labelled aptamer was incubated with varying concentrations of insulin in 10 mM $MgCl_2$ PBS at room temperature for 1 hour before measuring. The $m_2$ values extracted from $t_{av}$ were fit with Eq. (S29), and a mean and standard error on the mean of both repeats gives a $K_d = 98 \pm 16$ nM. In main text Fig. 6, we report the lowest $K_d$ value obtained, 87 nM.

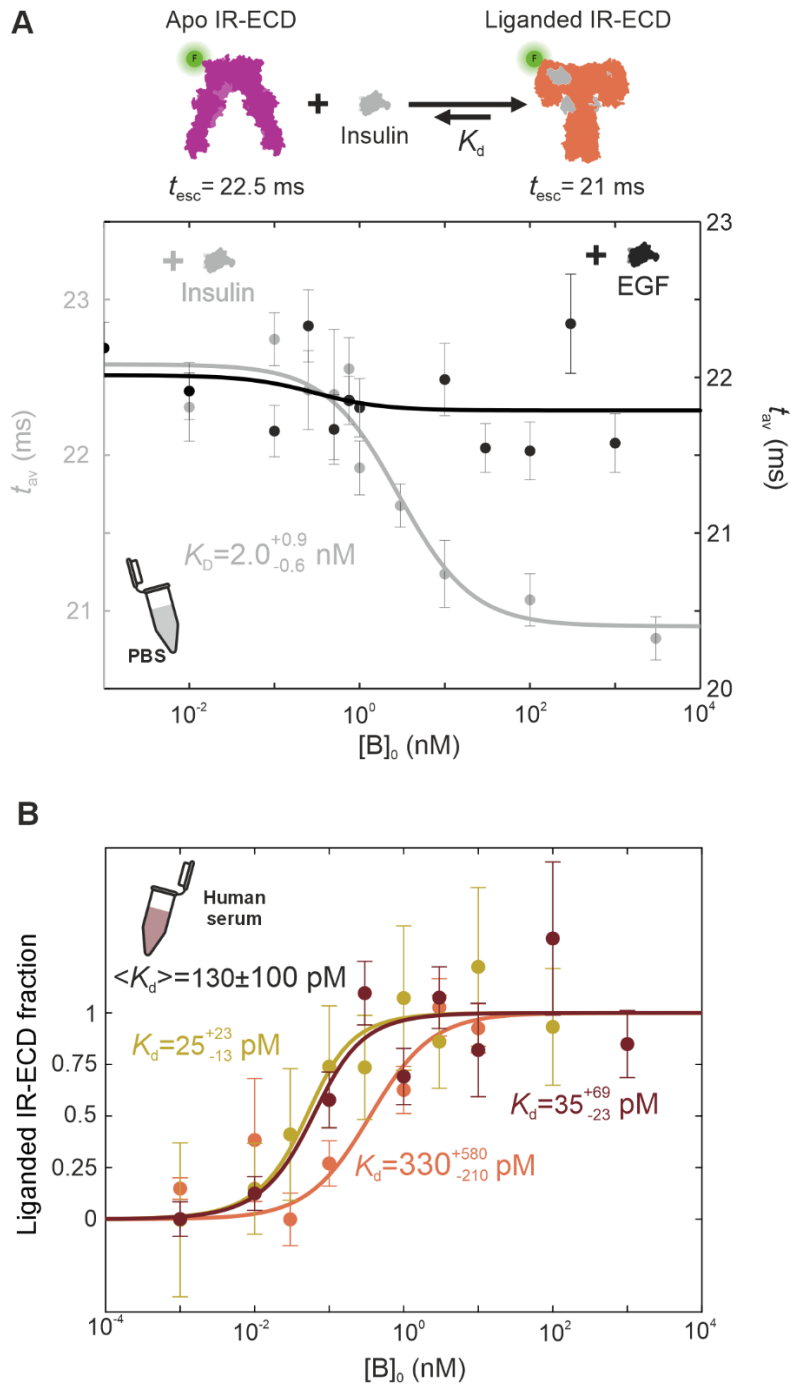

**Fig. S24. Measuring $K_d$ for the interaction of IR-ECD with insulin.** (**A**) Affinity measurement for the interaction of labelled IR-ECD with insulin (grey data), or epidermal growth factor (EGF) (black data). 0.08 nM of IR-ECD was incubated with varying concentrations of insulin or EGF in PBS with 0.03% Tween20 at room temperature for 1 hour before measuring. $m_2$ values were determined from $t_{av}$ and fit using Eq. (S29). We observe negligible change in $t_{av}$ upon addition of EGF showing the specificity of the IR-ECD to insulin. (**B**) Three repetitions of affinity measurements for the interaction of labelled IR-ECD with insulin in human serum. 0.05 nM of IR-ECD was incubated with varying concentrations of insulin with 0.03% Tween20 at room temperature for 1 hour before measuring. $m_2$ values were determined from $t_{av}$ and fit using Eq. (S29), and a mean and standard error on the mean of the three repeats gives a $K_d = 130 \pm 100$ pM. This value is a factor 10 lower than that measured in PBS or simulated serum.

## S9. Supplementary Tables

**Table S1** Nucleic acid sequences with color-coded complementary ssDNA sequences.

| Sample | Sequence (5' to 3') | Vendor | HPLC purified? |
|---|---|---|---|
| Insulin aptamer | GGT GGT GGG GGG GGT TGG TAG GGT GTC TTC T /3ATTO532N/ | IDT | Single |
| 15bp DNA | /5ATTO532N/TAG GAT CCA TAG TAG<br>CTA CTA TGG ATC CTA | IDT | Single |
| 24bp DNA | /5ATTO532N/ TAG GAT CCA TAG AAT TCA GCG TAG<br>/5ATTO532N/ CTA CGC TGA ATT CTA TGG ATC CTA | IDT | Double |
| 30bp DNA | /5ATTO532N/ GGA TGG GAC GGA CCC GGA CAC AGA CAG TGC<br>/5ATTO532N/ GCA CTG TCT GTG TCC GGG TCC GTC CCA TCC | IDT | Double |
| 40bp DNA | /5ATTO532N/ CGC TCA AGG TGG ATG GGA CGG ATT CGG ACA TAG ACA GTG C<br>/5ATTO532N/ GCA CTG TCT ATG TCC GAA TCC GTC CCA TCC ACC TTG AGC G | IDT | Double |
| 50bp DNA | /5ATTO532N/ CGG TCC AGT TGC ACT GTC TAT GTC CGA ATC CGT CCC ATC CAC CTT GAG CG<br>/5ATTO532N/ CGC TCA AGG TGG ATG GGA CGG ATT CGG ACA TAG ACA GTG CAA CTG GAC CG | IDT | Double |
| 56bp DNA | /5ATTO532N/ TA GAA CTA GTG GAT CCC CCG GGC TGC AGG AAT TCG ATA TCG CAA TGT ATA GCA ACT<br>/5ATTO532N/ A GTT GCT ATA CAT TGC GAT ATC GAA TTC CTG CAG CCC GGG GGA TCC ACT AGT TCT A | Microsynth | Double |
| 57bp DNA | /5ATTO532N/ TA GAA CTA GTG GAT CCC CCG GGC TGC AGG AAT TCG ATA TCG CAA TGT ATA GCA ACT T<br>/5ATTO532N/ AA GTT GCT ATA CAT TGC GAT ATC GAA TTC CTG CAG CCC GGG GGA TCC ACT AGT TCT A | IDT | Double |
| 58bp DNA | /5ATTO532N/TAG AAC TAG TGG ATC CCC CGG GCT GCA GGA ATT CGA TAT CGC AAT GTA TAG CAA CTT G<br>/5ATTO532N/ CAA GTT GCT ATA CAT TGC GAT ATC GAA TTC CTG CAG CCC GGG GGA TCC ACT AGT TCT A | IDT | Double |
| 59bp DNA | /5ATTO532N/ TAG AAC TAG TGG ATC CCC CGG GCT GCA GGA ATT CGA TAT CGC AAT GTA TAG CAA CTT GG<br>/5ATTO532N/ CCA AGT TGC TAT ACA TTG CGA TAT CGA ATT CCT GCA GCC CGG GGG ATC ACT AGT TCT A | IDT | Double |
| 60bp DNA | /5ATTO532N/ TAG AAC TAG TGG ATC CCC CGG GCT GCA GGA ATT CGA TAT CGC AAT GTA TAG CAA CTT GGG<br>/5ATTO532N/ CCC AAG TTG CTA TAC ATT GCG ATA TCG AAT TCC TGC AGC CCG GGG GAT CCA CTA GTT CTA | IDT | Double |
| 30bp RNA | /5ATTO532N/ GGA UGG GAC GGA CCC GGA CAC AGA CAG UGC<br>/5ATTO532N/ GCA CUG UCU GUG UCC GGG UCC GUC CCA UCC | IDT | Double |
| 40bp RNA | /5ATTO532N/ CGC UCA AGG UGG AUG GGA CGG AUU CGG ACA UAG ACA GUG C<br>/5ATTO532N/ GCA CUG UCU AUG UCC GAA UCC GUC CCA UCC ACC UUG AGC G | IDT | Double |
| 50bp RNA | /5ATTO532N/ CGC UCA AGG UGG AUG GGA CGG AUU CGG ACA UAG ACA GUG CAA CUG GAC CG | IDT | Double |
| | /5ATTO532N/ CGG UCC AGU UGC ACU GUC UAU GUC CGA AUC CGU CCC AUC CAC CUU GAG CG | | |
| 60bp RNA | /5ATTO532N/ UAG AAC UAG UGG AUC CCC CGG GCU GCA GGA AUU CGA UAU CGC AAU GUA UAG CAA CUU GGC /5ATTO532N/ GCC AAG UUG CUA UAC AUU GCG AUA UCG AAU UCC UGC AGC CCG GGG GAU CCA CUA GUU CUA | IDT | Double |
| 5ss DNA | /5ATTO532N/TAGAA | IBA | Double |
| 7ss DNA | TTGTGAG /3ATTO532N/ | IDT | Double |
| 8ss DNA | GTTGTGAG /3ATTO532N/ | IDT | Double |
| 9ss DNA | TGTTGTGAG /3ATTO532N/ | IDT | Double |
| 10ss DNA | C TGT TGT GAG /3ATTO532N/ | IDT | Double |
| 11ss DNA | CC TGT TGT GAG /3ATTO532N/ | IDT | Double |
| 11ss DNA chaser | CC TGT TGT GAG | IDT | No |
| 200ss DNA comp to 7-11 | CTC ACA ACA GG CTACTTACA ATGAAGATCA GGAACCCCCG GGGTGCAGGA ATTCGATATC GCAATGTATA GCAACTTGGC TAGAAGTAGT GGATCCCCCG GGCTGCAGGA ATTCGATATC GCAATGTATA GCAACTTGGC TAGAAGTAGT GGATCCCCCG GGCTGCAGGA ATTCGATATC GCAATGTATA GCAACTTGGC | IDT | No |
| 15ss DNA | /5Alexa532/ TAG GTT CCA TAG TAG | IDT | Single |
| 200ss DNA comp to 15ss | CTA CTA TGG AAC CTA TTACA ATGAACATCT CGAACCCCCG GGCTGCAGGA ATTCGATATC GCAATGTATA GCAACTTGGC TAGAAGTAGT GGATCCCCCG GGCTGCAGGA ATTCGATATC GCAATGTATA GCAACTTGGC TAGAAGTAGT GGATCCCCCG GGCTGCAGGA ATTCGATATC | IDT | No |
| 119ss RNA | /5ATTO532N/ GGUCAUGAGU GCCAGCGUCA AGCCCCGGCU UGCUGGCCGG CAACCCUCCA ACCGCGGUGG GGUGCCCCGG GUGAUGACCA GGUUGAGUAG CCGUGACGGC UACGCGGCAA GCGCGGGUC | N/A | Single |
| 119ss DNA | /5ATTO532N/ TATAACTAAT GTATCATCCT AACTGCAGTA ACTCTTTATC GCAATGTATA TCAACTTGTC TAGAACTACC TAATCTCTCG TATTGAATGA ATTTTATATC TTAATGTATA GCAACTTAC | N/A | Single |

**Table S2** Disordered Protein sequences.

| Protein | Sequence | Molecular Weight (kDa) |
|---|---|---|
| 13-mer proline rich polypeptide | PPPPPPPD PPPPPPPD PPPPPPPD PPPPPPPD PPPPPPPD PPPPPPPD PPPPPPPD PPPPPPPD PPPPPPPD PPPPPPPD PPPPPPP | 9.44 |
| ProTα | AHHHHHHSAA LEVLFQGPSD AAVDTSSEIT TKDLKEKKEV VEEAENGRDA PANGNAENEE NGEQEADNEV DEECEEGGEE EEEEEEGDGE EEDGDEDEEA ESATGKRAAE DDEDDDVDTK KQKTDEDC | 14.02 |
| Stm-l | GSMKNSDDESKE TSKCAAEVKT TEKTAENPDA TDEPSEETSE GEELDSADDS QDSTDEPSQE TAETKETESC DKSEDKSSAA EETDEEASDS ADVSEAVSES EETSSSPEVR GSKTNNTEDK ISFEEVEDQS DEMSMDKDKK SEESSMRPTD SVKGAEDSAS LESDEMFEES DEQSEEEKST SAPTSEDAGA DSEDESDQSS QDLSKEDDSK GQDAVQSDD GTTSKADLMD DDLGAEDDKD EEKEEAADGD AKDLGVMIEK DEEEKVEVNK EQVDSNEDQP AKENSDEKDE ADKEKDASDK KDLSMSEEEP DWEEEMGMDD SLEEEEKANK KTSDEKGSFD ETTSDQLDQP DNSTPPADAA M | 39.58 |

**Table S3** Details of protein labeling.

| Protein | Molar excess (dye) | Degree of labeling (DOL) | Post-labeling concentration after purification [μM] |
|---|---|---|---|
| INS | 1.5x | 0.44 | 15.5 |
| Ub | 2.5x | 0.48 | 177.5 |
| TRX | 2.5x | 0.36 | 13.56 |
| RNase A | 2.5x | 0.55 | 162.67 |
| MB | 2.5x | 0.49 | 115.59 |
| LGB | 2.5x | 1.08 | 76.99 |
| CA | 2.5x | 1.1 | 24.2 |
| TF | 2.5x | 2.4 | 14.5 |
| FER | 5x | 0.7 | 22 |
| IgG | 5x | 3 | 4.3 |
| HLA | 3x | 2.2 | 15.8 |

**Table S4** Measured and structural estimates of Stokes' radii, $r_{H,ETs}$, $r_{H,FCS}$ and $r_H$ of proteins. Molecular weight values include one ATTO532 dye (0.64 kDa), while PDB files used to estimate $r_H$ did not take the dye into consideration.

| Molecular Weight (kDa) (+ ATTO532) |سample | PDB ID | $r_H$ (nm) | $r_{H,ETs}$ (nm) | $r_{H,FCS}$ (nm) |
|---|---|---|---|---|---|
| 6.44 | INS | 3I40 | 1.455 | 1.32 ± 0.02 | 1.1 ± 0.2 |
| 9.13 | Ub | 3H1U | 1.690 | 1.46 ± 0.04 | 1.5 ± 0.2 |
| 12.29 | TRX | 1F6M | 1.797 | 1.83 ± 0.02 | 1.1 ± 0.1 |
| 14.32 | RNase A | 1FS3 | 1.954 | 2.03 ± 0.02 | 2.2 ± 0.2 |
| 17.62 | MB | 1DWR | 2.102 | 2.17 ± 0.04 | 1.4 ± 0.1 |
| 19.24 | LGB | 6GE7 | 2.186 | 2.29 ± 0.04 | 2.4 ± 0.1 |
| 32.07 | CA | 4CNX | 2.459 | 2.35 ± 0.06 | 2.5 ± 0.1 |
| 75.77 | TF | 3QYT | 3.647 | 3.69 ± 0.04 | 2.8 ± 0.9 |
| 465.18 | FER | 4V1W | 6.435 | 6.68 ± 0.06 | 4.5 ± 0.1 |
| 202.65 | IR-ECD (Apo) | 4ZXB | 5.85 | 5.70 | 4.8 ± 0.7 |
| 226.08 | IR-ECD (Liganded) | 6SOF | 5.92 | 5.45 | 5.1 ± 0.3 |

**Table S5** Measured Stokes' radius, $r_{H,FCS}$, of fluorescently labeled nucleic acids using dual-focus fluorescence correlation spectroscopy (2f-FCS), and their respective calculated values from cylindrical model (dsRNA & dsDNA) or oxDNA (tile & bundle)

| Sample | base pairs | $r_{H,FCS}$ (nm) | $r_H$ (nm) |
|---|---|---|---|
| dsRNA | 30 | 2.13 ± 0.04 | 2.28 |
| | 40 | 2.38 ± 0.01 | 2.63 |
| | 50 | 2.63 ± 0.03 | 2.99 |
| | 60 | 3.04 ± 0.02 | 3.34 |
| dsDNA | 30 | 2.64 ± 0.05 | 2.52 |
| | 40 | 3.03 ± 0.05 | 2.97 |
| | 50 | 3.50 ± 0.03 | 3.42 |
| | 56 | 3.82 ± 0.04 | 3.68 |
| | 57 | 4.04 ± 0.02 | 3.73 |
| | 58 | 3.90 ± 0.04 | 3.77 |
| | 59 | 4.04 ± 0.05 | 3.82 |
| | 60 | 3.99 ± 0.09 | 3.86 |
| Tile | 240 | 6.7 ± 0.2 | 4.39 ± 0.08 |
| Bundle | 240 | 6.6 ± 0.1 | 5.02 ± 0.05 |

## S10. Supplementary Movies

### Movie S1.
A representative movie of an ensemble-averaging ETs experiment, overlaid on a scanning electron micrograph of the nanostructured system of slits. A solution of 1 nM of 60 base-pair DNA in PBS was measured under vacuum-driven flow (right to left). The raw movie was recorded at 100 Hz and is exported at 20 Hz for visualization here. Pixel intensities are binarized and presented in false colour.

### Movie S2.
A representative movie of a single-molecule ETs experiment, overlaid on a scanning electron micrograph of the nanostructured system of slits, with single-molecule trajectories shown (coloured tracks). A 100 pM solution of the SAM-IV riboswitch (in PBS + 5 mM $MgCl_2$) was measured under vacuum-driven flow (left to right). The raw movie was recorded at 100 Hz and is exported at 20 Hz for visualization and presented in false colour. Coloured trajectories link residence events of individual molecules in pockets, and are filtered further to exclude any non-lattice localizations before $t_{av}$ determined for each trajectory.